\documentclass[aps,prb,print,twocolumn,groupedaddress,showkeys]{revtex4-2}
\usepackage{graphicx}
\usepackage{amssymb}
\usepackage{amsfonts}
\usepackage{amsmath}
\usepackage{natbib}
\usepackage{hyperref}
\usepackage{color}
\usepackage{bm}
\usepackage{mathrsfs}
\usepackage{amsthm}

\begin{document}

\title{Nonlinear topological edge states, topological gap solitons, and self-induced topological edge states in 
nonlinear Su-Schrieffer-Heeger circuit lattices}
\author{Rujiang Li$^{1}$}
\thanks{Corresponding author: {rujiangli@xidian.edu.cn}}
\author{Wencai Wang$^{1}$}
\author{Xiangyu Kong$^{1}$}
\author{Ce Shang$^{2}$}
\author{Yongtao Jia$^{1}$}
\author{Gui-Geng Liu$^{3,4}$}
\thanks{Corresponding author: {liuguigeng@westlake.edu.cn}}
\author{Huibin Tao$^{5}$}
\author{Ying Liu$^{1}$}
\author{Baile Zhang$^{6,7}$}
\thanks{Corresponding author: {blzhang@ntu.edu.sg}}

\affiliation{$^1$National Key Laboratory of Radar Detection and Sensing, School of Electronic 
Engineering, Xidian University, Xi'an 710071, China}

\affiliation{$^2$Aerospace Information Research Institute, Chinese Academy of Sciences, Beijing 100094, China}

\affiliation{$^3$Research Center for Industries of the Future, Westlake University, Hangzhou, 310030, China}

\affiliation{$^4$Department of Electronic and Information Engineering, School of Engineering, Westlake University,
Hangzhou, 310030, China}

\affiliation{$^5$School of Microelectronics, Xi'an Jiaotong University, Xi'an, China}

\affiliation{$^6$Division of Physics and Applied Physics, School of Physical and Mathematical Sciences, 
Nanyang Technological University, 21 Nanyang Link, Singapore 637371, Singapore}

\affiliation{$^7$Centre for Disruptive Photonic Technologies, The Photonics Institute, Nanyang Technological
University, 50 Nanyang Avenue, Singapore 639798, Singapore}

\begin{abstract}

Topological edge states typically arise at the boundaries of topologically nontrivial structures or at interfaces between regions with 
different topological invariants. When topological systems are extended into the nonlinear regime, linear topological edge states bifurcate into nonlinear counterparts, and topological gap solitons emerge in the bulk of the structures. 
Extensive studies of nonlinear topological edge states and topological gap solitons have been carried out. Following recent experimental observations
in photonic systems, we leverage the strong and tunable nonlinearity of electric circuits and systematically investigate the localized states in 
nonlinear Su-Schrieffer-Heeger (SSH) circuit lattices. Besides revisiting the nonlinear topological edge states and topological gap solitons, we uncover a 
new type of self-induced topological edge states which exhibit the hallmark features of linear topological edge states, 
including sublattice polarization, phase jumps, and decaying tails that approach zero. A distinctive feature of these states is the boundary-induced power threshold for existence. Our work unveils new opportunities for exploring novel nonlinear topological states, and paves the way for the development of nonlinear topological circuits.

\end{abstract}

\maketitle

\section{Introduction}

Topological insulators are physical structures that behave as conventional insulators in the bulk, but conducting
on their surfaces due to the existence of topologically protected edge states \cite{RMP82-3045,RMP83-1057,QF2-22}. 
The realization of topological
insulators has been demonstrated in diverse physical platforms, and the immunity of topological edge states to
local deformations and disorders are of great significance for the exciting potential applications \cite
{NRP4-184,QF3-21,NRP1-281,NRM7-974,RMP91-015005,NRP5-483,NRP3-520,
RMP96-021002,nphoton8-821,nphoton11-763,PQE55-52,RMP91-015006,LSA9-1,PR1093-1,arXiv:2502.18563,QF1-10,QF3-26}. 
Extending topological insulators into the nonlinear regime such as by considering the nonlinear response of optical materials
under high field intensity \cite{APR7-021306,NP20-905}, the interplay between
topology and nonlinearity leads to the formation of the nonlinear topological
edge states which bifurcate from the linear edge states
and inherit topological protection from the linear counterparts \cite{PRA90-023813,PRA94-021801,optica3-1228,
PRL119-253904,PRL121-163901,PRB102-115411,OL45-6466,nphys18-678,PRL128-093901,PRE104-054206,PRB104-235420,
LSA9-147,science372-72,CP8-342}. In two-dimensional topological systems, due to the modulation instability,
the nonlinear edge states develops into the edge solitons where dispersion is
balanced with nonlinearity \cite{PRA90-023813,PRA94-021801,optica3-1228,PRL117-143901,
ACSPhoton7-735,ncommun11-1902,PRX11-041057,PRA103-053507,ACSPhoton8-1077,
PRB106-195423}. The bifurcation from the linear
topological states to the nonlinear ones is also applicable to the nonlinear higher-order
topological systems, where nonlinear topological corner states are proposed \cite{PRB104-235420,nphys17-995,LSA10-164,
CSF207-118044}. 

Topological gap solitons,
which are self-localized topological states in the bulk with their spectra
lie within the topological band gap, are also discovered
in nonlinear topological systems \cite{PRB102-115411,nphys18-678,CP8-342,PRL111-243905,PRL118-023901,
PRA98-013827,LPR13-1900223,arxiv1904-10312,science368-856,CP5-275}.
Although the topological gap solitons have no direct linear counterparts, they exhibit the
similar proprieties of the linear topological edge states, 
such as the sublattice polarization \cite{nphys18-678,PRL118-023901,LPR13-1900223} 
and unidirectional transport \cite{PRL111-243905,science368-856}. 
In contrast to the conventional lattice solitons which are self-localized due to the balance
between coupling (commonly referred as diffraction in photonic systems) and nonlinearity,
and thus are topologically trivial \cite{RMP83-247,PR463-1,QF4-9}, 
the topological gap solitons are localized at the nonlinearity-induced 
topological interfaces and their formation may be interpreted as the
Jackiw-Rebbi Dirac boundary modes where mass inversion occurs across the interface
\cite{PRA98-013827,LPR13-1900223,CP5-275}.

\begin{figure*}[tbp]
\includegraphics[width=17.7cm]{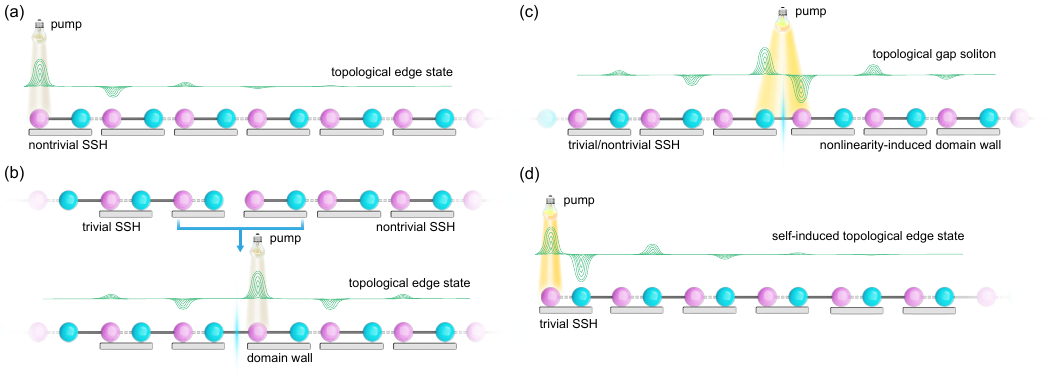}
\caption{Schematics of the nonlinear SSH lattices and the corresponding localized states.
(a) A topologically nontrivial SSH lattice terminates with a weak bond. 
When the lattice is excited by a low-intensity pump, 
i.e., an external continuous source with a small signal, topological edge states appear at the left edge.
(b) A structure connected by two SSH lattices with different topological properties.
Due to the formation of the domain wall, topological edge state also appears under a low-intensity pump. 
With the increasing of pump intensity, the topological edge states in both  (a) and (b) bifurcate to nonlinear ones.
(c) For a topologically trivial or nontrivial SSH lattice, under the excitation from an external pump, a 
nonlinearity-induced domain wall appears in the bulk and supports the formation of topological gap solitons.
(d) A topologically trivial SSH lattice ending with strong bond. Under the action of
onsite nonlinearity, self-induced topological edge states appear at the physical termination.}
\label{fig_findings}
\end{figure*}

Following the experimental observations of nonlinear topological edge states and topological gap solitons 
in photonic systems \cite{nphys18-678,nanophotonics14-769}, this study investigates localized states in 
nonlinear SSH models with onsite nonlinearity using electric circuits \cite{PRL42-1698}. Electric circuits are 
particularly well-suited for probing nonlinear topological physics due to the flexibility in constructing circuit lattices 
and the capability of phase-resolved measurement techniques \cite{nelectron1-178, ncommun10-1102,
PRL123- 053902,PNAS118-e2106411118,PRResearch5-L012041,arXiv:2411.07522}. We systematically explore 
the nonlinear topological edge states and topological gap solitons in SSH circuit lattices, providing comprehensive 
verification of sublattice polarization and phase jumps, as well as observing the full existence curves and the 
nonlinearity-induced effects on state localization, whether weakening or strengthening it. Such thorough experimental 
validation is difficult to achieve in other platforms. Moreover, we uncover a novel type of self-induced topological 
edge states that reside at the edge of a topologically trivial lattice yet exhibit hallmark features of linear topological 
edge states, including sublattice polarization, phase jumps, and decaying tails approaching zero. A distinctive 
characteristic of these states is the existence of a boundary-induced power threshold. These self-induced topological 
edge states have not been reported previously. Our work unveils new opportunities for exploring novel nonlinear 
topological states. It also paves the way for the development of nonlinear topological circuits.

Our main findings are summarized in Fig. \ref{fig_findings}.
As depicted in Fig. \ref{fig_findings}(a), under a low-intensity pump, i.e., under excitation from an external 
continuous source with a small signal, the system operates in the linear limit, and topological edge states 
appear at the edge of the topologically nontrivial SSH lattice that terminates with a weak bond.
With the increasing of pump intensity, we observe the nonlinear topological edge states
bifurcated from the linear counterpart.
We experimentally validate the inherited properties of sublattice polarization and phase jump, 
and find that nonlinearity weakens the localization of the topological edge states.
We then implement a structure by connecting two SSH lattices
with different topological properties, as shown in Fig. \ref{fig_findings}(b). Due to the formation of the domain
wall, under a weak pump topological edge state appears as well. We experimentally reveal the 
similar state properties under increased pump intensity, because the nonlinear topological 
edge states have the same physical origin.

Different to the nonlinear topological edge states, topological gap solitons have no direct
linear counterparts. For an SSH lattice shown in Fig. \ref{fig_findings}(c), no matter the lattice is topologically 
trivial or nontrivial in the linear limit, under the external pump, a nonlinearity-induced domain 
wall appears in the 
bulk and supports the formation of topological gap solitons. We experimentally observe that 
the left and right tails of the topological gap solitons have opposite sublattice polarizations, 
and reveal that nonlinearity increases the localization of the topological gap solitons.
These features are in contrast to those of nonlinear topological edge states because 
the physical origins are totally different.

Besides the nonlinear topological edge states and topological gap solitons, we discover
both theoretically and experimentally a new type of topological states residing at the edge of a semi-infinite lattice 
ending with a strong bond (Fig. \ref{fig_findings}(d)).
Although the lattice is topologically trivial in the linear limit,  under the action of onsite
nonlinearity, it supports the self-induced topological
edge states that reside at the physical termination of the structure \cite{PRL127-184101}.
These self-induced topological edge states exhibit sublattice polarization and 
phase jumps starting from the second site close to the edge.
They are the variant of the topological gap solitons
under the breaking of the discrete translational symmetry at the edge of the structure,
along with a boundary-induced power threshold for their existence.
Specifically, the self-induced topological edge states 
can be approximately mapped to 
the linear topological edge state of a semi-infinite SSH lattice.
Unlike the previously reported self-induced topological transitions driven by nonlinear 
couplings \cite{nelectron1-178}, 
which are conceptually straightforward but less common in realistic interacting systems, 
our self-induced topological edge states are realized in a lattice with onsite nonlinearity 
and feature decaying tails that approach zero, in contrast to the nonlocalized distributions 
that maintain non-zero plateau levels \cite{nelectron1-178}.
Our results are broadly applicable and can be readily extended to photonic and cold atomic 
systems, where onsite nonlinearities naturally arise from interparticle interactions. 

This paper is organized as follows. In Section \ref{model}, we construct a theoretical model based on the Kirchhoff circuit equations. Section \ref{edge_states} presents the experimental observation of the nonlinear topological edge states, accompanied by the theoretical discussion. In Section \ref{gap_solitons}, we focus on the topological gap solitons. In Section \ref{self_induced}, we discuss the results for the newly discovered self-induced topological edge states. Section \ref{diss} presents a discussion on the topological properties. Finally, Section \ref{con} is the conclusion.

\section{Model\label{model}}

\begin{figure}[tbp]
\includegraphics[width=8.6cm]{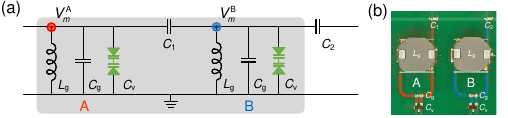}
\caption{Schematics of the nonlinear SSH circuit lattices. 
(a) Circuit implementation of the unit cell of the nonlinear SSH lattices.
The back-to-back varactor diodes provides the onsite nonlinearity of the SSH model.
(b) Unit cell of the experimentally realized nonlinear SSH circuit lattices.
}
\label{fig_circuit}
\end{figure}

All the nonlinear SSH models shown in Fig. \ref{fig_findings} can be realized 
using electric circuit lattices. Fig. \ref{fig_circuit}(a) shows the circuit implementation of a unit cell of the nonlinear
SSH lattices, where A and B denote the two sublattice sites. For one
sublattice, the inductor $L_{\text{g}}$, capacitor $C_{\text{g}}$, and
back-to-back varactor diode $C_{\text{v}}$ are wired in parallel. The
varactor diode acts as a voltage-dependent variable capacitor with $C_{\text{%
v}}=C_{\text{L}}+C_{\text{NL}}$, where the linear part $C_{\text{L}}$ is the
capacitance at the zero voltage and the nonlinear part can be
phenomenologically written as $C_{\text{NL}}=-C_{\text{L}}+\frac{C_{\text{L}}%
}{\left( 1+\left\vert v/v_{0}\right\vert \right) ^{M}}$
(see Appendix \ref{app_A} for the modeling of back-to-back varactor diodes). 
Here, $v_{0}$ and $M$
are constants, and $v$ is the voltage amplitude. The
sublattice sites are wired with each other through the intracell coupling
capacitor $C_{1}$ and intercell coupling capacitor $C_{2}$.

For a typical nonlinear SSH circuit that emulates the lattice shown in Fig. \ref{fig_findings}(a), 
it can be described by the discretized Gross-Pitaevskii (GP) equation (also called as the
discrete nonlinear Schr\"{o}dinger equation):
\begin{eqnarray}
\text{i}\frac{d}{dt}\left[ 
\begin{array}{c}
V_{m}^{\text{A}} \\ 
V_{m}^{\text{B}}%
\end{array}%
\right] &=&E_{0}\left[ 
\begin{array}{c}
V_{m}^{\text{A}} \\ 
V_{m}^{\text{B}}%
\end{array}%
\right] +J_{1}\left[ 
\begin{array}{c}
V_{m}^{\text{B}} \\ 
V_{m}^{\text{A}}%
\end{array}%
\right] +J_{2}\left[ 
\begin{array}{c}
V_{m-1}^{\text{B}} \\ 
V_{m+1}^{\text{A}}%
\end{array}%
\right]  \notag \\
&&+\left[ 
\begin{array}{c}
g\left(V_{m}^{\text{A}}\right){V_{m}^{\text{A}}} \\ 
g\left(V_{m}^{\text{B}}\right){V_{m}^{\text{B}}}%
\end{array}%
\right] ,  \label{gp_main}
\end{eqnarray}%
where $\left[ V_{m}^{\text{A}} \left(t\right),
V_{m}^{\text{B}} \left(t\right) \right]^{\text{T}}$ are the time-dependent voltages
on sites A and B in the $m$th unit cell,
$E_{0}={\omega _{0}+}\Delta E$ is equivalent to the constant onsite energy,
which includes the resonant frequency of the linear oscillators
$\omega _{0}=1/\sqrt{L_{\text{g}}\left( C_{%
\text{g}}+C_{\text{L}}\right) }$ and the frequency shift induced by the coupling capacitors
$\Delta E=-\frac{{C}_{1}+{C}_{2}}{2\left( C_{\text{g}}+C_{\text{L}%
}\right) }{\omega _{0}}$, $J_{1}=\frac{{C}_{1}}{2\left( C_{\text{g}}+C_{%
\text{L}}\right) }{\omega _{0}}$ and $J_{2}=\frac{{C}_{2}}{2\left( C_{\text{g}}+C_{%
\text{L}}\right) }{\omega _{0}}$ are the intracell and intercell coupling coefficients,
respectively, and $g\left( {V_{m}^{\sigma }}\right) =-%
\frac{C_{\text{NL}}\left( {V_{m}^{\sigma }}\right) }{2\left( C_{\text{g}}+C_{%
\text{L}}\right) }{\omega _{0}}$ $\left( \sigma = \text{A}, \text{B} \right)$ is the 
voltage-dependent 
onsite energy. Eq. (\ref{gp_main}) is valid under $C_{1,2}\ll C_{\text{g}}+C_{\text{L}}$
and $C_{\text{NL}}\ll C_{\text{g}}+C_{\text{L}}$, where the slowly-varying 
envelope approximation holds (see Appendix \ref{app_B} for the derivation from
the Kirchhoff circuit equations). Thus, the nonlinear circuit lattice realizes
the SSH model with onsite nonlinearity. The four lattice configurations shown in Fig. \ref{fig_findings} 
all can be implemented based on the circuit dimer in Fig. \ref{fig_circuit}(a).
Fig. \ref{fig_circuit}(b) shows the unit cell of the experimentally realized circuit lattices.

The circuit parameters in the four nonlinear SSH circuits are $C_{\text{g}}=4.7~\text{nF}$,  
$L_{\text{g}}=15~\mu\text{H}$, $C_{\text{L}}=73.48~\text{pF}$, $v_{0}=2.1935$,
and $M=0.4548$. For the first lattice shown in Fig. \ref{fig_findings}(a), the intracell and intercell
coupling capacitors are $C_{1}=180~\text{pF}$ and $C_{2}=560~\text{pF}$,
respectively. For the second lattice shown in Fig. \ref{fig_findings}(b), the topologically nontrivial part has
$C_{1}=180~\text{pF}$ and $C_{2}=560~\text{pF}$, while the trivial part
has $C_{1}=560~\text{pF}$ and $C_{2}=180~\text{pF}$. An additional capacitance 
$C_{2}-C_{1}$ is added to the grounding capacitance of the interface circuit node.
For the third and fourth lattices shown in Figs. \ref{fig_findings}(c)-(d),  the intracell and intercell
coupling capacitors are $C_{1}=560~\text{pF}$ and $C_{2}=180~\text{pF}$,
respectively.

\section{Nonlinear topological edge states\label{edge_states}}

\begin{figure*}[tbp]
\includegraphics[width=17.5cm]{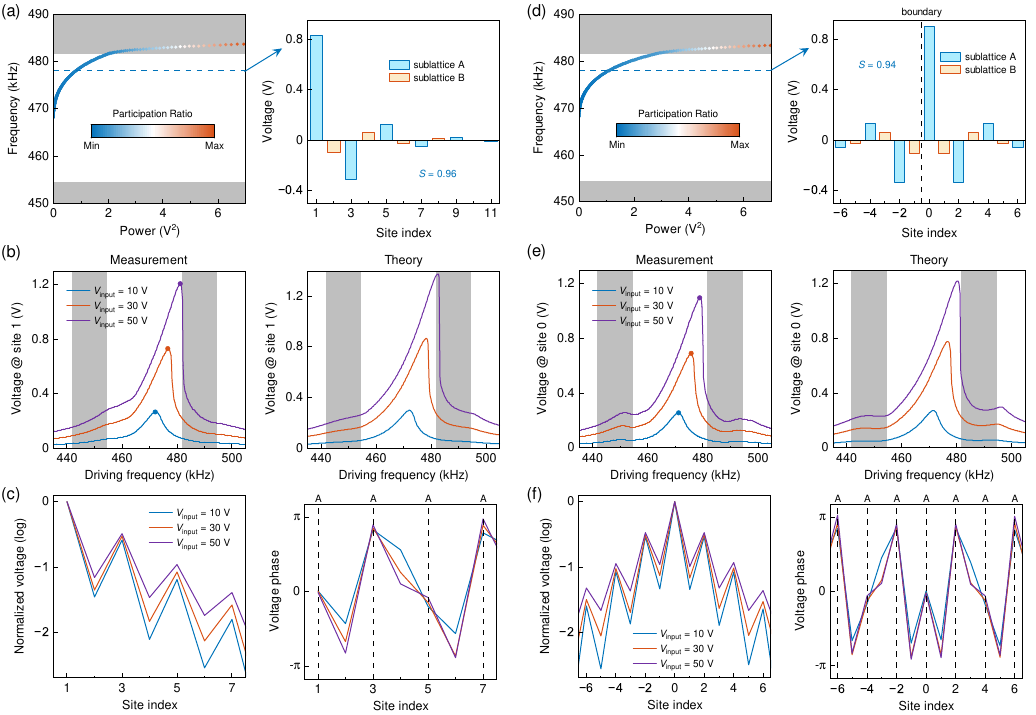}
\caption{Nonlinear topological edge states.
(a) Nonlinear topological edge states residing at the edge of a semi-infinite 
SSH lattice that ends with a weak bond (Fig. \ref{fig_findings}(a)). 
Participation ratio measures the state localization,
with a large participation ratio signifying weak localization.
The inset shows a typical voltage distribution of the nonlinear topological edge states.
(b) At different input voltages, the voltage spectra measured experimentally and calculated theoretically.
The driving frequency corresponds to the output frequency of the external voltage source.
(c) The normalized amplitudes (left) and phases (right) of the experimental voltage distributions
at the resonant frequencies indicated by the dots in (b).
(d)-(f) Results for the nonlinear topological edge states in a structure formed by connecting two lattices
with different topological properties in the linear limit (Fig. \ref{fig_findings}(b)). }
\label{fig2}
\end{figure*}

We first study the nonlinear topological edge
states in the first SSH lattice (Fig. \ref{fig_findings}(a)). The configuration is a typical semi-infinite 
SSH lattice that ends with a weak bond. In the linear limit, due to the nontrivial topology with 
the winding number $\mathcal{W}=1$, topological edge state residing at the left edge 
exists \cite{PRL42-1698}. Under the action of the onsite nonlinearity, by bifurcating from 
the linear edge state,
nonlinear topological edge states with the topological protection inherited from the 
linear counterpart appear (see Appendix \ref{app_C} for the calculation algorithm and
stability analysis of the nonlinear topological edge states). As depicted in Fig. \ref{fig2}(a),
with the power increasing (defined as the sum of the squares of all the site voltages),
the frequency of the nonlinear edge state exhibits a blue shift due to the
decreased grounding capacitance, and the state localization becomes weak with a larger 
participation ratio defined as $( \sum\limits_{m,\sigma} \vert  v_{m}^{\sigma} \vert ^{2} )^{2}
/\sum\limits_{m,\sigma} \vert  v_{m}^{\sigma} \vert ^{4}$ ($\sigma = \text{A}, \text{B}$), 
particularly when the frequency enters the linear bulk band (denoted by the shaded gray areas).
Here, the participation ratio is introduced to measure state localization, with a large participation ratio signifying weak localization.
The inset of Fig. \ref{fig2}(a) shows a typical profile of the nonlinear topological edge states. Since the 
nonlinear edge states are bifurcated from the linear counterpart which satisfies
$v_{m}^{\text{A}} = \left(- \frac{J_{1}}{J_{2}} \right)^{\vert m \vert} v_{0}^{\text{A}}$
and $v_{m}^{\text{B}}=0$, their voltages mainly distribute on
sublattice site A with the sublattice pseudospin 
\begin{equation}
S=\frac{\sum\limits_{m} \left( \left\vert v_{m}^{\text{A}}\right\vert^{2}-\left\vert v_{m}^{\text{B}}\right\vert ^{2} \right)}
{\sum\limits_{m} \left( \left\vert v_{m}^{\text{A}}\right\vert ^{2}+\left\vert v_{m}^{\text{B}}\right\vert ^{2} \right)}
\label{sp}
\end{equation}%
nearly equal to $1$ and the phase jump of $\pi $ among the neighboring cells. 

Experimentally, we excite site $1$ (i.e., the leftmost site) of the circuit lattice using an external input voltage with
frequency sweep to observe the nonlinear topological edge states
(see Appendix \ref{app_D} for the detailed experimental implementation).
The output frequency of the external voltage source corresponds to the driving frequency of the lattice
(see further explanations in Appendix \ref{app_B}, section 1).
From the experimental and theoretical results shown in Fig. \ref{fig2}(b), under a small input voltage,
the voltage spectrum exhibits a nearly symmetric peak with respect to the resonant frequency, 
indicating the presence of the topological
edge state in the linear limit. As the input voltage increases, the spectrum
peak becomes asymmetric and the resonant frequency exhibits a blue shift
(see Fig. \ref{fig_edge_spectra} in Appendix \ref{app_D} for the complete voltage spectra).
We further extract the resonant frequencies under different input voltages and measure
the voltage distributions in the circuit lattice. The normalized amplitudes and phases of the 
experimental voltage distributions are shown in Fig. \ref{fig2}(c). At higher input voltages,
the localization of the nonlinear edge state decreases, although the characteristics of 
sublattice polarization and phase jump remain intact (see Fig. \ref{fig_edge_PR_S} and Appendix \ref{app_D} for 
the quantitative analysis). This observation aligns with the theoretical prediction and 
confirms that nonlinearity weakens the localization of the nonlinear topological edge states.

We then study the second configuration shown in Fig. \ref{fig_findings}(b). The structure is connected 
by two lattices with different topological properties defined 
in the linear limit (winding number $\mathcal{W}=0$ for the left topologically trivial
part and $\mathcal{W}=1$ for the right nontrivial part \cite{PRB84-195452}).
Without nonlinearity, a topological edge state 
is localized at the boundary between the two parts \cite{ncommun6-6710,ncommun3-882}. 
Similarly, with onsite nonlinearity,
nonlinear topological edge states which are bifurcated from the linear edge state 
appear, as shown in Fig. \ref{fig2}(d) (see Fig. \ref{fig_interface_stability} and Appendix \ref{app_E} for the stability analysis). 
Due to the same physical origin, the nonlinear topological edge states in this lattice also 
exhibit the sublattice polarization and phase jump, and nonlinearity also weakens the 
state localization. These properties are confirmed by the experimental measurements 
presented in in Figs. \ref{fig2}(e)-(f) (see Appendix \ref{app_F} for the experimental 
implementation, complete voltage spectra, and quantitative analysis). 
Note that the nonlinear topological edge states are fundamentally different from conventional 
topologically trivial solitons, which do not exhibit these characteristics (see Fig. \ref{fig_interface_nonsaturable} 
and Appendix \ref{app_E} for a discussion on topologically trivial edge states).

\section{Topological gap solitons \label{gap_solitons}}

\begin{figure}[tbp]
\includegraphics[width=8.6cm]{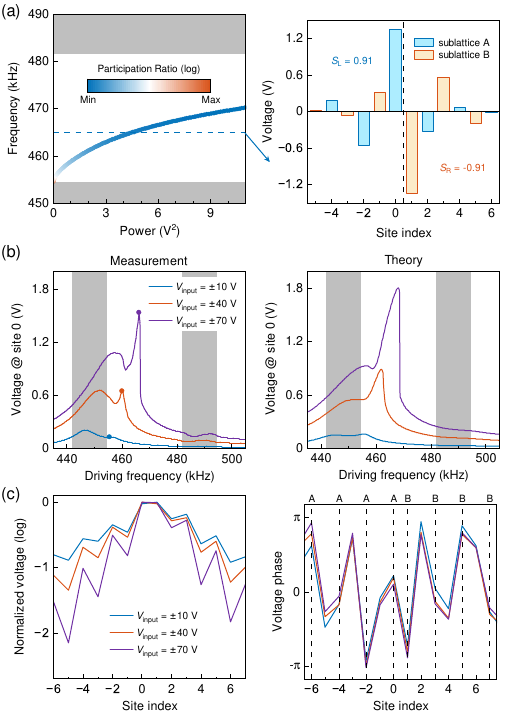}
\caption{Topological gap solitons.
(a) Frequencies and participation ratios of the topological gap solitons in the nonlinear SSH lattice shown in Fig. \ref{fig_findings}(c).
The inset shows a typical voltage distribution of the topological gap solitons.
(b) Voltage spectra measured experimentally and calculated theoretically.
(c) The normalized amplitudes (left) and phases (right) of the experimental voltage distributions
at the resonant frequencies indicated by the dots in (b).
}
\label{fig3}
\end{figure}

We will now discuss the topological gap
solitons in the third kind of the lattice (Fig. \ref{fig_findings}(c)). In contrast to the nonlinear topological edge states,
the topological gap solitons lack a direct linear counterpart, although they reside spectrally within 
the SSH gap. From Fig. \ref{fig3}(a), the topological 
gap solitons bifurcate from the edge of the linear Bloch band and converge to the bulk state 
in the linear limit (see Figs. \ref{fig_bulk_mode}-\ref{fig_bulk_stability} and Appendix \ref{app_G} for the bifurcation and stability).
Under the influence of onsite nonlinearity, a topological interface can be induced at the center 
unit cell of the
lattice \cite{PRA98-013827,LPR13-1900223}. The Dirac mass defined as $m_{\text{Dirac}} = 
\frac{g\left( V_{m}^{A} \right) - g\left( V_{m}^{B} \right)}
{2}$ exhibits an inversion ($m_{\text{Dirac}} >0$ for the left part and $m_{\text{Dirac}}<0$ 
for the right part),
and the topological gap solitons emerge as the Jackiw-Rebbi-type Dirac boundary modes
(see Appendix \ref{app_G} for the physical interpretation). The nonlinearity-induced
interface can also be interpreted as an impurity potential that splits the the original lattice into
two topological regions, and the topological gap soliton represents a combination of the two 
topological edge states \cite{PRB102-115411}.

The global sublattice polarizations of the topological gap solitons vanish because the left 
and right tails exhibit opposite chiralities \cite{nphys18-678}. From the inset in Fig. \ref{fig3}(a),
the left tail is primarily confined to the sublattice site $\text{A}$, exhibiting a positive 
local sublattice pseudospin $S_{\text{L}}$, while the right tail displays a negative $S_{\text{R}}$.
Experimentally, the topological gap solitons are excited using two out-of-phase input voltages,
revealing the properties of both sublattice polarization and phase jumping, as shown 
in Figs. \ref{fig3}(b)-(c) (see Appendix \ref{app_H} for the comprehensive analysis and excitation
of topologically trivial gap solitons). 
In Fig. \ref{fig3}(b), the experimental measurement results deviate from the theoretical predictions 
at higher input voltages, primarily due to increased series resistance of the inductors and decreased 
driving voltages provided by the voltage source. In Fig. \ref{fig3}(c), the small deviations in the voltage 
distributions from the perfectly symmetric profiles may result from imperfections in the circuit lattice, 
discrepancies between actual and set input voltage levels, and the limited temporal resolution in the 
measurements (see Appendix \ref{app_D}).
Furthermore, since the topological gap solitons are nonlinearity-induced states, their localization 
is enhanced under stronger input voltages, provided that the delocalized gap solitons within the 
linear bulk band are not excited (see Appendix \ref{app_G}).

\section{Self-induced topological edge states \label{self_induced}}

The nonlinear topological states we currently implement rely on either the physical edge of a 
topologically nontrivial lattice or a nonlinearity-induced topological interface between two lattices. 
In this section, we shift our focus to a semi-infinite lattice terminating with a strong bond and 
demonstrate the emergence of a new type of nonlinear topological state that resides at the edge 
of a nonlinear topologically trivial lattice.

\begin{figure}[tbp]
\includegraphics[width=8.6cm]{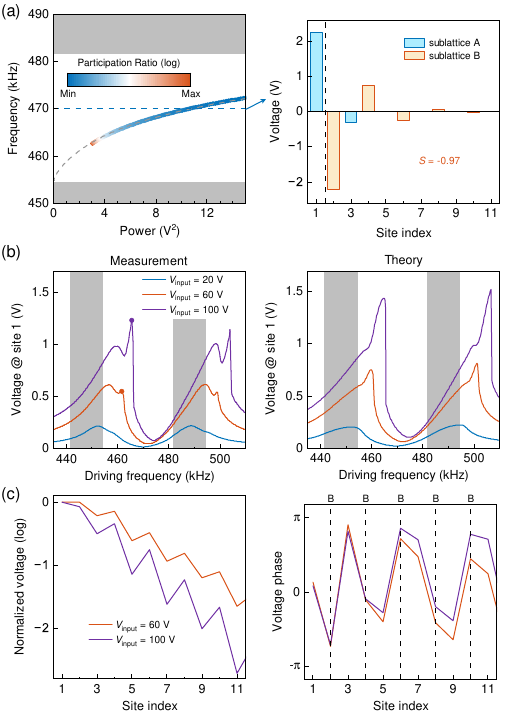}
\caption{Self-induced topological edge states.
(a) Frequencies and participation ratios of the self-induced topological edge states in the 
nonlinear SSH lattice shown in Fig. \ref{fig_findings}(d).
For comparison, the frequencies of the topological gap solitons from Fig. \ref{fig3}(a) are also plotted.
The inset shows a typical voltage distribution of the self-induced topological edge states.
(b) Voltage spectra measured experimentally and calculated theoretically.
(c) The normalized amplitudes (left) and phases (right) of the experimental voltage distributions
at the resonant frequencies indicated by the dots in (b).
}
\label{fig4}
\end{figure}

Figure \ref{fig4}(a) shows the frequencies and participation ratios of the self-induced topological 
edge states. Although topological edge states do not exist in the linear limit, above a certain power
threshold, localized states with their maxima reside at the leftmost unit cell
appear in the SSH gap (see the inset of Fig. \ref{fig4}(a)). Neglecting the voltage at the first site, the self-induced topological edge state exhibits
the similar profile with the nonlinear topological edge state (the inset of Fig. \ref{fig2}(a)) and
half of the topological gap soliton (the inset of Fig. \ref{fig3}(a)). Considering the similarity, the self-induced
topological edge states at the left edge can be created by the discrete translational transformation
$\mathcal{T} V_{m}^{\text{A}, \text{B}} = V_{m-1}^{\text{A}, \text{B}}$, where 
$\mathcal{T}$ is the translational operator. The power threshold is induced by the 
breaking of discrete translational symmetry when moving the topological gap solitons towards 
the edge (see Appendix \ref{app_I}, section 3, for the 
relation between self-induced topological edge states and topological gap solitons).
Since the circuit nonlinearity adds perturbations to the onsite energy, the self-induced topological 
edge states
can be approximated using the solutions of Eq. (\ref{gp_main}) with $g \ne 0$ for the leftmost unit cell 
and $g=0$ for the other sites. When the state frequency equals to the constant 
onsite energy $E_{0}$, i.e., the frequency of the linear topological edge state, the 
Gross-Pitaevskii equation reduces to
\begin{eqnarray}
J_{1}v_{1}^{\text{B}}+g\left( v_{1}^{\text{A}}\right) v_{1}^{\text{A}} &=&0, \label{ac1}\\
J_{1}v_{1}^{\text{A}}+g\left( v_{1}^{\text{B}}\right) v_{1}^{\text{B}} &=&0, \label{ac2}
\end{eqnarray}
with $v_{m}^{\text{B}} = \left(- \frac{J_{1}}{J_{2}} \right)^{\vert m \vert -1} v_{1}^{\text{B}}$
and $v_{m}^{\text{A}}=0$ for $m \geq 2$ (see Appendix \ref{app_I}). 
Eqs. (\ref{ac1})-(\ref{ac2}) govern the voltage distributions
in a single dimer and allow for an antisymmetric solution: $v_{1}^{\text{A}} = -v_{1}^{\text{B}}$
with $g \left( v_{1}^{\text{A}} \right) = J_{1}$, which is consistent with the profile shown 
in the inset of Fig. \ref{fig4}(a).
The relationship between the voltages at the other sites is precisely analogous to that of a linear 
topological edge state.
Thus, the self-induced topological edge states residing at the edge of a topologically trivial lattice represent a continuation of 
the antisymmetric state in the anti-continuum limit (see Appendix \ref{app_I}, section 3), and they can 
be approximately mapped to the linear topological edge state of a semi-infinite SSH lattice.

To experimentally investigate the self-induced topological edge states, we excite the leftmost site, 
specifically site 1 of the circuit lattice. Figure \ref{fig4}(b) displays the experimental and theoretical 
voltage spectra. The experimental results show deviations from the theoretical predictions at higher 
input voltages, primarily due to increased series resistance in the inductors and reduced driving 
voltages supplied by the voltage source. From Fig. \ref{fig4}(b), no peak is observed in the SSH gap 
for $V_{\text{input}} = 20~\text{V}$, indicating the existence of a power threshold for the self-induced 
topological edge states.
The threshold value obtained from the excitation spectra deviates from the result in 
Fig. \ref{fig4}(a) possibly due to the excitation
of other nonlinear states (see Appendix \ref{app_J}). At high input voltages, the voltage
distributions in Fig. \ref{fig4}(c) reveal that the self-induced topological edge states exhibit
sublattice polarization and phase jump starting from the second site near the physical termination. 
Meanwhile, similar to the topological gap solitons, nonlinearity can enhance the localization of 
self-induced topological edge states, unless the resonant frequency is driven into the linear 
bulk band, which is consistent with the theoretical prediction in Fig. \ref{fig4}(a). Note that 
the resonant peak in the semi-infinite gap in Fig. \ref{fig4}(b) corresponds to the excitation of 
topologically trivial conventional solitons (see Appendix \ref{app_J} for the discussion on 
topologically trivial edge solitons). 

In comparison to the self-induced topological states reported by \textit{Hadad} et al. \cite{nelectron1-178}, 
our self-induced topological edge states offer two main advantages. First, our self-induced topological edge states 
feature decaying tails that approach zero, which indicates that the powers of the edge states are finite. In contrast, 
the self-induced topological states reported by \textit{Hadad} et al. exhibit non-zero plateau levels in the limit 
of $m \rightarrow \infty$ \cite{nelectron1-178}. Such extended modes carry infinite power and are physically unrealizable.
Second, our self-induced topological edge states are realized in a lattice with onsite nonlinearity, as opposed to 
the topological edge states induced by nonlinear couplings. In realistic interacting systems, such as photonic and 
cold atomic systems, onsite nonlinearity is more readily achievable \cite{RMP83-247,PR463-1,RMP78-179,NRP1-185}. 
For instance, in a cold atomic system, the 
interaction between particles at the same site is significantly stronger than the interactions between particles at 
neighboring sites, leading to the emergence of onsite nonlinearity (see further discussions in Appendix \ref{app_B}, section 3). 
These two advantages imply that our results are physically realizable and broadly applicable.

\section{Discussion on the topological properties\label{diss}}

In the previous sections, we studied nonlinear topological edge states, topological gap solitons, 
and self-induced topological edge states. Here, we aim to address the topological properties of 
these types of nonlinear states.

In the absence of onsite nonlinearity, a linear SSH model belongs to a one-dimensional (1D) chiral 
symmetric system in class BDI. The traditional 1D winding number can be well defined as the 
topological invariant \cite{book1}, as introduced in Section \ref{edge_states}. For a topologically 
nontrivial SSH lattice with a winding number of $\mathcal{W}=1$, topological edge states emerge 
at the physical termination of an open chain. These topological edge states can also appear at the 
domain wall between two chains with $\mathcal{W}=0$ and $\mathcal{W}=1$, respectively. The 
nonlinear topological edge states shown in Figs. \ref{fig2}(a)-(c) bifurcate from the topological edge 
state at the physical termination, and those depicted in Figs. \ref{fig2}(d)-(f) bifurcate from the topological 
edge state located at the domain wall. In contrast, the topological gap solitons shown in Fig. \ref{fig3} 
and the self-induced topological edge states illustrated in Fig. \ref{fig4} do not have linear counterparts, 
as they are localized states purely induced by nonlinearity.

When nonlinearity is introduced to the original SSH model, the ability to define a topological invariant 
depends on the nature of the nonlinearity \cite{PRB104-235420}. If the introduced nonlinearity does not break the symmetry 
operations associated with the original topological classification—such as in nonlinear SSH models with 
nonlinear coupling coefficients—the original definition of the winding number can still be utilized or modified to 
describe the topological properties of the edge states \cite{PRB93-15512,nelectron1-178,ncommun13-3379,FP18-33311,
ncommun16-422}. In our study, we add nonlinear onsite energies to the original SSH model. When we consider
the contribution from the nonlinear states to the onsite energies, the chiral symmetry, which is one of the symmetry operators 
in class BDI, is broken. Consequently, the nonlinear SSH model cannot be classified into any of the classes in the periodic table of 
topological insulators. Although the band structure can still be defined by using the Bloch ansatz, Zak phase is generally
not quantized \cite{PRB102-115411}. Even the spectral localizer, a local topological invariant defined in real space, 
becomes inapplicable in this scenario \cite{AP356-383,nano11-4765,ncommun14-3071,PRL133-116602}. These previous findings 
indicate that it remains challenging to identify an appropriate topological invariant that accurately describes a 
nonlinear system where nonlinearity break the symmetries necessary for topological classification, 
unless the topological protection of the original linear system does not rely on any symmetry operations, 
as exemplified by Chern insulators \cite{nphys20-1164}.

Considering this ongoing challenge, our study approaches the determination of whether a nonlinear state 
is topologically nontrivial or trivial from a phenomenological perspective. We assert that the nonlinear edge states 
shown in Fig. \ref{fig2} are topologically nontrivial because they bifurcate from their linear counterparts, which are 
topological edge states residing at the edge of a topologically nontrivial lattice. The term ``nonlinear topological edge states" 
aligns with the terminology used in previous studies \cite{optica3-1228,PRL119-253904,OL45-6466,nphys18-678,PRL128-093901,
PRE104-054206,CP8-342}.
The gap solitons depicted in Fig. \ref{fig3} are classified as topological gap solitons, as they reside in the topological bandgap and
represent nonlinearity-induced Jackiw-Rebbi-type Dirac boundary modes \cite{nphys18-678,PRL118-023901,PRA98-013827,LPR13-1900223,
arxiv1904-10312,science368-856,CP5-275}. 
For the self-induced topological edge state shown in Fig. \ref{fig4}, we argue that
it can be considered a topological edge state due to its hallmark features of a linear topological edge state, 
including sublattice polarization, phase jumps, and a decaying tail that approaches zero. Additionally, it can be regarded 
as a variant of the topological gap soliton resulting from the breaking of discrete translational symmetry at the edge of the structure.
Even so, a precise definition of the topological invariants and the establishment of a general bulk-boundary correspondence are still needed.

Very recently, \textit{Sone} et al. developed an elegant framework for the nonlinear SSH model with onsite Kerr nonlinearity. 
By applying the Bloch ansatz, they demonstrated that the model preserves space-inversion symmetry in momentum space \cite{arxiv}. 
Within this setting, the nonlinear Berry phases become quantized, establishing a well-defined bulk-boundary correspondence in the nonlinear regime. 
Building upon this insightful and highly relevant approach, we performed supplementary calculations for our model with onsite saturable nonlinearity. 
Our results confirm that the nonlinear topological edge states presented in Fig.~\ref{fig2} can be described within this theoretical framework (see Appendix \ref{app_K} for further details).

\section{Conclusion\label{con}} 

In conclusion, we experimentally and theoretically reveals the 
existence of the nonlinear topological edge states and topological gap solitons in the nonlinear 
SSH circuit lattices. Contrary to the conventional understanding of the bulk-boundary 
correspondence, we demonstrate the formation of the self-induced topological edge 
states that reside at the physical termination (rather than the nonlinearity-induced 
interface) of a nonlinear topologically trivial lattice.
This exotic type of nonlinear topological states may pave the way to exploring intriguing 
topological states in more complex nonlinear topological systems, such as high-dimensional 
or higher-order systems, and the platform of nonlinear electric circuits may be used 
to explore the interacting topological physics \cite{ncommun13-2392}. \newline

\begin{acknowledgments}

The authors thank Terry A. Loring, Di Zhou, Meng Xiao, Kai Bai, Weixuan Zhang, and Fengxiao Di for fruitful discussions. 
The authors also thank Yongle Lou for providing access to the experimental facilities.
R.L., W.W., and X.K. was sponsored by the National Key Research and Development Program of China
(Grant No. 2022YFA1404902), the National Natural Science Foundation of China (Grant No. 12104353), and the 
Fundamental Research Funds for the Central Universities (Grant No. QTZX25086). 
Y.L. was sponsored by the National
Natural Science Foundation of China (NSFC) under Grant No. 61871309 and the 
111 Project. B.Z. acknowledges support from Singapore National Research Foundation 
Competitive Research Program Grant No. NRF-CRP23-2019-0007, Singapore Ministry of 
Education Academic Research Fund Tier 2 Grant No. MOE-T2EP50123-0007, and Tier 1 
Grant No. RG81/23.

\end{acknowledgments}

\appendix

\section{Modeling of back-to-back varactor diodes\label{app_A}}

\begin{figure*}[tbp]
\includegraphics[width=13.3cm]{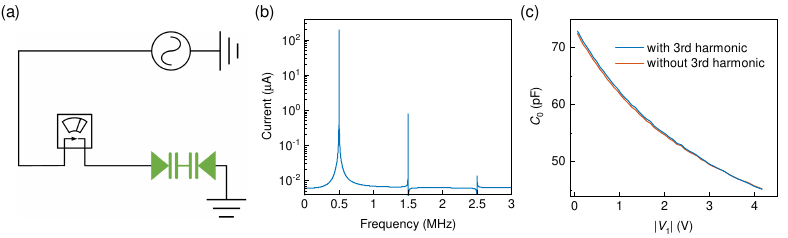}
\caption{Modeling of back-to-back varactor diodes. 
(a) Circuit diagram of the simulation, where the current response of a
back-to-back varactor diode under the excitation from a voltage source is probed. 
(b) The probed current signal displayed in the frequency domain. 
(c) The experimentally measured capacitance-voltage relation of the back-to-back
varactor diode, where the blue and orange curves denote the results calculated with
and without the third-harmonic current components.}
\label{fig_varactor}
\end{figure*}

We use the back-to-back varactor diodes to provide the onsite nonlinearity 
of the nonlinear SSH models. Such topology of double diode with a common cathode ensures the 
same response of the two half-cycles of an AC voltage and avoids the use of a bias voltage \cite{nelectron1-178}.
To model the capacitance of the back-to-back varactor diodes, we simulate the current response of a variable capacitance
double diode with a common cathode (BB201) under the excitation from a
voltage source using Advanced Design System (ADS). The circuit diagram is
shown in Fig. \ref{fig_varactor}(a) and the parameters for the Spice
model are taken from the manufacturer website. The voltage source has an
amplitude of $1~\text{V}$ and the frequency is $500~\text{kHz}$. Note that
there is no DC biasing in the simulation. Based on the circuit configuration, 
after a Fourier transform of the temporal current signal, we find that the current contains the high-harmonic
components, as shown in Fig. \ref{fig_varactor}(b). The even-order
harmonics are suppressed due to the symmetric configuration of the
back-to-back varactor diode \cite{ncommun10-1102}.

The existence of the high-harmonic currents implies that the capacitance of
the back-to-back varactor diode is time-dependent. In this paper, we only consider the
first- and third-harmonic current components, and neglect the higher-order
harmonics. The voltage and current in the circuit can be written as $V\left( t\right) =%
\frac{1}{2}V_{1}e^{-\mathrm{i}\omega t}+\text{c.c.}$ and $I\left( t\right) =\frac{1}{2%
}I_{1}e^{-\mathrm{i}\omega t}+\frac{1}{2}I_{3}e^{-3\mathrm{i}\omega t}+\text{c.c.}$,
respectively, where \text{c.c.} denotes the complex conjugates of the terms to
the left. From the expressions, $V_{1}$ and $I_{1}$ are the amplitudes of the first-harmonic
components, and $I_{3}$ is the amplitude of the third-harmonic current. 
Then the capacitance of the back-to-back varactor diode can be
expressed as 
\begin{equation}
C\left( t\right) =C_{0}+\frac{1}{2}C_{2}e^{-2\mathrm{i}\omega t}+\text{c.c.},
\label{eq_C}
\end{equation}%
where%
\begin{eqnarray}
C_{0}&=&\frac{I_{1}+I_{3}\frac{V_{1}^{\ast }}{V_{1}}}{-\mathrm{i}\omega V_{1}},
\label{eq_C0} \\
C_{2}&=&\frac{2I_{3}}{-\mathrm{i}\omega V_{1}}.  \label{eq_C2}
\end{eqnarray}
Specifically, the capacitance of the back-to-back varactor diode further
reduces to 
\begin{equation}
C =C_{0}=\frac{I_{1}}{-\mathrm{i}\omega V_{1}},  \label{eq_C_reduce}
\end{equation}%
when the third-harmonic current component is also neglected. To characterize the capacitance-voltage relation of the back-to-back varactor diode, we
experimentally measure the voltage and current using an oscilloscope, and calculate the value of $C_{0}$ when considering or neglecting the
third-harmonic current component, respectively. The capacitance-voltage relationship of the back-to-back varactor diode
is symmetric with respect to $V_{1} = 0$, as shown in Fig. \ref{fig_varactor}(c). The
two curves for $C_{0}$ calculated with and without the third harmonics agree
well with each other within the whole voltage range. The small discrepancy
near the zero voltage is due to the inaccuracy in measuring the small
voltage signals. The result implies that the third-harmonic current
component can be neglected in our study, and Eq. (\ref{eq_C_reduce}) is
sufficient to characterize the capacitance of the back-to-back varactor diodes.
To facilitate the theoretical modeling, we fit the experimental
capacitance-voltage curve with the formula 
\begin{equation}
C \left( v \right) = \frac{C_{\text{L}}} {\left(1+|\frac{v}{v_{0}}| \right)
^{M}},  \label{C_model}
\end{equation}
where $C_{\text{L}}$ is the capacitance at the zero voltage, $v_{0}$ and $M$ are
constants, and $v$ is the amplitude of the applied voltage. Thus, the
back-to-back varactor diode acts as a variable capacitor, whose capacitance
depends on the voltage amplitude. We have also measured the back-to-back
varactor diode at other frequencies experimentally and Eq. (\ref{C_model})
is valid within the whole parameter range of our study. Extending the above
conclusion, when the applied
voltage is quasi-monochromatic with $V\left( t\right) =\frac{1}{2}v \left( t
\right) e^{-\mathrm{i}\omega t}+\text{c.c.}$, the capacitance of the back-to-back
varactor diode can be written as 
\begin{equation}
C \left( v \right) = \frac{C_{\text{L}}} {\left[1+|\frac{v \left( t \right) 
}{v_{0}}| \right] ^{M}},  \label{C_model_1}
\end{equation}
where $v \left( t \right )$ is the slowly-varying amplitude envelope. In
this case, the capacitance of the varactor diode varies with the voltage
envelope. This formula will be used to study the localized states in the nonlinear SSH 
circuit lattices. Specifically, when the applied voltage
is near zero, Eq. (\ref{C_model_1}) can be approximated as
\begin{equation}
C \left( v \right) = C_{\text{L}} \left(1-M|\frac{v}{v_{0}}| \right).  
\label{C_approximate}
\end{equation}

Based on the discussion above, for a fixed type of back-to-back varactor diodes, we can determine 
the parameters $C_{\mathrm{L}}$, $v_{0}$, and $M$ by fitting the experimental capacitance-voltage 
relationship. For our specific type (BB201), we measured the capacitance-voltage relationship at different 
frequencies and found that the parameter values remain unchanged within the range of $100~\mathrm{kHz}$ 
to $1~\mathrm{MHz}$.

\section{Derivation of GP equation from Kirchhoff circuit equations \label{app_B}}

This section is organized as follows. First,
we give the derivation of the GP equation from the Kirchhoff circuit equations, and
discuss the applicable conditions of the GP equation. In the second subsection, we prove the
validity of the GP equation by developing an accurate nonlinear circuit model.
Finally, in the last subsection, we show that the nonlinear electric circuit lattices can be 
used to explore the many-body interacting topological physics.

\subsection{Derivation of GP equation}

\begin{figure}[tbp]
\includegraphics[width=8.4cm]{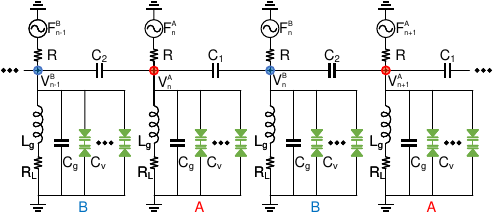}
\caption{Diagram of the nonlinear SSH circuit lattice. For generality, we show
the schematic of the driven-dissipative nonlinear SSH circuit, where the voltage sources 
$F_{n}^{\text{A},\text{B}}$ with 
the shunt resistors $R$ are connected to the circuit nodes and the series resistance of 
the inductors $R_{L} $ is considered. To enhance the onsite nonlinearity, $\protect\eta$ 
varactor diodes are wired in parallel.}
\label{fig_GP}
\end{figure}

For generality, we derive the GP equation 
of a nonlinear SSH circuit lattice with the external driving sources and circuit dissipations.
As shown in Fig \ref{fig_GP}, the circuit lattice is composed of the coupled nonlinear
$LC$ oscillators. There are two sublattice sites $\text{A}$ and $\text{B}$ in one unit cell.
For one sublattice, the inductor $L_{\text{g}}$, capacitor $C_{\text{g}}$, and
back-to-back varactor diode $C_{\text{v}}$ are wired in parallel. The
varactor diode acts as a voltage-dependent variable capacitor with $C_{\text{%
v}}=C_{\text{L}}+C_{\text{NL}}$, where the linear part $C_{\text{L}}$ is the
capacitance at the zero voltage and the nonlinear part can be usually
phenomenologically written as $C_{\text{NL}}=-C_{\text{L}}+\frac{C_{\text{L}}%
}{\left( 1+\left\vert v/v_{0}\right\vert \right) ^{M}}$. Here, $v_{0}$ and $M$
are constants, and $v$ is the amplitude of the applied voltage. The
sublattice sites are wired with each other through the intracell coupling
capacitor $C_{1}$ and intercell coupling capacitor $C_{2}$.
Experimentally, we need to add voltage sources to excite the circuit lattice
and the Ohmic losses of the circuit components should be considered.
From Fig. \ref{fig_GP}, the circuit lattice is driven by the external continuous voltage
sources with $F_{m}^{\sigma }=\frac{1}{2}{f_{m}^{\sigma }e^{-\mathrm{i}{\omega }t}}+%
\text{c.c.}$ through the shunt resistors $R$, where $f_{m}$ and $\omega$ are the voltage amplitude
and output angular frequency of the voltage source, respectively. Under such excitation, the
frequency of the voltages at the circuit nodes is always equal to the
frequency of the external voltage sources, because here we neglect the small
high-harmonic components induced by the varactor diodes. We assume that the shunt resistors
are large enough such that the circuit lattice is excited by the equivalent
current sources (please refer to the additional discussion in Appendix \ref{app_F}, section 1). 
For the Ohmic losses of the circuit components, here we
only consider the series resistance of inductors $R_{L}$ because the
varactor diodes that we use have very low series resistance.
Besides, to enhance the nonlinearity of the circuit we use $\eta $ varactor
diodes which are wired in parallel. 

In this circuit, the nonlinear part of the varactor diode $C_{\text{NL}}$ is usually small,
and the total grounding capacitance can be approximated as $C_{\text{g}}+ \eta C_{%
\text{L}}$. We require that the coupling capacitances are much smaller than
the grounding capacitance, i.e. $C_{1,2}\ll C_{\text{g}}+ \eta C_{\text{L}}$, then
the voltages at the lattice sites of the $m$th cell can be written as 
$V_{m}^{\sigma }\left( t\right) =\frac{1}{2}{v_{m}^{\sigma }}\left( t\right) {%
e^{-\mathrm{i}{\omega _{0}}t}}+\text{c.c.}$ with $\sigma =\text{A},\text{B}$, where ${v_{m}^{\sigma
}}$ are the amplitude envelopes (slowly-varying complex amplitudes), 
$\omega _{0}=1/\sqrt{L_{\text{g}}\left( C_{%
\text{g}}+\eta C_{\text{L}}\right) }$ is the carrier frequency of the voltages, 
and $\text{c.c.}$ denotes
the complex conjugate of the term to the left. Using this expression, we
have separated the slowly-varying and fast-varying parts. Accordingly, the
capacitance of the varactor diode in the $m$th cell varies with the voltage
envelope and we have $C_{\text{NL},m}^{\sigma }=-C_{\text{L}}+\frac{C_{\text{L}}}{%
\left( 1+\left\vert {v_{m}^{\sigma }}/v_{0}\right\vert \right) ^{M}}$.
For the $m$th unit cell, at node $\text{A}$ where the voltage is $V_{m}^{\text{A}}$, 
we have the following relations: 
\begin{eqnarray}
L_{\text{g}}\frac{{d{I_{L_{\text{g}}}}}\left( t\right) }{{dt}} &=&{V_{m}^{\text{A}}}%
\left( t\right) {-V}_{R,L}\left( t\right) ,  \label{r1} \\
I_{L_{\text{g}}}\left( t\right) {R}_{L} &=&V_{R,L}\left( t\right) ,
\label{r2} \\
C_{\text{g}}\frac{{d{V_{m}^{\text{A}}}}\left( t\right) }{{dt}} &=&I_{C_{\text{g}%
}}\left( t\right) ,  \label{r3} \\
C_{\text{v}}\left[ {v_{m}^{\text{A}}}\left( t\right) \right] \frac{{d{V_{m}^{\text{A}}}}\left(
t\right) }{{dt}} &=&I_{C_{\text{v}}}\left( t\right) ,\text{ }  \label{r4} \\
{C}_{2}\frac{{d\left[ {{V_{m}^{\text{A}}}}\left( t\right) {-{V_{m-1}^{\text{B}}}}\left(
t\right) \right] }}{{dt}} &=&{I_{C_{2}}}\left( t\right) ,  \label{r5} \\
{C}_{1}\frac{{d\left[ {{V_{m}^{\text{A}}}}\left( t\right) {-{V_{m}^{\text{B}}}}\left(
t\right) \right] }}{{dt}} &=&{I_{C_{1}}}\left( t\right) {.}  \label{r6}
\end{eqnarray}%
\begin{widetext}
Considering the Kirchhoff's current law with 
\begin{equation}
I_{C_{1}}\left( t\right)
+I_{C_{2}}\left( t\right) +I_{{C_{\text{g}}}}\left( t\right)
+I_{L_{g}}\left( t\right) +\eta I_{C_{\text{v}}}\left( t\right)
=\frac{F_{m}^{\text{A}}\left( t\right) -V_{m}^{\text{A}}\left( t\right) }{R},  \label{new_eq}
\end{equation}
we have%
\begin{eqnarray}
\left[ {C}_{1}+{C}_{2}+C_{\text{g}}+\eta C_{\text{v}}\left( {v_{m}^{\text{A}}}\right) %
\right] \frac{{d{{V_{m}^{\text{A}}}}}}{{dt}} 
-{C}_{1}\frac{{d{{V_{m}^{\text{B}}}}}}{{dt}} 
-{C}_{2}\frac{{d{{V_{m-1}^{\text{B}}}}}}{{dt}} 
+\frac{V_{R,L}}{{R}_{L_{\text{g}}}}
&=&\frac{1}{R}F_{m}^{\text{A}}-\frac{1}{R}V_{m}^{\text{A}},  \label{K1} \\
\left[ {C}_{1}+{C}_{2}+C_{\text{g}}+{\eta} C_{\text{v}}\left( {v_{m}^{\text{A}}}\right) %
\right] \frac{{d}^{2}{{{V_{m}^{\text{A}}}}}}{{dt}^{2}} 
-{C}_{1}\frac{{d}^{2}{{{V_{m}^{\text{B}}}}}}{{dt}^{2}}
-{C}_{2}\frac{{d}^{2}{{{V_{m-1}^{\text{B}}}}}}{{dt}^{2}}+%
\frac{{V_{m}^{\text{A}}-V}_{R,L}}{L_{\text{g}}}
&=&\frac{1}{R}\frac{dF_{m}^{\text{A}}}{dt}-\frac{1}{R}\frac{dV_{m}^{\text{A}}}{dt},  \label{K2}
\end{eqnarray}%
where the approximation $d{v_{m}^{\text{A}}}/dt=0$ is used since ${v_{m}^{\text{A}}}$ is
slowly-varying against time. From Eqs. (\ref{K1})-(\ref{K2}), we get the
following relation: 
\begin{eqnarray}
&&L_{\text{g}}\left( C_{\text{g}}+\eta C_{L}\right) \frac{{d}^{2}{{{V_{m}^{\text{A}}%
}}}}{{dt}^{2}} 
+L_{\text{g}}\left[ {C}_{1}+{C}_{2}+{\eta}C_{\text{NL}}\left( {%
v_{m}^{\text{A}}}\right) \right] \frac{{d}^{2}{{{V_{m}^{\text{A}}}}}}{{dt}^{2}} 
-L_{\text{g}}{C}_{1}\frac{{d}^{2}{{{V_{m}^{\text{B}}}}}}{{dt}^{2}}-L_{\text{g}}{C}_{2}\frac{{d}%
^{2}{{{V_{m-1}^{\text{B}}}}}}{{dt}^{2}}  \notag \\
&&+{R}_{L_{\text{g}}}\left( C_{\text{g}}+\eta C_{\text{L}}\right) \frac{%
dV_{m}^{\text{A}}}{dt} 
+{R}_{L}\left[ {C}_{1}+{C}_{2}+ \eta C_{\text{NL}}\left( {%
v_{m}^{\text{A}}}\right) \right] \frac{dV_{m}^{\text{A}}}{dt}
+\frac{L_{\text{g}}}{R}\frac{%
dV_{m}^{\text{A}}}{dt}-{R}_{L}{C}_{1}\frac{{d{{V_{m}^{\text{B}}}}}}{{dt}}-{R}_{L}{C}_{2}%
\frac{{d{{V_{m-1}^{\text{B}}}}}}{{dt}}  \notag \\
&&+V_{m}^{\text{A}}+\frac{{R}_{L}}{R}V_{m}^{\text{A}}  \notag \\
&=&\frac{L_{\text{g}}}{R}\frac{dF_{m}^{\text{A}}}{dt}+\frac{{R}_{L}}{R}F_{m}^{\text{A}}.
\label{eq-1}
\end{eqnarray}%
\end{widetext}
From the expressions for the voltages $V_{m}^{\text{A},\text{B}}\left( t\right) $, we can get
their first and second derivatives. Considering that $C_{\text{g}}+\eta C_{%
\text{L}}\gg {C}_{1,2,\text{NL}}$ and $R\gg {R}_{\text{L}}$, we adopt the slowly-varying 
envelop approximation. In the first line of Eq. (\ref{eq-1}), we use 
\begin{eqnarray}
\frac{{{d^{2}}{V_{m}^{\text{A},\text{B}}}}}{{d{t^{2}}}}&=&
-\mathrm{i}{\omega _{0}}\frac{{dv{_{m}^{\text{A},\text{B}}%
}}}{{dt}}{e^{-\mathrm{i}{\omega _{0}}t}}-\frac{{\omega }_{0}^{2}}{2}{v_{m}^{\text{A},\text{B}}}{%
e^{-\mathrm{i}{\omega _{0}}t}} \notag \\
&&+\text{c.c.}  \label{v3}
\end{eqnarray}%
for the first term and 
\begin{equation}
\frac{{{d^{2}}{V_{m}^{\text{A},\text{B}}}}}{{d{t^{2}}}}=-\frac{{\omega }_{0}^{2}}{2}{%
v_{m}^{\text{A},\text{B}}}{e^{-\mathrm{i}{\omega _{0}}t}}+\text{c.c.}  \label{v4}
\end{equation}%
for the other terms. In the second line, we use%
\begin{equation}
\frac{dV_{m}^{\text{A},\text{B}}}{dt}=-\frac{\mathrm{i}{\omega _{0}}}{2}{v_{m}^{\text{A},\text{B}}e^{-\mathrm{i}{\omega
_{0}}t}}+\text{c.c.}  \label{v5}
\end{equation}%
for the first and third terms, and the other terms are directly omitted. In
the third line, the second term is also omitted. Similarly, in the last line
we omit the second term. Thus, Eq. (\ref{eq-1}) reduces to%
\begin{eqnarray}
\mathrm{i}\frac{{dv{_{m}^{\text{A}}}}}{{dt}} &=&-\frac{{C}_{1}+{C}_{2}+{\eta}C_{\text{NL}%
}\left( {v_{m}^{\text{A}}}\right) }{2\left( C_{\text{g}}+\eta C_{\text{L}}\right) }{\omega }_{0}{v_{n}^{\text{A}}} \notag \\
&&-\mathrm{i}\left[ \frac{{R}_{L}}{2L_{\text{g}}}+\frac{1}{%
2R\left( C_{\text{g}}+ \eta C_{\text{L}}\right) }\right] {v_{m}^{\text{A}}} \notag \\
&&+\frac{{C}_{1}}{2\left( C_{\text{g}} +\eta C_{L}\right) }{\omega }_{0}{v_{m}^{\text{B}}}  \notag \\
&&+\frac{{C}_{2}}{2\left( C_{\text{g}}+\eta C_{\text{L}}\right) }{\omega }_{0}{v_{m-1}^{\text{B}}}\notag \\
&&+\mathrm{i}\frac{{\omega }}{2R\left( C_{\text{g}}+\eta C_{\text{L}
}\right) {\omega _{0}}}{f_{m}^{\text{A}}e^{-\mathrm{i}\left( {\omega -\omega _{0}}\right) t}.}  \label{eq-2}
\end{eqnarray}%
The slowly varying envelope approximation
simplifies the second-order differential equations into the first-order ones.
We define ${v_{m}^{\text{A},\text{B}}}\left( t\right) =
{V_{m}^{\text{A},\text{B}}}\left( t\right) \exp
\left( {\mathrm{i}{\omega _{0}}t}\right) $ and ${f_{m}^{\text{A},\text{B}}}\left( t\right) ={%
F_{m}^{\text{A},\text{B}}}\left( t\right) \exp \left( {\mathrm{i}{\omega }t}\right) $, where $%
\omega $ is the external driving frequency. Substituting the two expressions
into Eq. (\ref{eq-2}), we get%
\begin{eqnarray}
\mathrm{i}\frac{dV{_{m}^{\text{A}}}}{dt} &=&\left[ 1-\frac{{C}_{1}+{C}_{2}+\eta C_{\text{NL}%
}\left( {V_{m}^{\text{A}}}\right) }{2\left( C_{\text{g}}+\eta C_{\text{L}}\right) }%
\right] {\omega }_{0}{V_{m}^{\text{A}}} \notag \\
&&-\mathrm{i}\left[ \frac{{R}_{L}}{2L_{\text{g}}}+\frac{%
1}{2R\left( C_{\text{g}}+\eta C_{\text{L}}\right) }\right] {V_{m}^{\text{A}}}\notag \\
&&+\frac{{C}_{1}}{2\left( C_{\text{g}}+\eta C_{\text{L}}\right) }{\omega }_{0}{%
V_{m}^{\text{B}}}  \notag \\
&&+\frac{{C}_{2}}{2\left( C_{\text{g}}+\eta C_{\text{L}}\right) }{\omega }_{0}{V_{m-1}^{\text{B}}} \notag \\
&&+\mathrm{i}\frac{{\omega }}{2R\left( C_{\text{g}}+\eta C_{\text{L}%
}\right) {\omega _{0}}}{F_{m}^{\text{A}}.}  \label{eq-3}
\end{eqnarray}%
For the convenience of theoretical calculation, we introduce the frequency $\omega_{n}$ which 
is a constant with an arbitrary value. We normalize $t$
by defining $T=\omega _{n}t$ and get%
\begin{eqnarray}
\mathrm{i}\frac{{dV{_{m}^{\text{A}}}}}{{dT}} &=&\left[ 1-\frac{{C}_{1}+{C}_{2}+\eta C_{\text{%
NL}}\left( {V_{m}^{\text{A}}}\right) }{2\left( C_{\text{g}}+\eta C_{\text{L}%
}\right) }\right] \frac{{\omega _{0}}}{\omega _{n}}{V_{m}^{\text{A}}} \notag \\
&&-\mathrm{i}\left[ \frac{{R}_{L}}{2\omega _{n}L_{\text{g}}}+\frac{1}{2R\omega _{n}\left( C_{\text{g}%
}+\eta C_{\text{L}}\right) }\right] {V_{m}^{\text{A}}} \notag \\
&&+\frac{{C}_{1}}{2\left( C_{%
\text{g}}+\eta C_{\text{L}}\right) }\frac{{\omega _{0}}}{\omega _{n}}{%
V_{m}^{\text{B}}}  \notag \\
&&+\frac{{C}_{2}}{2\left( C_{\text{g}}+\eta C_{\text{L}}\right) }\frac{{%
\omega _{0}}}{\omega _{n}}{V_{m-1}^{\text{B}}} \notag \\
&&+\mathrm{i}\frac{1}{2R{\omega _{0}}\left( C_{%
\text{g}}+\eta C_{\text{L}}\right) }\bar{\omega}{F_{m}^{\text{A}},}  \label{eq-4}
\end{eqnarray}%
where $\bar{\omega}={\omega /}\omega _{n}$ is the normalized external
driving frequency. This equation can be rewritten as%
\begin{eqnarray}
\mathrm{i}\frac{{dV{_{m}^{\text{A}}}}}{{dT}}&=&\left( E_{0}-\mathrm{i}\gamma \right) {V_{m}^{\text{A}}+g\left( 
{V_{m}^{\text{A}}}\right) V_{m}^{\text{A}}}+J_{1}{V_{m}^{\text{B}}} \notag \\
&&+J_{2 }{V_{m-1}^{\text{B}}}+\mathrm{i}d{F_{m}^{\text{A}},}  \label{eq-5}
\end{eqnarray}%
where%
\begin{equation}
E_{0} =\left[ 1-\frac{{C}_{1}+{C}_{2}}{2\left( C_{\text{g}}+\eta C_{\text{L%
}}\right) }\right] \frac{{\omega _{0}}}{\omega _{n}}  \label{E0}
\end{equation}
is the constant onsite energy,
\begin{equation}
\gamma =\gamma _{L}+\gamma _{R}=\frac{{R}_{L}}{2\omega _{n}L_{\text{g}}}+%
\frac{1}{2R\omega _{n}\left( C_{\text{g}}+\eta C_{\text{L}}\right) } \label{gamma}
\end{equation}
is the dissipation coefficient,
\begin{equation}
J_{1} =\frac{{C}_{1}}{2\left( C_{\text{g}}+\eta C_{\text{L}}\right) }\frac{%
{\omega _{0}}}{\omega _{n}}  \label{J} \\
\end{equation}
and
\begin{equation}
J_{2} =\frac{{C}_{2}}{2\left( C_{\text{g}}+\eta C_{\text{L}}\right) }\frac{%
{\omega _{0}}}{\omega _{n}}  \label{Jp} \\
\end{equation}
are the intracell and intercell coupling coefficients, respectively,
\begin{equation}
g\left( {V_{m}^{\text{A}}}\right) =-\frac{\eta C_{\text{NL}}\left( {V_{m}^{A}}%
\right) }{2\left( C_{\text{g}}+\eta C_{\text{L}}\right) }\frac{{\omega _{0}}%
}{\omega _{n}}  \label{gA} 
\end{equation}
is the voltage-dependent onsite energy, and
\begin{equation}
d =\frac{1}{2R{\omega _{0}}\left( C_{\text{g}}+\eta C_{\text{L}}\right) }%
\bar{\omega}  \label{d}
\end{equation}
is the coupling coefficient between the excitation source and circuit node.
Note that $\gamma _{L}$ and $\gamma _{R}$ denote the dissipations induced by the
series resistance of the inductors and shunt resistors, respectively. Following
the similar procedure, the equation for $V_{m}^{\text{B}}$ is written as
\begin{eqnarray}
\mathrm{i}\frac{{dV{_{m}^{\text{B}}}}}{{dT}}&=&\left( E_{0}-\mathrm{i}\gamma \right) {V_{m}^{\text{B}}+g\left( 
{V_{m}^{\text{B}}}\right) V_{m}^{\text{B}}}+J_{1}{V_{m}^{\text{A}}} \notag \\
&&+J_{2}{V_{m+1}^{\text{A}}}+\mathrm{i}d{F_{m}^{\text{B}},}  \label{eq-6}
\end{eqnarray}%
where%
\begin{equation}
g\left( {V_{m}^{\text{B}}}\right) =-\frac{\eta C_{\text{NL}}\left( {V_{m}^{\text{B}}}%
\right) }{2\left( C_{\text{g}}+\eta C_{\text{L}}\right) }\frac{{\omega _{0}}%
}{\omega _{n}}.  \label{gB}
\end{equation}
Thus, starting from the Kirchhoff circuit equations, we have analytically derived the discretized
GP equation. Due to the external driving and circuit dissipation, Eqs. (\ref{eq-5}) and
(\ref{eq-6}) contain the driven and dissipative terms. If we are interested in the mode
properties of the nonlinear
topological states in the SSH circuit, we can disconnect the voltage sources and shunt resisters,
and neglect the series resistance of the inductors. With $R = \infty$ and $R_{L} = 0$, 
Eqs. (\ref{eq-5}) and (\ref{eq-6}) reduce to
\begin{eqnarray}
\mathrm{i}\frac{{dV{_{m}^{\text{A}}}}}{{dT}}&=& E_{0} {V_{m}^{\text{A}}+g\left( 
{V_{m}^{\text{A}}}\right) V_{m}^{\text{A}}}+J_{1}{V_{m}^{\text{B}}}\notag \\
&&+J_{2 }{V_{m-1}^{\text{B}}}, 
\label{eq-7} \\
\mathrm{i}\frac{{dV{_{m}^{\text{B}}}}}{{dT}}&=& E_{0}{V_{m}^{\text{B}}+g\left( 
{V_{m}^{\text{B}}}\right) V_{m}^{\text{B}}}+J_{1}{V_{m}^{\text{A}}} \notag \\
&&+J_{2}{V_{m+1}^{\text{A}}},  
\label{eq-8}
\end{eqnarray}%
with the parameters the same as those defined in Eqs. (\ref{E0}), (\ref{J})-(\ref{gA}), and (\ref{gB}).
Eqs. (\ref{eq-7})-(\ref{eq-8}) are exactly the discretized GP equation introduced in the main text.

\begin{figure}[tbp]
\includegraphics[width=7.8cm]{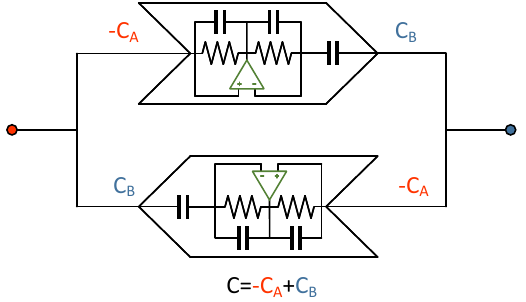}
\caption{Circuit diagram of the negative capacitor. The upper component has a negative 
capacitance $-C_{\text{A}}$ at the left side and positive capacitance $C_{\text{B}}$ at the 
right side. While the lower component has a positive 
capacitance $C_{\text{B}}$ at the left side and negative capacitance 
$-C_{\text{A}}$ at the right side. Under such configuration, the total capacitance $C$ is
negative if $\vert C_{\text{A}} \vert >  \vert C_{\text{B}} \vert$.}
\label{fig_NIC}
\end{figure}

We would like to note that, the coupling coefficients $J_{1}$ and $J_{2}$ 
in Eqs. (\ref{eq-5}) and (\ref{eq-6}) are both
positive. However, the coupling phases can be tuned by changing the coupling capacitors. 
For instance, $J_{1}$ and $J_{2}$ would be negative if we use the negative capacitors.
Fig. \ref{fig_NIC} shows the circuit diagram of the negative capacitor.  
It is composed of two parallelly connected negative impedance converters (NICs) \cite{PRL122-247702}. 
The upper NIC creates the negative capacitance $-C_{\text{A}}$ at the left side and positive 
capacitance $C_{\text{B}}$ at the right side. Reversely, the lower NIC creates the positive 
capacitance $C_{\text{B}}$ at the left side and negative capacitance 
$-C_{\text{A}}$ at the right side. Thus, we have $C = C_{\text{A}} + C_{\text{B}}$, and the total capacitance is 
negative if $\vert C_{\text{A}} \vert >  \vert C_{\text{B}} \vert$.

The validity of Eqs. (\ref{eq-5}) and (\ref{eq-6}), as well as Eqs. (\ref{eq-7})-(\ref{eq-8}), 
require that $C_{\text{g}}+\eta
C_{\text{L}}\gg {C}_{1,2,\text{NL}}$ and $R\gg {R}_{L}$. These conditions
can be easily satisfied by properly determining the component values. Thus,
considering the realistic experimental realization, the circuit lattice is
captured by the discretized GP equation.
Besides, we would like to note that Eqs. (\ref{eq-5}) and (\ref{eq-6}) can also be treated
as the temporal coupled-mode equations in the circuit lattices \cite{IEEE40-1511}.

\subsection{Validity of GP equation}

To validate the correctness of the GP equation, we have also developed an accurate
nonlinear circuit model which can be used to study the steady states in the circuits. 
We study the SSH circuit lattice shown in Fig. \ref%
{fig_GP} and consider the case where the external excitation is
monochromatic. For the $m$th unit cell, at node $\text{A}$ where the voltage is $%
V_{m}^{\text{A}} \left( t \right)$, Eqs. (\ref{r1})-(\ref{r6}) are still valid when the circuit
reaches the steady state, and the only difference is that ${v_{m}^{\text{A}}}$ in
Eq. (\ref{r4}) is now time independent. Transforming the relations into the
frequency domain, we have%
\begin{eqnarray}
-\mathrm{i}\omega L_{\text{g}}{i{_{L_{\text{g}}}}} &=&{v_{m}^{\text{A}}-v}_{R,L},  \label{r7}
\\
i_{L_{\text{g}}}{R}_{L}&=&v_{R,L},  \label{r8} \\
-\mathrm{i}\omega C_{\text{g}}{v{_{m}^{\text{A}}}} &=&i_{C_{\text{g}}},  \label{r9} \\
-\mathrm{i}\omega C_{v}\left( {v_{m}^{\text{A}}}\right) {v{_{m}^{\text{A}}}} &=&i_{C_{v}},
\label{r10} \\
-\mathrm{i}\omega {C}_{2}\left( {{{v_{m}^{\text{A}}}-v{_{m-1}^{\text{B}}}}}\right) &=&{i_{C_{2}}},
\label{r11} \\
-\mathrm{i}\omega {C}_{1}\left( {{{v_{m}^{\text{A}}}-v{_{m}^{\text{B}}}}}\right)&=&{i_{C_{1}}}.
\label{r12}
\end{eqnarray}%
The Kirchhoff's current law is also valid in the frequency domain with $%
i_{C_{1}}+i_{C_{2}}+i_{{C_{\text{g}}}}+i_{L_{\text{g}}}+\eta i_{C_{v}}=\frac{%
f_{m}^{\text{A}}-v_{n}^{\text{A}}}{R}$. Then we have%
\begin{eqnarray}
&&-\mathrm{i}\omega {C}_{1}\left( {{{v_{m}^{\text{A}}}-v{_{m}^{\text{B}}}}}\right) -\mathrm{i}\omega {C}%
_{2}\left( {{{v_{m}^{\text{A}}}-v{_{m-1}^{\text{B}}}}}\right) -\mathrm{i}\omega C_{\text{g}}v%
_{m}^{\text{A}} \notag \\
&&+\frac{{v_{m}^{\text{A}}}}{-\mathrm{i}\omega L_{\text{g}}{{+}R}_{L}}-\mathrm{i}\omega \eta
C_{v}\left( {v_{m}^{\text{A}}}\right) v_{m}^{\text{A}}\notag \\
&=&\frac{f_{m}^{\text{A}}-v_{m}^{\text{A}}}{R},
\end{eqnarray}%
and the equation can be rewritten as%
\begin{eqnarray}
&&-\mathrm{i}\omega \left[ {C}_{1}+C_{2}+C_{\text{g}}+\eta C_{v}\left( {v_{m}^{\text{A}}}%
\right) \right] {{{v_{m}^{\text{A}}}+}}\frac{{v_{m}^{\text{A}}}}{-\mathrm{i}\omega L_{\text{g}}{{+}R%
}_{L}} \notag \\
&&+\frac{v_{m}^{\text{A}}}{R}+\mathrm{i}\omega {C}_{1}{{v{_{m}^{\text{B}}}}}
+\mathrm{i}\omega {C}_{2}v{_{m-1}^{\text{B}}} \notag \\
&=&\frac{f_{m}^{\text{A}}}{R}.
\end{eqnarray}%
Similarly, the equation for $v_{m}^{\text{B}}$ is%
\begin{eqnarray}
&&-\mathrm{i}\omega \left[ {C}_{1}+C_{2}+C_{\text{g}}+\eta C_{v}\left( {v_{m}^{\text{B}}}%
\right) \right] {{{v_{m}^{\text{B}}}+}}\frac{{v_{m}^{\text{B}}}}{-\mathrm{i}\omega L_{\text{g}}{+R%
}_{L}} \notag \\
&&+\frac{v_{m}^{\text{B}}}{R}+\mathrm{i}\omega {C}_{1}{{v{_{m}^{\text{A}}}}}+\mathrm{i}\omega {C}_{2}v{%
_{m+1}^{\text{A}}} \notag \\
&=&\frac{f_{m}^{\text{B}}}{R}.
\end{eqnarray}%
The circuit equations can be further rewritten as%
\begin{eqnarray}
&&\omega \left[ {C}_{1}+C_{2}+C_{\text{g}}+ \eta C_{v}\left( {v_{m}^{\text{A}}}%
\right) \right] {{{v_{m}^{\text{A}}}-}}\frac{\omega L_{g}}{\omega ^{2}L_{\text{g}%
}^{2}+{R}_{L}^{2}}v_{m}^{\text{\text{A}}} \notag \\
&&+\mathrm{i}\frac{{R}_{L}}{\omega ^{2}L_{\text{g}}^{2}+{R}%
_{L}^{2}}{v_{m}^{\text{A}}}+\frac{\mathrm{i}}{R}v_{m}^{\text{A}}-\omega {C}_{1}{{v{_{m}^{\text{B}}}}}%
-\omega {C}_{2}{{v{_{m-1}^{\text{B}}}}}  \notag \\
&=&\frac{{\mathrm{i}}}{R}f_{m}^{\text{A}}, \\
&&\omega \left[ {C}_{1}+C_{2}+C_{\text{g}}+ \eta C_{v}\left( {v_{m}^{\text{B}}}%
\right) \right] {v_{m}^{\text{B}}}-\frac{\omega L_{\text{g}}}{\omega ^{2}L_{%
\text{g}}^{2}+{R}_{L}^{2}}v_{m}^{\text{B}} \notag \\
&&+\mathrm{i}\frac{{R}_{L}}{\omega ^{2}L_{g}^{2}+{R}%
_{L}^{2}}v_{m}^{\text{B}}+\frac{\mathrm{i}}{R}v_{m}^{\text{B}}-\omega {C}_{1}{{v{_{m}^{\text{A}}}}}%
-\omega {C}_{2}{{v{_{m+1}^{\text{A}}}}} \notag \\
&=&\frac{\mathrm{i}}{R}f_{m}^{\text{B}}.
\end{eqnarray}%
To normalize the equations, we multiply the above equations with $\omega L_{\text{g}%
} $. The resulting new equations are 
\begin{eqnarray}
&&\omega ^{2}L_{\text{g}}\left( C_{\text{g}}+\eta C_{\text{L}}+{C}%
_{1}+C_{2}\right) v_{m}^{\text{A}}+\omega ^{2}L_{\text{g}} \eta C_{\text{NL}%
}\left( {v_{m}^{\text{A}}}\right) v_{m}^{\text{A}} \notag \\
&&-\frac{\omega ^{2}L_{\text{g}%
}^{2}}{\omega ^{2}L_{\text{g}}^{2}+{R}_{L}^{2}}v_{m}^{\text{A}}+\mathrm{i}\frac{\omega L_{%
\text{g}}{R}_{L}}{\omega ^{2}L_{\text{g}}^{2}+{R}_{L}^{2}}{v_{m}^{\text{A}}}+\mathrm{i}\frac{%
\omega L_{\text{g}}}{R}v_{m}^{\text{A}}  \notag \\
&&-\omega ^{2}L_{\text{g}}{C}_{1}{v_{m}^{\text{B}}}-\omega ^{2}L_{g}{C}_{2}{{v{%
_{m-1}^{\text{B}}}}-}\mathrm{i}\frac{\omega L_{\text{g}}}{R}f_{m}^{\text{A}}  \notag \\
&=&0,  \label{circuit_eq_1} \\
&&\omega ^{2}L_{\text{g}}\left( C_{\text{g}}+\eta C_{\text{L}}+{C}%
_{1}+C_{2}\right) {v_{m}^{\text{B}}+}\omega ^{2}L_{\text{g}} \eta C_{\text{NL}%
}\left( v_{m}^{\text{B}}\right) v_{m}^{\text{B}} \notag \\
&&-\frac{\omega ^{2}L_{\text{g}}^{2}%
}{\omega ^{2}L_{\text{g}}^{2}+{R}_{L}^{2}}v_{m}^{\text{B}}+\mathrm{i}\frac{\omega L_{\text{g}%
}{R}_{L}}{\omega ^{2}L_{\text{g}}^{2}+{R}_{L}^{2}}v_{m}^{\text{B}}+\mathrm{i}\frac{\omega
L_{\text{g}}}{R}v_{m}^{\text{B}}  \notag \\
&&-\omega ^{2}L_{\text{g}}{C}_{1}{{v{_{m}^{\text{A}}}}}-\omega ^{2}L_{\text{g}}{C}%
_{2}{v{_{m+1}^{\text{A}}}}-\mathrm{i}\frac{\omega L_{\text{g}}}{R}f_{m}^{\text{B}}  \notag \\
&=&0.  \label{circuit_eq_2}
\end{eqnarray}%
Similarly, we can also introduce the normalized frequency $\bar{\omega}={\omega /}\omega _{n}$
into the circuit equations. From Eqs. (\ref{circuit_eq_1})-(\ref{circuit_eq_2}), the first and second
terms are related to the grounding capacitors, coupling capacitors, and
varactor diodes. The third and fourth terms are related to the inductors,
which includes both the inductance and series resistance of the inductors.
The fifth term is related to the shunt resistors which are wired to the
driving voltage sources. The sixth and seventh terms are related to the
coupling capacitors. And the last term at the left of the equal sign is
related to the driving voltage sources. Note that Eqs. (\ref{circuit_eq_1})-(\ref{circuit_eq_2})
can also be used to study the mode properties of the nonlinear topological states, as long 
as we let
$R = \infty$ and $R_{L} = 0$. Thus, Eqs. (\ref{circuit_eq_1})-(\ref{circuit_eq_2}) governs
the steady states in the circuit lattices without any approximation and we can validate
the results from the GP equation using the accurate nonlinear circuit model. In this paper,
we primarily utilize the GP equation to investigate the nonlinear topological states, including
the mode properties and excitations, because the mathematical form of GP equation is
simple and GP equation has be widely studied in the field of nonlinear physics. 

\subsection{Interacting topological physics}

Electric circuits have been proposed as a versatile platform to
study the various topological phases in the non-interacting systems and few-body
interacting systems. Due to the flexibility of constructing the circuit
lattices and the site-resolved, phase-resolved, time-resolved, and
frequency-resolved measurement techniques, electric circuits have shown
significant values in exploring the high-dimensional, higher-order,
non-Hermitian, non-Abelian and nonlinear topological physics. 
Specifically, by mapping the physical dimension to the particle number, 
electric circuits have been used to reveal
the novel topological phases in few-body interacting systems \cite%
{ncommun11-1436}. This mapping faces a great challenge when the
number of the interacting particles is large. However,
the derivation of the GP equation provides a direct link between the
many-body interacting systems and nonlinear electric circuits, and it is feasible to
explore the many-body interacting topological physics in nonlinear circuit lattices.

\begin{figure}[tbp]
\includegraphics[width=8cm]{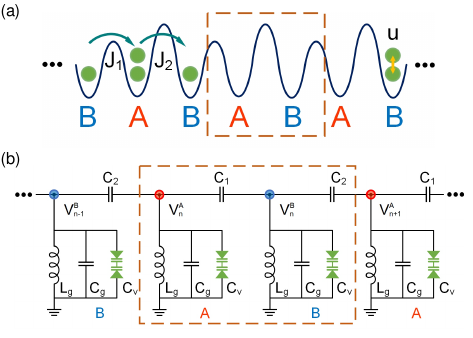}
\caption{Implementation of the interacting SSH lattice. (a) Many
interacting bosons in an SSH lattice with the chemical potential $E_{0}$,
tunneling coefficients $J_{1,2}$, and onsite interaction strength $u$. The red rectangle denotes one unit
cell in the lattice. (b) Schematic of a nonlinear electric circuit lattice
that emulates the interacting SSH lattice.}
\label{fig_interacting}
\end{figure}

As a minimal model
featuring the interacting topological physics, we consider many interacting
bosons in a dimerized SSH lattice, illustrated in Fig. %
\ref{fig_interacting}(a). Each minimum of the potential corresponds to an individual
site, which is called $\text{A}$ or $\text{B}$. The bosons at sites $\text{A}$ and $\text{B}$
experience the same chemical potential $E_{0}$. Since the barriers between the sites have the 
different heights, the tunneling coefficients $J_{1}$ and $J_{2}$ are also
different. A pair of particles occupying the same site experience an onsite
interaction strength $u$, and the weak interactions between the particles from the
neighbor sites are neglected. The Hamiltonian describing this system reads 
\begin{eqnarray}
\hat{H} =&& -J_{1}\sum_{m}\left( c_{m}^{\text{B}\dag }c_{m}^{\text{A}}+\text{H.c.}\right)
-J_{2}\sum_{m}\left( c_{m+1}^{\text{A}\dag }c_{m}^{\text{B}}+\text{H.c}\right)  \notag \\
&&+\frac{u}{2}\sum_{m}\left[ n_{m}^{\text{A}}\left( n_{m}^{\text{A}}-1\right)
+n_{m}^{\text{B}}\left( n_{m}^{\text{B}}-1\right) \right]  \notag \\
&&+E_{0}\sum_{m}\left( n_{m}^{\text{A}}+n_{m}^{\text{B}} \right),  \label{H1}
\end{eqnarray}
where $c_{m}^{\sigma\dag }$ and $c_{m}^{\sigma}$ are the creation and
annihilation operators for the particles at the site $\text{A}$ ($\sigma = \text{A}$) 
or $\text{B}$ ($%
\sigma = \text{B}$) of the lattice cell $m$, the summation enumerated with the index $m$
is performed over all the cells in the lattice, $\text{H.c.}$ denotes the Hermitian
conjugate of the term to the left, and $n_{m}^{\sigma}=c_{m}^{\sigma\dag
}c_{m}^{\sigma}$ are the particle number operators. This Hamiltonian is the
extended version of the Bose-Hubbard Hamiltonian \cite{PR129-959}.
Under the mean-field approximation, the macrooccupied state $\left\vert \Phi
\right\rangle $ is chosen as the tensor product of the Glauber coherent states 
\cite{PRL80-2189}. With $H=\left\langle \Phi \right\vert \hat{H}\left\vert
\Phi \right\rangle $, the semiclassical Hamiltonian is%
\begin{eqnarray}
H =&& -J_{1}\sum_{m}\left( \psi _{m}^{\text{B}\ast }\psi _{m}^{\text{A}}+\text{H.c.}\right)
-J_{2}\sum_{m}\left( \psi _{m+1}^{\text{A}\ast }\psi _{m}^{\text{B}}+\text{H.c.}\right) \notag \\
&&+\frac{u}{2}\sum_{m}\left( \left\vert \psi _{m}^{\text{A}}\right\vert
^{4}+\left\vert \psi _{m}^{\text{B}}\right\vert ^{4}\right) \notag \\
&&+E_{0}\sum_{m}\left( \left\vert \psi _{m}^{\text{A}}\right\vert
^{2}+\left\vert \psi _{m}^{\text{B}}\right\vert ^{2}\right) .  \label{H2}
\end{eqnarray}%
Using the Hamiltonian's equation $\mathrm{i}\hbar \dot{\psi}_{m}^{\sigma}=\delta
H/\delta \psi _{m}^{\sigma\ast }$, we get the discretized GP equation:
\begin{eqnarray}
\mathrm{i}\hbar \frac{d}{dt}\left[ 
\begin{array}{c}
\psi _{m}^{\text{A}} \\ 
\psi _{m}^{\text{B}}%
\end{array}%
\right] &=& E_{0}\left[ 
\begin{array}{c}
\psi _{m}^{\text{A}} \\ 
\psi _{m}^{\text{B}}%
\end{array}%
\right]
-J_{1}\left[ 
\begin{array}{c}
\psi _{m}^{\text{B}} \\ 
\psi _{m}^{\text{A}}%
\end{array}%
\right] -J_{2}\left[ 
\begin{array}{c}
\psi _{m-1}^{\text{B}} \\ 
\psi _{m+1}^{\text{A}}%
\end{array}%
\right]  \notag \\
&&+u\left[ 
\begin{array}{c}
\left\vert \psi _{m}^{\text{A}}\right\vert ^{2}\psi _{m}^{\text{A}} \\ 
\left\vert \psi _{m}^{\text{B}}\right\vert ^{2}\psi _{m}^{\text{B}}%
\end{array}%
\right] .  \label{eq1}
\end{eqnarray}%
This approach reduces a many-body interacting problem to the single-particle
nonlinear one. The GP equation, also known as the nonlinear Schr\"{o}dinger
equation, provides a simplified treatment of the complex many-body systems.

Considering the GP equation for the nonlinear electric circuit lattices,
Eqs. (\ref{eq-7})-(\ref{eq-8} can be rewritten as
\begin{eqnarray}
\mathrm{i}\frac{d}{dt}\left[ 
\begin{array}{c}
V_{m}^{\text{A}} \\ 
V_{m}^{\text{B}}%
\end{array}%
\right] =&& E_{0}\left[ 
\begin{array}{c}
V_{m}^{\text{A}} \\ 
V_{m}^{\text{B}}%
\end{array}%
\right] +J_{1}\left[ 
\begin{array}{c}
V_{m}^{\text{B}} \\ 
V_{m}^{\text{A}}%
\end{array}%
\right] +J_{2}\left[ 
\begin{array}{c}
V_{m-1}^{\text{B}} \\ 
V_{m+1}^{\text{A}}%
\end{array}%
\right]  \notag\\
&&+\left[ 
\begin{array}{c}
g{V_{m}^{\text{A}}} \\ 
g{V_{m}^{\text{B}}}%
\end{array}%
\right] .  \label{gp}
\end{eqnarray}%
Since Eq. (\ref{eq1}) describes the many interacting bosons in an SSH lattice and
Eq. (\ref{gp}) is mathematically equivalent to Eq. (\ref{eq1}),
we can use the nonlinear circuit lattice shown in Fig. \ref{fig_interacting}(b) to explore
the many-body interacting topological physics in an SSH lattice.
By comparing the two equations, the voltage at a circuit node
corresponds to the wavefunction of the particles occupied at a lattice site. 
In Eq. (\ref{gp}), the first term on the right side of the equal sign corresponds to the onsite potential 
term in Eq. (\ref{eq1}), and it only introduces an offset to the band structure.
The coupling terms in Eq. (\ref{gp}) correspond to the tunneling terms in Eq. (\ref{eq1}), and the
coupling phases can be tuned by changing the coupling capacitors. 
The last term, which is a nonlinear term, is
equivalent to the onsite interaction term in Eq. (\ref{eq1}). Although the
nonlinear term for a circuit lattice exhibits the saturable nonlinearity,
which is different to the cubic Kerr nonlinearity in Eq. (\ref{eq1}), this
difference does not affect the topological physics that we study here. For
an interacting system, the particle interaction leads to the increase or
decrease of the effective onsite energy, depending on whether the particle
interaction is attractive or repulsive \cite{nphys8-267}. While for the
circuit lattice in Fig. \ref{fig_interacting}(b), the interaction term always increases
the onsite energy since the capacitance of the varactor diode decreases with an
increasing voltage. Besides, in an interacting system the total particle
number $N=\sum_{m} \left(n_{m}^{A} + n_{m}^{B} \right)$ is preserved but the
interaction strength $u$ may change, e.g. through the Feshbach resonance 
\cite{RMP82-1225}. In a circuit lattice, the parameters $C_{%
\text{L}}$, $\nu_{0}$, and $M$ are fixed, but we can change the nonlinearity strength
by tuning the voltages \cite{nelectron1-178}. Consequently, under the slowly-varying 
envelope approximation, the voltages for a circuit lattice composed
of the coupled nonlinear $LC$ oscillators are also governed by the GP equation.
Due to the mathematical correspondence, it is feasible to explorer the
many-body interacting topological physics in nonlinear electric circuits.

\section{Theoretical results of the nonlinear topological edge states\label{app_C}}

In this section, we give more theoretical results of the nonlinear topological edge states in the typical 
semi-infinite SSH lattice that ends with a weak bond (Fig. \ref{fig_findings}(a)). 

Generally speaking, to calculate the frequencies and voltage distributions of the localized states in the nonlinear SSH
circuit lattices, we solve the GP equation (i.e., Eq. (\ref{gp})) using the ansatz $%
V_{m}^{\text{A,B}}=v _{m}^{\text{A,B}}\exp \left( -\mathrm{i}\omega t\right)$. 
We employ Newton's method to solve the eigenvalue
equation for each $\omega$. Open boundary conditions are used to truncate 
the nonlinear circuit lattice. For the calculation of the nonlinear topological 
edge states, the linear topological edge states with scaling factors are taken 
as initial guesses. To calculate the topological gap solitons, 
self-induced topological edge states, and other topologically trivial solitons,
we employ the anti-continuum (AC) approach. Once we obtain the soliton solution 
at a given $\omega$, solutions at other frequencies can be obtained iteratively.
The stability of the nonlinear states is analyzed using the standard linear 
stability technique and subsequently confirmed through temporal evolution 
based on the Runge-Kutta algorithm.

The excitation spectra are calculated by solving the driven-dissipative 
GP equaiton (i.e., Eqs. (\ref{eq-5}) and (\ref{eq-6})). Dissipation is characterized by $\gamma$ at the input
node and $\gamma_{L}$ at the other nodes. Newton's method is employed to 
find the complex solutions. We calculate the voltage distributions at two frequencies 
away from the resonant frequencies, and use these solutions as initial guesses to 
obtain the entire spectrum. Finally, for a fixed driving voltage, we obtain two voltage 
spectra: one curve is obtained by increasing the driving frequency, while the other 
corresponds to a frequency sweep in the opposite direction. The bistable response 
can be observed by comparing the two voltage spectra.

In the theoretical calculation of 
the mode properties of the nonlinear topological edge states, we consider an SSH lattice with 
$N=120$ unit cells, i.e., 240 sites. In the linear limit, the linear topological edge states residing at the 
opposite two edges are decoupled, thereby avoiding the finite size effect. For simplicity, we only study 
the nonlinear topological edge states that reside at the left edge of the lattice. 

\begin{figure}[tbp]
\includegraphics[width=8.6cm]{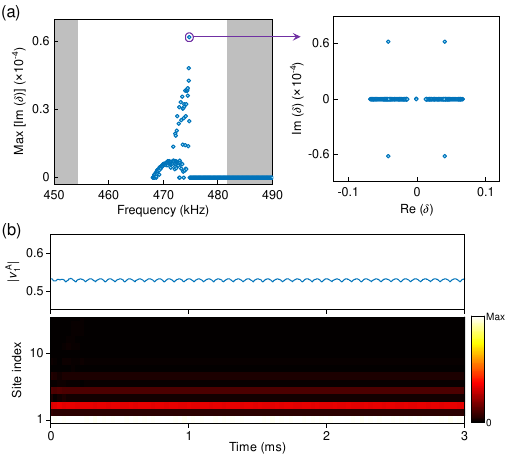}
\caption{Stability analysis of the nonlinear topological edge states. (a)
The maximum growth rates of the perturbed solutions of the nonlinear topological edge states. 
The gray areas denote the linear bulk bands. The inset shows the linear stability spectrum
at the frequency which has the largest
$\text{Max} \left[ \text{Im} \left(\delta \right) \right]$.
(b) The temporal evolution of the nonlinear topological edge state which has the largest 
$\text{Max} \left[ \text{Im} \left(\delta \right) \right]$ [indicated in (a)].
Noises with $\pm 5\%$ random perturbations are added to the initial input.}
\label{fig_edge_stability}
\end{figure}

We then study the stability analysis of the nonlinear topological edge states. 
Figure \ref{fig_edge_stability}(a) shows the maximum growth rates at the 
different frequencies
of the nonlinear topological edge states. From the figure, the maximum growth rates are in the
order of $10^{-5}$, implying that the nonlinear topological edge states can be considered 
linearly stable. We further find the frequency which has the largest $\text{Max} \left[ \text{Im} \left(
\delta \right) \right]$ and plot the linear stability spectrum in the inset of Fig. \ref{fig_edge_stability}(a). 
Compared with the real parts of $\delta$, the imaginary parts are negligible. Thus, at least within the 
experimental measurement period, the nonlinear topological edge states do not exhibit the 
instability, such as the exponential increase or significant oscillation. To confirm the results from
the linear stability analysis, we further add the noises with $\pm 5\%$ random perturbations 
to the amplitude of the nonlinear topological edge state which has the largest
$\text{Max} \left[ \text{Im} \left(
\delta \right) \right]$, and study the temporal evolution. From the voltage distribution 
and edge voltage shown in Fig. \ref{fig_edge_stability}(b), the nonlinear topological edge
state is stable at least up to $3~\text{ms}$. 

\section{Experimental measurement of the nonlinear topological edge states\label{app_D}}

In this section, we give more experimental results of the nonlinear topological edge states in the typical 
semi-infinite SSH lattice that ends with a weak bond (Fig. \ref{fig_findings}(a)). 

Generally speaking, to ensure the observation of the nonlinear topological states, the circuit components should have
minimal parasitic parameters, and their tolerance should be as low as possible. 
For this purpose, we utilize capacitors with low ESL and $\pm 1\%$ tolerance. 
We also employ inductors with magnetic shielding and low DCR (SPM12565VT-150M-D), and delicately 
select the components using an LCR meter (HIOKI IM3536). 
After performing Monte Carlo simulations, the tolerance for the inductance of the magnetically shielded inductors is set to $\pm 1\%$.
Although the series resistance of these inductors increases 
with higher frequencies, we maintain the tolerances for series resistance at $\pm 2\%$ across all frequencies
of interest in our study (see error analysis in Appendix \ref{app_F}, section 3). 
Our measurements indicate that the average series resistance of the magnetically shielded inductors 
is approximately $600~\text{m}\Omega$. 
It is important to note that our measurements were conducted 
under a fixed current, and the series resistance of the inductors increases with larger currents.
To characterize the voltage response of
varactor diodes (BB201), we measure the $C$-$V$ curves and the parameters $%
C_{\text{L}}$, $v_{0}$, and $M$ are obtained by fitting these curves with the
phenomenological formula. 
We use standard PCB techniques to fabricate the
lattice, ensuring that the inductors are sufficiently spaced to prevent mutual coupling.
The PCB traces have a relatively large width of $0.75~\text{mm}$ to accommodate 
high currents, and the layouts are carefully optimized to minimize parasitic parameters 
and coupling with other circuit components.

To excite the circuit lattices, SubMiniature
version A (SMA) connectors are soldered onto the PCB for signal injection. 
Since the output voltage of an arbitrary function generator (Tektronix AFG31022) is typically low, 
we use a high-voltage amplifier (Aigtek ATA-2022B) to amplify the voltage signals. The amplified 
voltage signals are then injected into the circuit sample through the SMA connectors.
Note that the leakage current of the high-voltage amplifier is negligible compared to the operating current of the circuit sample.
Due to bandwidth limitations, the output voltage of the high-voltage amplifier may fall below the 
preset value at high operating frequencies. Additionally, the two channels of the high-voltage amplifier 
may not provide strictly equal amplification factors. To detect the voltage signals at the circuit nodes, 
custom connectors on the PCB are connected to an oscilloscope (Tektronix MDO34).
For the excitation spectrum measurement, we fix the input voltage and conduct
program-controlled frequency sweeps, during which the voltage signals at the 
circuit nodes are recorded.

\begin{figure}[tbp]
\includegraphics[width=8.6cm]{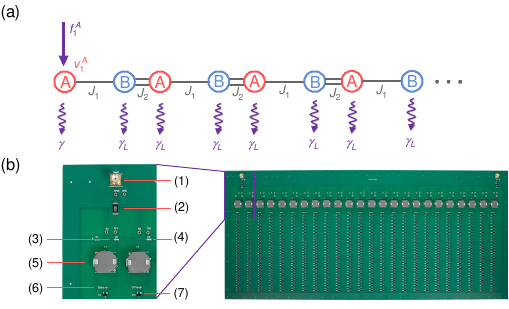}
\caption{Experimental implementation of the SSH lattice which is topologically nontrivial
in the linear limit (Fig. \ref{fig_findings}(a)).
(a) Schematic of the nonlinear SSH lattice which is topologically nontrivial in the 
linear limit. The external driving voltage is $f_{1}^{\text{A}}$ and the voltage at the 
edge node is $v_{1}^{\text{A}}$.
(b) Fabricated PCB of the driven-dissipative nonlinear circuit
lattice with the enlarged figure showing the circuit components:
(1) SMA connector, (2) shunt resistor $R$, (3) coupling capacitor $C_{1}$,
(4) coupling capacitor $C_{2}$, (5) grounding inductor $L_{\text{g}}$, (6)
grounding capacitor $C_{\text{g}}$, and (7) varactor diode $C_{\text{v}}$.}
\label{fig_edge_PCB}
\end{figure}

To theoretically demonstrate the excitation and observation of the nonlinear topological
edge states, we first solve the driven-dissipative GP equation 
with the driven terms ${F_{m}^{\text{A}}}\left( t\right) =\delta _{m,1}{f_{0}^{\text{A}}}\exp \left( -%
{\mathrm{i}\bar{\omega}T}\right) $ and $F_{m}^{\text{B}} \left( t \right) = 0$, 
i.e., only the edge site is excited.
Then the equations are numerically solved by following the same procedure. 
In order to experimentally observe the nonlinear topological edge states, we excite 
the leftmost edge site with the external driving voltage $f_{1}^{\text{A}}$. Similarly, the
dissipations induced both from the series resistance of the inductors and shunt
resistors are considered. Figure \ref{fig_edge_PCB}(a) schematically shows the
nonlinear SSH lattice with the nearest neighbour couplings $J_{1,2}$, the
dissipations $\protect\gamma_{L}$ and $\protect\gamma$, and the external
driving voltage $f_{1}^{\text{A}}$. In the linear limit, this lattice is a topologically nontrivial 
lattice since $J_{1} < J_{2}$.
Figure \ref{fig_edge_PCB}(b) shows the fabricated PCB of the
driven-dissipative nonlinear circuit lattice, and the inset shows the
enlarged figure with the circuit components: (1) SMA connector, (2) shunt resistor $R$, 
(3) coupling capacitor $%
C_{1}$, (4) coupling capacitor $C_{2}$, (5) grounding inductor $L_{\text{g}}$%
, (6) grounding capacitor $C_{\text{g}}$, and (7) varactor diode $C_{\text{v}}$.
The SMA connector is connected to an external voltage source with the
voltage amplitude $f_{1}^{\text{A}}$. Here, the experimental circuit lattice consists 
of $N = 12$ unit cells. We determine this value of $N$ by noting that the voltage spectra for lattices 
with $N = 12$ and $N = 120$ are identical (for example, see the discussion on nonlinear topological 
interface states in Appendix \ref{app_F}, section 3).

\begin{figure*}[tbp]
\includegraphics[width=17.3cm]{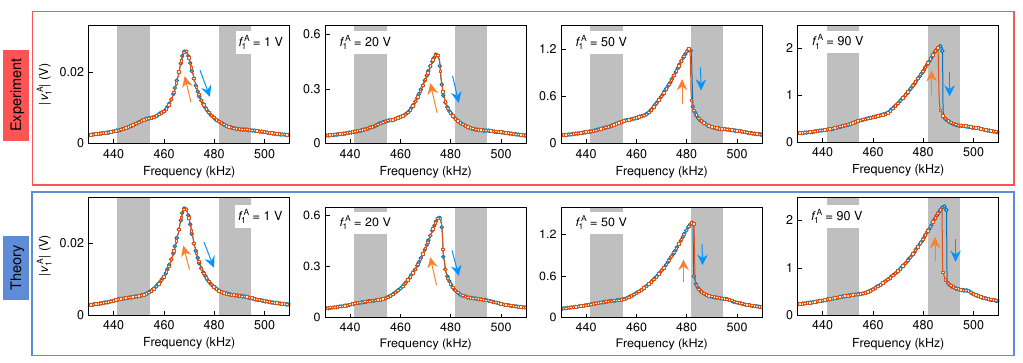}
\caption{Voltage spectra at the edge node when the circuit is excited with the
driving voltage $f_{1}^{\text{A}}$. The first and second rows show the results 
from the experiment and GP equation, respectively. In both rows, the blue and orange
curves correspond to the frequency sweeps along the two opposite directions
(denoted by the blue and orange arrows). The gray areas correspond to the linear 
bulk bands. }
\label{fig_edge_spectra}
\end{figure*}

Experimentally, by following the same procedure in the observation of the nonlinear
topological interface states (see Appendix \ref{app_F}), we excite and observe the nonlinear topological
edge states in the SSH lattice which is topologically nontrivial in the linear limit. 
Figure \ref{fig_edge_spectra} shows the
voltage spectra at the edge node when the circuit is excited with the
driving voltage $f_{1}^{\text{A}}$. The first and second rows show the experimental
and theoretical results, respectively. The directions of frequency sweep are
denoted by the blue and orange arrows. From the first column, when the driving
voltage is small with $f_{1}^{\text{A}} = 1~\text{V}$, the voltage spectra for
the frequency sweeps along the two opposite directions coincide with each other. The
symmetric peaks exhibited by the voltage spectra are the signature of the
excitation of the linear topological edge state in the linear limit. With the 
increasing of the driving voltage, the spectrum peaks become asymmetric 
with respect to the resonant frequency and the resonant frequency increases,
as shown by the figures in the second and third columns.
When the driving voltage is large enough, the spectrum peaks become highly
asymmetric and exhibit the bistable response. 
From the fourth column where the driving voltage is $%
f_{1}^{\text{A}} = 90~\text{V}$, the voltage spectra for the frequency sweep 
along the two
opposite directions are no longer coincide with each other and we observe the
bistable response which is induced by the nonlinear topological edge state in an
SSH lattice which is topologically nontrivial in the linear limit (i.e., in the nonlinear 
SSH lattice shown in Fig. \ref{fig_findings}(a)). 
Note that in the theoretical calculation, 
we set $N=120$, indicating that the lattice consists of $120$ unit cells.

\begin{figure*}[tbp]
\includegraphics[width=12.9cm]{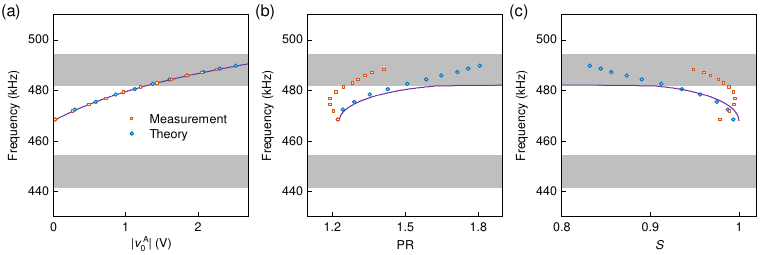}
\caption{Edge voltages $\vert v_{1}^{\text{A}} \vert$, participation ratios (PRs), and 
sublattice polarizations $S$ of the nonlinear topological edge states.
The orange squares and blue circles correspond to the experimental and theoretical 
results, respectively. The purple curves denote the result
calculated from the GP equation without the driven-dissipative terms,
and the gray areas correspond to the linear bulk bands.  }
\label{fig_edge_PR_S}
\end{figure*}

To quantitatively study the properties of the nonlinear topological edge states,
we measure the voltage distributions at the resonant frequencies of
the voltage spectra. The nonlinear topological edge states are all mainly 
confined to the sublattice site $\text{A}$ of the unit cells, and the phase jump of 
$\pi$ among the neighboring cells still holds. For larger driving 
voltages, the nonlinear topological edge states exhibit the decreased sublattice 
pseudospin $S$, agreeing well with the theoretical prediction. Meanwhile, the 
localization of the nonlinear topological edge states decreases. 
These properties are confirmed by the 
results shown in Fig. \ref{fig_edge_PR_S}. From Fig. \ref{fig_edge_PR_S}(a), 
in the linear limit, the frequency of the nonlinear topological edge state
resides in the middle of the SSH bandgap. With the increasing of the circuit
nonlinearity, the frequency of the nonlinear edge state exhibits a blue shift, 
similar to the phenomenon observed for the nonlinear topological interface states (see Appendix \ref{app_F}). 
From Fig. \ref{fig_edge_PR_S}(b), with the increasing of the onsite nonlinearity, 
the localization of the nonlinear topological edge state becomes weak. The deviations 
between the experimental and theoretical values are due to the circuit dissipation 
and measurement errors. Figure \ref{fig_edge_PR_S}(c) shows the dependence between 
the frequency and sublattice pseudospin of the nonlinear topological edge states. 
The experimental and theoretical
frequencies correspond to the resonant frequencies of the voltage
spectra under the external voltage driving. The purple curve denotes the result
calculated from the GP equation without the driven-dissipative terms. From the figure, 
the behaviors of the nonlinear topological edge states are very similar to those of the nonlinear
topological interface states (see Appendix \ref{app_F}). 
In the linear limit, we have $S \approx 1$ for both the experimental and theoretical
results. These results agree with the prediction from the GP equation without 
the driven-dissipative
terms. For a larger driving voltage, i.e., for a larger frequency, the nonlinear topological edge state
exhibits a decreased sublattice pseudospin $S$. The experimental result
deviates from the theoretical data because of the increased circuit dissipation
and decreased driving voltage. The small discrepancy near the linear limit is due to the 
inaccuracy of the measurement data recorded by the oscilloscope.

\section{Theoretical results of the nonlinear topological interface states\label{app_E}}

In this section, we give more theoretical results of the nonlinear topological edge states in the structure 
connected by two lattices with the different topological properties in the linear limit. Since the 
nonlinear topological edge states reside at the interface, we simply call them nonlinear topological 
interface states for clarity.
This section is organized as follows. In the first subsection, we introduce the procedure to find
the nonlinear topological interface states and discuss their properties. Then we study the stability analysis of the
nonlinear topological interface states. In the last subsection, we discuss the nonlinear topological trivial
interface states. 

\subsection{Existence of the nonlinear topological interface states}

\begin{figure}[tbp]
\includegraphics[width=8.8cm]{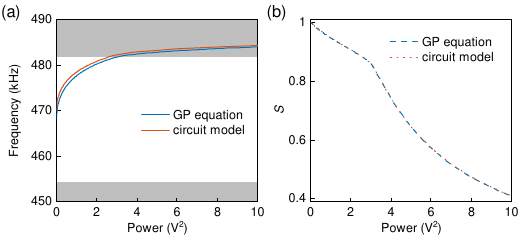}
\caption{Nonlinear topological interface states. (a)
Eigenfrequencies of the nonlinear topological interface states calculated from the 
GP equation and circuit model. The gray areas correspond to the linear 
bulk bands. 
(b) Sublattice pseudospin of of the nonlinear topological interface 
states indicated in (a).}
\label{fig_interface_GP_circuit}
\end{figure}

We focus on the lattice configuration shown in Fig. \ref{fig_findings}(b).
This configuration can be treated as an SSH lattice with an interface defect at the center.
The chain has $N$ unit cells for the left topologically trivial part with the intracell coupling larger than the
intercell coupling, and $N+1/2$ unit cells for the right nontrivial part with the intracell coupling smaller 
than the intercell coupling. To avoid the finite size effect, here we let $N = 60$ and there are
$241$ lattice sites (circuit nodes) in total.
Based on Eqs. (\ref{eq-7})-(\ref{eq-8}), we can get the governing equations of all the
lattice sites and solve the equations using the Newton's method. For the nonlinear topological interface
states which bifurcate from the linear topological interface state, we take the linear topological 
interface state
as the initial guess solution. In order to find the nonlinear topologically trivial interface states 
which are conventional lattice solitons localized at the interface, we solve the GP equation 
using the AC approach.
In the AC limit, we have $C_{1}=0$ and Eqs. (\ref{eq-7})-(\ref{eq-8}) reduce to
\begin{equation}
E_{0}{+g\left( {{v }_{0}^{\text{A}}}\right) =\bar{\omega},}
\end{equation}%
where 
\begin{eqnarray}
E_{0} &=&\left[ 1-\frac{{C}_{2}}{2\left( C_{\text{g}}+\eta C_{\text{L}}\right) }\right] \frac{{%
\omega _{0}}}{\omega _{n}}, \\
g\left( {{v }_{0}^{\text{A}}}\right) &=&-\frac{{\eta}C_{\text{NL}}\left( {v_{0}^{\text{A}}}%
\right) }{2\left( C_{\text{g}}+ \eta C_{\text{L}}\right) }\frac{{\omega _{0}}}{\omega _{n}}.
\end{eqnarray}%
After solving this equation, we get the value of ${{v }_{0}^{\text{A}}}$. Then we seek for 
the solutions for the nonlinear topologically trivial interface states by gradually
increasing $C_{1} $ to the original value. Meanwhile, since the GP equation is derived under the 
slowly-varying approximation, to validate the existence of the nonlinear topologically trivial interface 
states, we also
seek for the solutions for the nonlinear topologically nontrivial and trivial interface 
states based on the accurate 
nonlinear circuit model. Eqs. (\ref{circuit_eq_1})-(\ref{circuit_eq_2}) with $R = \infty$ and $R_{L} = 0$
are numerically solved using the Newton's method with the suitable initial guess solutions.
The calculation results from the GP equation and circuit model are shown in 
Fig. \ref{fig_interface_GP_circuit}.

Figure \ref{fig_interface_GP_circuit}(a) shows the frequencies of the nonlinear topological 
interface states. 
The blue and red curves correspond to the results calculated from the GP equation and circuit model,
respectively, and the shaded regions denote the linear bulk bands.
We define $P = \sum_{m}\left(
\left\vert v_{m}^{\text{A}}\right\vert ^{2}+\left\vert v_{m}^{\text{B}}\right\vert
^{2}\right) $ as the equivalent power in the circuit lattice to measure the strength of nonlinearity.
From the figure, the results from the GP equation and circuit model agree well with each other.
Figure \ref{fig_interface_GP_circuit}(b) further shows the sublattice pseudospin of the 
nonlinear topological interface states. In the linear limit, the topological interface 
states are perfectly localized 
on the sublattice site $\text{A}$ with $S = 1$. This property can be revealed from the solution of the
GP equation. With $g = 0$ and $V_{m}^{\text{A,B}} \left( T \right)= v_{m}^{\text{A,B}} e^{-i \bar{\omega}T}$, 
the governing equations reduce to
\begin{eqnarray}
&\ldots& ,\\
\bar{\omega} v_{-1}^{\text{B}}&=& E_{0}v_{-1}^{\text{B}}+J_{2}{v_{-1}^{\text{A}}}+J_{1}{v_{0}^{\text{A}}},   \\
\bar{\omega} v_{0}^{\text{A}}&=& E_{0} v_{0}^{\text{A}}+J_{1}{v_{0}^{\text{B}}}+J_{1 }{v_{-1}^{\text{B}}},  \\
\bar{\omega} v_{0}^{\text{B}}&=& E_{0}v_{0}^{\text{B}}+J_{1}{v_{0}^{\text{A}}}+J_{2}{v_{1}^{\text{A}}},  \\
&\ldots&  ,
\end{eqnarray}%
where the interface defect is located at the site $\text{A}$ of the $0$th cell.
Since the linear topological interface state has $\bar{\omega} = E_{0}$, the voltage 
distribution satisfies
$v_{m}^{\text{A}} = \left(- \frac{J_{1}}{J_{2}} \right)^{\vert m \vert} v_{0}^{\text{A}}$
and $v_{m}^{\text{B}}=0$. With the increasing of the power, $S$ decreases implying that the 
site $\text{B}$ has the nonzero voltages. Besides, the results from the GP equation and 
circuit model shown in Fig. \ref{fig_interface_GP_circuit}(b) again agree well with each other. 

\subsection{Stability analysis of the nonlinear topological interface states}

\begin{figure}[tbp]
\includegraphics[width=8.6cm]{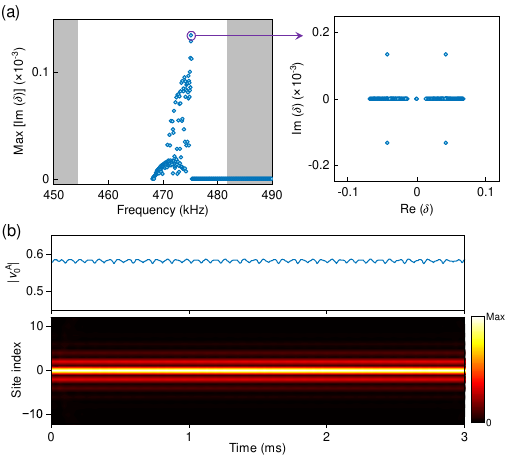}
\caption{Stability analysis of the nonlinear topological interface states. (a)
The maximum growth rates of the perturbed solutions of the nonlinear topological interface states. 
The gray areas denote the linear bulk bands. The inset shows the linear stability spectrum
at the frequency which has the largest
$\text{Max} \left[ \text{Im} \left(\delta \right) \right]$.
(b) The temporal evolution of the nonlinear topological interface state which has the largest
$\text{Max} \left[ \text{Im} \left(\delta \right) \right]$ [indicated in (a)].
Noises with $\pm 5\%$ random perturbations are added to the initial input.}
\label{fig_interface_stability}
\end{figure}

We study the dynamical stability/instability of the nonlinear topological interface states. 
The stability or instability of nonlinear states and solitons is crucial. Typically, under initial excitation, 
only stable solitons can be effectively excited and observed after a period of temporal evolution.
For a general nonlinear SSH circuit lattice,  the linear stability/instability of the nonlinear states
can be evaluated by substituting the following perturbed solutions
\begin{eqnarray}
V_{m}^{\text{A}} &=&e^{-\mathrm{i}\bar{\omega}T}\left( v _{m}^{\text{A}}+\varepsilon
_{m}^{\text{A}}e^{-\mathrm{i}\delta T}+\mu _{m}^{\text{A}\ast }e^{\mathrm{i}\delta ^{\ast }T}\right) , \label{per_sol_1}\\
V_{m}^{\text{B}} &=&e^{-\mathrm{i}\bar{\omega}T}\left( v _{m}^{\text{B}}+\varepsilon
_{m}^{\text{B}}e^{-\mathrm{i}\delta T}+\mu _{m}^{\text{B}\ast }e^{\mathrm{i}\delta ^{\ast }T}\right) , \label{per_sol_2}
\end{eqnarray}
into Eqs. (\ref{eq-7})-(\ref{eq-8}) and performing a standard linearization procedure.
In Eqs. (\ref{per_sol_1})-(\ref{per_sol_2}), $v_{m}^{\text{A,B}} e^{-\mathrm{i}\bar{\omega}T}$ 
are the unperturbed 
solutions of the nonlinear topological interface states, $\varepsilon_{m}^{\text{A,B}}$ and
$\mu_{m}^{\text{A,B}}$ are the infinitesimal amplitudes of the perturbations, and $\delta$
is the eigenfrequency. Obviously, the nonlinear topological interface states are linearly stable 
if $\delta$ is real, and they are linearly unstable if the imaginary part of $\delta$, namely the 
growth rate, is positive. After solving the linearized equations regarding to $\varepsilon_{m}^{\text{A,B}}$
and $\mu_{m}^{\text{A,B}}$, we can get the maximum growth rates, i.e. $\text{Max} \left[ 
\text{Im} \left( \delta \right) \right]$, at the different powers or frequencies. For simplicity,
we let $\omega_{n} = \omega_{0}$ in the calculations and the growth rates are thus normalized.
Fig. \ref{fig_interface_stability}(a) shows the maximum growth rates at the 
different frequencies
of the nonlinear topological interface states. From the figure, the maximum growth rates are in the
order of $10^{-4}$, implying that the nonlinear topological interface states can be considered 
linearly stable. We further find the frequency which has the largest $\text{Max} \left[ \text{Im} \left(
\delta \right) \right]$ and plot the linear stability spectrum in the inset of Fig. \ref{fig_interface_stability}(a). 
Compared with the real parts of $\delta$, the imaginary parts are negligible. Thus, at least within the 
experimental measurement period, the nonlinear topological interface states do not exhibit the 
instability, such as the exponential increase or significant oscillation. To confirm the results from
the linear stability analysis, we further add the noises with $\pm 5\%$ random perturbations 
to the amplitude of the nonlinear topological interface state which has the 
largest $\text{Max} \left[ \text{Im} \left(
\delta \right) \right]$, and study the temporal evolution. From the voltage distribution 
and interface voltage shown in Fig. \ref{fig_interface_stability}(b), the nonlinear topological interface 
state is stable at least up to $3~\text{ms}$. 

\subsection{Nonlinear topologically trivial interface states}

\begin{figure}[tbp]
\includegraphics[width=8.6cm]{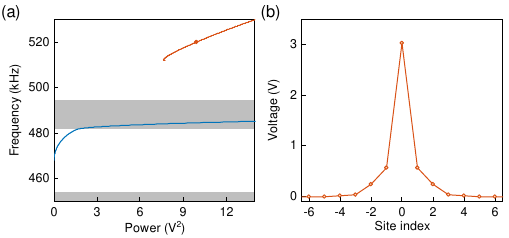}
\caption{Nonlinear interface states in a circuit lattice with the nonsaturable nonlinearity. 
(a) Frequencies of the nonlinear interface states, where the topologically nontrivial and trivial states
are denoted by the blue and red curves, respectively. The gray areas correspond to the linear bulk bands. 
(b) Voltage distribution of the nonlinear topologically trivial interface state with its frequency 
at $520~\text{kHz}$ (marked by the dot in (a)).}
\label{fig_interface_nonsaturable}
\end{figure}

Here we discuss the nonlinear topologically trivial interface states. Usually, when the nonlinearity
dominates the intracell and intercell couplings, the conventional lattice solitons appear. However, in this
circuit lattice such kind of nonlinear topologically trivial interface states do not exist, because the 
capacitance of the varactor
diodes saturates and the circuit lattice can not provide enough nonlinearity to support the existence
of such states. 
For comparison, we calculate the nonlinear interface states of a nonsaturable circuit model 
where the capacitance of the varactor diodes is characterized by Eq. (\ref{C_approximate}),
and the 
results are shown in Fig. \ref{fig_interface_nonsaturable}. From Fig. \ref{fig_interface_nonsaturable}(a), 
the nonlinear topological interface states exist as well (blue curve), and their frequencies 
exhibit the blue shift more obviously than 
the frequencies of the nonlinear topological interface states in the circuit lattice with the saturable nonlinearity. 
Besides, in the nonsaturable model, the nonlinear topologically trivial interface 
states emerge in the upper semi-infinite
gap (denoted as red curve). In Fig. \ref{fig_interface_nonsaturable}(b), we plot the voltage distribution of the 
nonlinear topologically trivial interface state with its frequency at $520~\text{kHz}$. Since the nonlinear
topologically trivial interface state is an self-sustained state and emerges when the nonlinearity dominates 
the couplings, the state 
is confined to both the sublattice sites $\text{A}$ and $\text{B}$, and there is no phase jump 
among the neighboring cells or sites.

\section{Experimental measurement of the nonlinear topological 
interface states\label{app_F}} 

In this section, we give more experimental results of the nonlinear topological edge 
states in the structure connected by two lattices with different topological properties in the linear limit (Fig. \ref{fig_findings}(b)). 
Since the nonlinear edge states reside at the interface, we simply call them nonlinear topological 
interface states for clarity. This section is organized as follows.
In the first two subsections, we introduce the experimental principle and show the experimental result 
for the linear SSH lattice, respectively. Then we discuss the experimental measurement
of the nonlinear topological interface states in the third subsection.

\subsection{Experimental principle}

The GP equation (Eq. \ref{gp_main} or Eqs. (\ref{eq-7})-(\ref{eq-8})) describe the circuit lattices
without the external sources and circuit dissipation. However, in the realistic experiments,
the circuit lattice has to be excited by an external source and the circuit itself is inherently lossy. 
For a linear
topological circuit, we usually excite the circuit using a continuous voltage source and measure
the voltage distributions at the circuit nodes. The band structure and topological states
can be obtained using this measurement technique \cite{NSR8-nwaa192}. However, this 
technique can not be directed extended to study the nonlinear
topological states in the nonlinear topological circuits because of the two reasons. 
First, the response of a nonlinear system is sensitive to the external excitation and the 
complicated bistable behavior may emerge in the nonlinear system. 
Second, when a nonlinear system is driven by
a continuous source, the dissipation of the system has to be considered, otherwise the
steady states are not supported. Taking these factors into account, we use the 
discretized driven-dissipative GP equation, i.e. Eqs. (\ref{eq-5}) and (\ref{eq-6}), to 
study the excitation of the nonlinear topological states and compare the theoretical results
with the experimental ones.

\begin{figure}[tbp]
\includegraphics[width=8.5cm]{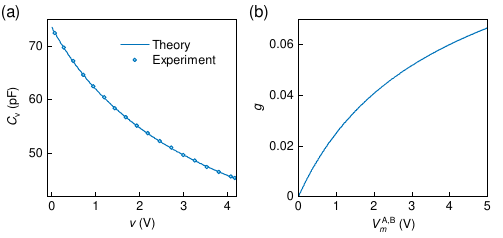}
\caption{Nonlinearity of the circuit lattice.
(a) The capacitance of a back-to-back varactor
diode, where the circles are the experimental measurement data and the solid
curve is calculated from the theoretical formula with the fitting parameters. 
(b) The voltage dependent onsite energy.}
\label{fig_nonlinearity}
\end{figure}

In this study, for the typical SSH lattice with the intracell coupling smaller
than the intercell coupling (Fig. \ref{fig_findings}(a)), the circuit parameters 
are $L_{\text{g}}=15~\mu \text{H}$, $%
C_{\text{g}}=4.7~\text{nF}$, $C_{1}=180$ pF, and $C_{2}=560$ pF. The
parameters of the varactor diodes are $C_{\text{L}}=73.48$ pF, $v_{0}=2.1935$%
, and $M=0.4548$, which are obtained by fitting the experimental measurement
results. Under these parameters,
we have $J_{1} < J_{2}$. Fig. \ref{fig_nonlinearity}(a) shows the
capacitance-voltage curve of a back-to-back varactor diode, where the
measurement frequency is 500 kHz. Since the capacitance decreases with an
increasing voltage, the onsite nonlinearity $g$ increases for large
voltages at the circuit nodes (Fig. \ref{fig_nonlinearity}(b)). Here we use 30
back-to-back varactor diodes, i.e. $\eta = 30$. Although the varactor diodes
exhibit the saturable nonlinearity, the capacitance does not reach saturation within our
experimental voltage range. Besides, the shunt resistors connected to the
external voltage sources are $R=100$ k$\Omega $, and the series resistance of
the inductors are $R_{L}=600~\text{m}\Omega $. These parameters 
ensure the validity of the GP equations, both with and without the driven-dissipative terms.
Meanwhile, for all circuit samples, our experimental measurements indicate that the 
maximum impedance between the excitation node and ground is $2.8~\mathrm{k}\Omega$. 
Therefore, the voltage source, combined with the shunt resistor, can be treated as an equivalent current source.
Note that to quantitatively test the validity of the GP equations, we
have also checked the results using the accurate nonlinear circuit model 
without the slowly-varying envelope approximation. The comparison shows 
that the nonlinear topological states in the nonlinear topological circuit lattices 
can be effectively captured by the GP equation.

\subsection{Linear SSH lattice}

Before we study the nonlinear topological states, we implement an SSH 
circuit lattice with the periodic boundary condition and measure 
the linear band structure.
The study of the linear SSH lattice is important because of the two reasons.
First, we would like to validate the GP equation by comparing the theoretical result
with the experimental data. If the results from the GP equation, nonlinear circuit model,
and experimental measurement agree with each other, the validity of the GP equation 
is well proved, at least in the linear limit.
Second, we need to get the experimental frequency gap
in the linear limit before we study the nonlinear topological states. Due to the 
experimental errors, there may be a frequency offset between the experimental gap and
theoretically predicted range. The prior knowledge of the experimental result in the
linear limit is helpful for the physical explanation of the nonlinear states, including both
the topologically nontrivial and trivial ones.

The nonlinear SSH circuit lattice reduces to the linear limit
when the external driving voltage is small. In this limit, the varactor diodes act
as the normal capacitors, and the circuit lattice is described by the
single-particle SSH Hamiltonian which leads to two bulk bands with a
topological bandgap between them \cite{PRL42-1698}. 
We start from Eqs. (\ref{eq-7})-(\ref{eq-8}) with $g=0$.
Under the periodic boundary condition, the solutions are in the form of the Bloch
functions ${V{_{m}^{\text{A},\text{B}}=\phi }}_{\text{A},\text{B}}
\exp \left(\mathrm{i}k m -\mathrm{i}\bar{\omega}T\right) $ and the normalized frequencies are calculated as 
\begin{equation}
\bar{\omega}{=E}_{0}\pm \sqrt{J^{2}+J^{\prime 2}+2JJ^{\prime }\cos k}.
\label{omega_GP}
\end{equation}
Then we calculate the band structure based on the circuit model. Under the
periodic boundary condition, Eqs. (\ref{circuit_eq_1})-(\ref{circuit_eq_2}) 
with $g=0$ can be written as 
\begin{widetext}
\begin{equation}
\left[ 
\begin{array}{cc}
L_{\text{g}}\left( {C}_{1}+C_{2}+C_{\text{g}}+\eta C_{L}\right) & -L_{\text{g%
}}\left( {C}_{1}+{C}_{2}e^{-ik}\right) \\ 
-L_{\text{g}}\left( {C}_{1}+{C}_{2}e^{ik}\right) & L_{\text{g}}\left( {C}%
_{1}+C_{2}+C_{\text{g}}+\eta C_{L}\right)
\end{array}%
\right] \left[ 
\begin{array}{c}
{\phi }_{\text{A}} \\ 
{\phi }_{\text{B}}%
\end{array}%
\right] =\frac{1}{\omega ^{2}}\left[ 
\begin{array}{c}
{\phi }_{\text{A}} \\ 
{\phi }_{\text{B}}%
\end{array}%
\right],
\end{equation}%
and the eigenfrequencies are%
\begin{equation}
\omega =\left[ L_{\text{g}}\left( {C}_{1}+C_{2}+C_{\text{g}}+\eta
C_{L}\right) \pm L_{\text{g}}\sqrt{C_{1}^{2}+C_{2}^{2}+2C_{1}C_{2}\cos k}%
\right] ^{-\frac{1}{2}}.  \label{omega_circuit}
\end{equation}%
\end{widetext}
Equation (\ref{omega_circuit}) reduces to Eq. (\ref{omega_GP}) under the
approximation $C_{1,2}\ll C_{\text{g}} +\eta C_{\text{L}}$, and this implies
that the GP equation is valid as long as the coupling capacitance is much smaller
than the grounding capacitance.

\begin{figure*}[tbp]
\includegraphics[width=17.1cm]{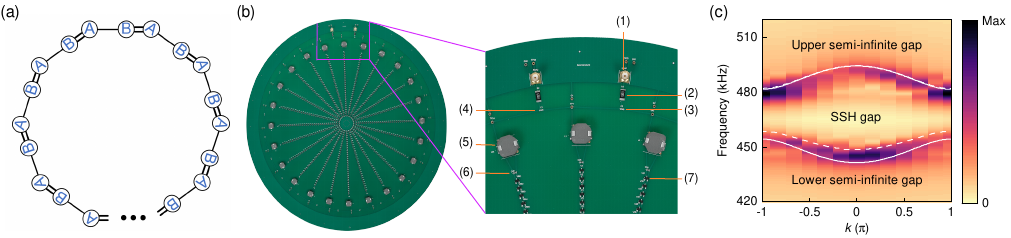}
\caption{Band structure for a linear SSH circuit lattice. (a)
Schematic of the linear SSH lattice with the periodic boundary
condition. (b) Fabricated PCB of the circuit lattice. The inset shows
the enlarged figure with the circuit components: (1) SMA connector, (2)
shunt resistor $R$, (3) coupling capacitor $C_{1}$, (4) coupling capacitor $%
C_{2}$, (5) grounding inductor $L_{\text{g}}$, (6) grounding capacitor $C_{%
\text{g}}$, and (7) varactor diode $C_{v}$. (c) Experimentally
measured band structure. For comparison, the band structures calculated from
the GP equation and circuit model are also plotted in sold and dashed white
curves, respectively.}
\label{fig_band_linear}
\end{figure*}

To experimentally probe the band structure for a linear SSH lattice, 
we fabricate an end-to-end circuit lattice
which satisfies the periodic boundary condition, as shown in Fig. %
\ref{fig_band_linear}(a). Since the circuit structure contains $N$ unit cells ($N$ is an
even number and $N=12$ is our experiment), $k$ has $N$ discrete values with $%
k=-\pi ,-\frac{N-2}{N}\pi ,\cdots ,0,\frac{N-2}{N}\pi$. The fabricated
circuit sample is shown in Fig. \ref{fig_band_linear}(b). From the inset, two circuit
nodes are wired to the shunt resistors $R$. In the experiment, we excite the circuit with $%
f_{1}^{\text{A}} = 1 ~\text{V}$ and $f_{N}^{\text{B}} = 1 ~\text{V}$, respectively, by
connecting the two SMA connectors to an arbitrary function generator. The
complex voltages at all the circuit nodes are measured using an
oscilloscope. Then by applying a Fourier transform and taking an average
between the results from the sites $\text{A}$ and $\text{B}$, we get the voltage distribution
in $k$ space, i.e., the band structure. The experimental band structure is
shown in Fig. \ref{fig_band_linear}(c). For comparison, the theoretical band structure
calculated from the GP equation and circuit model are also plotted in sold
and dashed white curves, respectively. From the figure, besides the middle SSH gap, the
band structure exhibits two semi-infinite gaps, one above the top band and
the other one below the bottom band. The existence of the semi-infinite gaps
are ensured because the $LC$ oscillators are single mode and such property
of topological circuits has be used in the observation of the
complicated band degeneracies \cite{NSR8-nwaa192}. 

From Fig. \ref{fig_band_linear}(c), the theoretical band structure calculated from the
GP equation and circuit model agree well with each other. Although there is
a small discrepancy between the bottom bands, this discrepancy does not
affect the nonlinear topological physics that we study in this
paper. Meanwhile, the experimental band structure agrees well with the
theoretical one. The comparison between the three results implies that the
experimental observation of a linear SSH circuit lattice can be
effectively described by the GP equation. 

\begin{figure}[tbp]
\includegraphics[width=8.6cm]{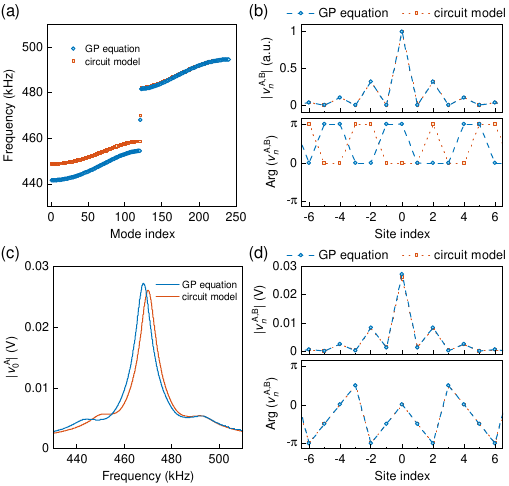}
\caption{Linear topological interface states.
(a) Eigenfrequencies of the states in the linear SSH lattice,
where the blue and orange symbols correspond to
the results from the GP equation and circuit model, respectively.
The middle frequencies correspond to the linear topological interface states.
(b) The voltage amplitudes and phases of the linear topological interface states. 
(c) Voltage spectra of the linear SSH lattice when the driving voltage 
is $f_{0}^{\text{A}}  = 1~\text{V}$.
The blue and orange curves correspond to
the results from the GP equation and circuit model, respectively.
(d) Amplitudes and phases of the voltage distributions at the respective resonant 
frequencies in (c). }
\label{fig_linear_interface}
\end{figure}

Based on the above result, we discuss the linear topological interface 
states and show that the linear topological 
interface states can be excited under the external driving voltage.
First, we neglect the external driving and circuit dissipation,
and study the linear topological interface state in a linear SSH lattice with $g = 0$. 
The circuit dissipation is also omitted.
The right end of the lattice is truncated with ${V_{N}^{\text{B}}=0}$
to avoid the coupling between the interface state residing at the interface
and the edge state residing at the right
topologically nontrivial part. 
Here, we set $N=60$, i.e., there are $60$ unit cells for the left topologically trivial 
part and $60.5$ unit cells for the right nontrivial part.
Starting from the GP equation, we seek for the solutions
with ${{V_{m}^{\text{A},\text{B}}=\psi _{m}^{\text{A},\text{B}}\exp }}
\left( -\mathrm{i}\bar{\omega}T\right) $, and the normalized frequency $\bar{\omega}$
can be directly obtained. Note that it is also possible to
study the linear topological interface state using the accurate circuit model. 
Figure \ref{fig_linear_interface}(a) shows the eigenfrequencies
of the linear SSH lattice, where the blue and orange symbols correspond to
the results from the GP equation and circuit model, respectively. The middle frequencies
correspond to the liner topological interface states and their voltage distributions 
are shown in Fig. \ref{fig_linear_interface}(b). The linear interface states have 
the well-defined sublattice 
pseudospin with $S=1$, and they exhibit the phase jump of $\pi$ among 
the neighboring cells.
Besides, in Figs. \ref{fig_linear_interface}(a)-(b), the results from the GP 
equation well agree with the results from the circuit model. 
Then we study the excitation of the linear topological interface state based on the
driven-dissipative GP equation. The dissipation is $\gamma = \gamma_{L} + \gamma_{R}$ at the
driven node and $\gamma_{L}$ at the other nodes. Similarly, we also study the 
excitation of the linear topological interface state based on the circuit model.
Fig. \ref{fig_linear_interface}(c) shows the voltage spectra of the linear SSH lattice when
the driving voltage is $f_{0}^{\text{A}} = 1 ~\text{V}$, where the blue and orange 
curves correspond to the results from the GP equation and circuit model, respectively. 
The peaks of the voltage spectra are symmetric with respect to their respective resonant frequencies.
We obtain the resonant frequencies
from the voltage spectra and plot the voltage distributions at 
the resonant frequencies. From Fig. \ref{fig_linear_interface}(d), the voltage distributions 
from the GP equation and circuit
model both exhibit the sublattice pseudospin of $S = 0.99$ and the property of phase jump  
still holds. These features imply that, under the driving of the external voltage
source, the linear topological interface states are excited at the resonant 
frequencies. 

\subsection{Nonlinear topological interface states}

To theoretically demonstrate the excitation and observation of the nonlinear topological
interface states, we first solve the driven-dissipative GP equation with the driven terms 
${F_{m}^{\text{A}}}\left( t\right) =\delta _{m,0}{f_{0}^{\text{A}}}\exp \left( -%
{\mathrm{i}\bar{\omega}T}\right) $ and $F_{m}^{\text{B}} \left( t \right) = 0$, i.e. only the 
interface site is excited.
Since the solutions are generally complex-valued, we separate the real and
imaginary parts of the solutions, and the new equations are solved using the Newton's method.
We then study the nonlinear circuit model with the driven-dissipative terms. Eqs. 
(\ref{circuit_eq_1})-(\ref{circuit_eq_2}) can be rewritten as%
\begin{eqnarray}
&&\left( E_{0}-i\gamma \right) {{v}_{m}^{\text{A}}+g\left( {{v}_{m}^{\text{A}}}\right)
v_{m}^{\text{A}}}+J{{v}_{m}^{\text{B}}}+J^{\prime }{{v}_{m-1}^{\text{B}}}
+id{f_{m}^{\text{A}}}\notag \\
&=&0,  \\
&&\left( E_{0}-i\gamma \right) {{v}_{m}^{\text{B}}+g\left( {{v}_{m}^{\text{B}}}\right)
v_{m}^{\text{B}}}+J{{v}_{m}^{\text{A}}}+J^{\prime }{{v}_{m+1}^{\text{A}}}
+id{f_{m}^{\text{B}}}\notag \\
&=&0.
\end{eqnarray}%
where 
\begin{eqnarray}
E_{0} &=&\omega ^{2}L_{\text{g}}\left( C_{\text{g}}+ \eta C_{\text{L}}+{C}_{1}+C_{2}\right)
\notag \\
 &&-\frac{\omega ^{2}L_{\text{g}}^{2}}{\omega ^{2}L_{\text{g}}^{2}+{R}_{L}^{2}}, \\
\gamma &=&{-}\frac{\omega L_{\text{g}}{R}_{L}}{\omega ^{2}L_{\text{g}}^{2}+{R}%
_{L}^{2}}-\frac{\omega L_{\text{g}}}{R}, \\
{g\left( {{v}_{m}^{\text{A},\text{B}}}\right) } &=&{{\omega ^{2}L_{\text{g}} \eta C_{\text{NL}}\left( {%
v_{m}^{\text{A}}}\right) ,}} \\
J &{=}&-\omega ^{2}L_{\text{g}}{C}_{1}, \\
J^{\prime } &=&-\omega ^{2}L_{\text{g}}{C}_{2}, \\
d &=&-\frac{\omega L_{\text{g}}}{R}.
\end{eqnarray}
These equations can also be solved using the Newton's method. Note that, for both the 
GP equation and the circuit model, we set all lattice sites (including the interface site) to 
have the same onsite energy $E_{0}$ in order to ensure the chiral symmetry of the linear 
SSH model. Experimentally, this equal onsite energy can be achieved by adding an additional 
capacitance $C_{2} - C_{1}$ to the grounding capacitance of the interface circuit node.
Furthermore, in the theoretical calculations of both the GP equation and the circuit model, we set $N = 60$.

\begin{figure*}[tbp]
\includegraphics[width=9.2cm]{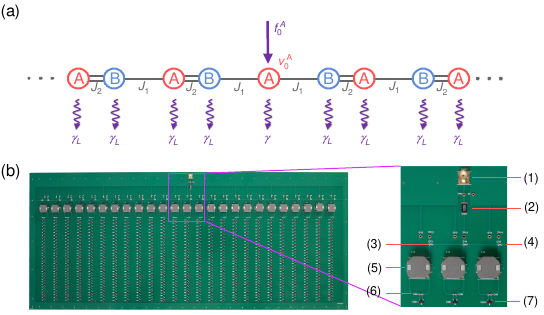}
\caption{Experimental implementation of the circuit sample connected by two lattices
with the different topological properties in the linear limit (Fig. \ref{fig_findings}(b)). (a) Schematic of the
nonlinear SSH lattice with the nearest neighbour couplings $J_{1,2}$, the
dissipations $\protect\gamma_{L}$ and $\protect\gamma$, and the external
driving voltage $f_{0}^{\text{A}}$. The voltage at the interface node is $v_{0}^{\text{A}}$.
(b) Fabricated PCB of the nonlinear SSH circuit
lattice. The inset shows the enlarged figure with the circuit components:
(1) SMA connector, (2) shunt resistor $R$, (3) coupling capacitor $C_{1}$,
(4) coupling capacitor $C_{2}$, (5) grounding inductor $L_{\text{g}}$, (6)
grounding capacitor $C_{\text{g}}$, and (7) varactor diode $C_{\text{v}}$.}
\label{fig_interface_PCB}
\end{figure*}

To experimentally observe the nonlinear topological interface states, we excite the interface
circuit node with the external driving voltage $f_{0}^{\text{A}}$, and consider the
dissipations induced both from the series resistance of the inductors and shunt
resistors. As schematically shown in Fig. \ref{fig_interface_PCB}(a), the interface circuit
node experiences the dissipation $\gamma$ and the other nodes have the
dissipation $\gamma_{L}$. The voltage at the interface node is $v_{0}^{\text{A}}$.
Fig. \ref{fig_interface_PCB}(b) shows the fabricated PCB of the
nonlinear circuit lattice, and the inset shows the
enlarged figure with the circuit components: (1) 
SMA connector, (2) shunt resistor $R$, (3) coupling capacitor $%
C_{1}$, (4) coupling capacitor $C_{2}$, (5) grounding inductor $L_{\text{g}}$%
, (6) grounding capacitor $C_{\text{g}}$, and (7) varactor diode $C_{\text{v}}$.
The SMA connector is connected to an external voltage source with the
voltage amplitude $f_{0}^{\text{A}}$. 
Here, the experimental circuit lattice has the parameter 
$N = 6$, which corresponds to a total of 25 lattice sites.

\begin{figure*}[tbp]
\includegraphics[width=17.5cm]{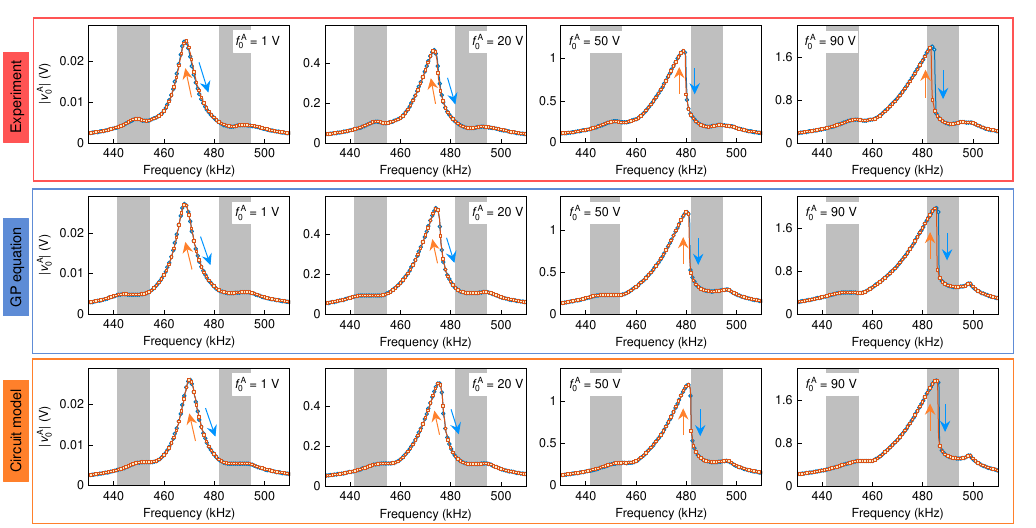}
\caption{Voltage spectra at the interface node when the circuit is excited with the
driving voltage $f_{0}^{\text{A}}$. The first, second and third rows show the results 
from the experiment,
GP equation and circuit model, respectively. In all the rows, the blue and orange
curves correspond to the frequency sweep along the two opposite directions
(denoted by the blue and orange arrows). The gray areas correspond to the linear 
bulk bands. }
\label{fig_interface_spectra}
\end{figure*}

Experimentally, for each driving voltage $f_{0}^{\text{A}}$, we measure the voltage
at the interface node $v_{0}^{\text{A}}$ when the driving frequency is swept from
low to high and from high to low, respectively. Figure \ref{fig_interface_spectra} shows
voltage spectra at the interface node when the circuit is excited with the
driving voltage $f_{0}^{\text{A}}$. The first, second and third rows show the results 
from the experiment,
GP equation and circuit model, respectively. The directions of the frequency sweep are
denoted by the blue and orange arrows. From the first column, when the driving
voltage is small with $f_{0}^{\text{A}} = 1~\text{V}$, the voltage spectra for the
frequency sweep along the two opposite directions coincide with each other. The
symmetric peaks exhibited by the voltage spectra are the signature of the
excitation of the linear topological interface state in the linear limit. When
the driving voltage increases to $f_{0}^{\text{A}} = 20~\text{V}$, the spectrum
peaks become asymmetric with respect to the resonant frequency (see the
second column). The voltage amplitude $\vert v_{0}^{\text{A}} \vert$ shows faster
variations at the higher frequencies and slower variations at the lower frequencies.
Although the resonant frequency increases due to the enhanced circuit
nonlinearity, the voltage spectra for the frequency sweep along the two opposite
directions still coincide with each other. When the driving voltage further
increases to $f_{0}^{\text{A}} = 50~\text{V}$, the spectrum peaks become highly
asymmetric, showing the sudden interface voltage change at the same driving
frequency for the frequency sweep along the two opposite directions (see the
third column). Above this driving voltage, the voltage spectra exhibit the
bistable response. From the last column where the driving voltage is $%
f_{0}^{\text{A}} = 90~\text{V}$, the voltage spectra for the frequency sweep 
along the two
opposite directions are no longer coincide with each other. A frequency
interval exists where one driving frequency corresponds to two interface
voltages. The bistable response observed in the nonlinear SSH circuit
lattice is a typical phenomenon in the driven-dissipative nonlinear systems. It
is worthwhile to note that the voltage spectra from the experiment, GP equation,
and circuit model agree well with each other, validating the correctness of the
GP equation.

\begin{figure}[tbp]
\includegraphics[width=8.6cm]{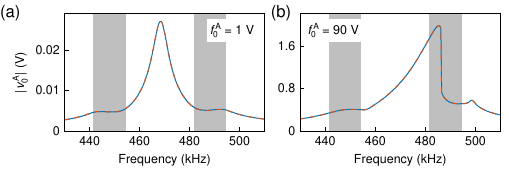}
\caption{Voltage spectra calculated from the driven-dissipative GP equation. Panels (a) 
and (b) correspond to $f_{0}^{\text{A}} = 1~\text{V}$ and $f_{0}^{\text{A}} = 90~\text{V}$, 
respectively. In both panels, the blue and red curves represent the lattices with $N = 6$ and 
$N = 60$, respectively. For clarity, we present only the results obtained when sweeping 
the frequency from low to high.}
\label{fig_interface_lattice_size}
\end{figure}

Before proceeding with further analysis, we would like to note that in Fig. \ref{fig_interface_spectra}, 
the value of $N = 6$ corresponds to the experimental results shown in the first row, while $N = 60$ applies 
to the theoretical results derived from the GP equation (presented in the second row) and the circuit 
model (shown in the third row). Experimentally, the circuit lattice with $N = 6$ is sufficiently large for 
our study. To support this claim, we have also conducted theoretical calculations for a lattice with $N = 6$. 
Figures \ref{fig_interface_lattice_size}(a) and \ref{fig_interface_lattice_size}(b) display the voltage spectra 
calculated from the driven-dissipative GP equation under $f_{0}^{\text{A}} = 1~\text{V}$ and 
$f_{0}^{\text{A}} = 90~\text{V}$, respectively. For clarity, we only present the results obtained 
when sweeping the frequency from low to high. The results indicate that the voltage spectra for lattices with 
$N = 6$ (represented by the blue curves) and $N = 60$ (represented by the red curves) are identical, 
demonstrating that fabricating a lattice with $N = 6$ is experimentally sufficient.

\begin{figure}[tbp]
\includegraphics[width=8.6cm]{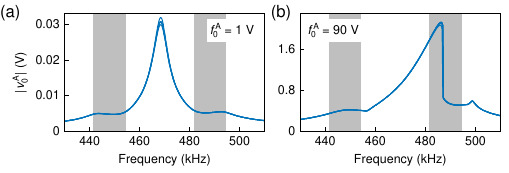}
\caption{Voltage spectra calculated from the driven-dissipative GP equation with 
$\pm 5\%$ variations in the series resistance of inductors. Panels (a) 
and (b) correspond to $f_{0}^{\text{A}} = 1~\text{V}$ and $f_{0}^{\text{A}} = 90~\text{V}$, 
respectively. For clarity, we present only the results obtained by sweeping the frequency from low to high, 
and we focus on five random samples to represent this variation.}
\label{fig_interface_disorder}
\end{figure}

Furthermore, we conducted an error analysis by considering the variations due to the series resistance of the 
inductors. In the study of topological circuits, the tolerances for capacitors and inductors are typically set at $\pm 1\%$. 
Here, we explain why we set the tolerance for the series resistance of inductors at $\pm 2\%$. Since the series resistance 
depends on frequency, we fit the experimental relationship between series resistance and frequency using the formula 
$R_{\mathrm{L}} = p_{1} f + p_{2}$, where $f$ is the working frequency, and $p_{1}$ and $p_{2}$ are parameters. 
Our fitting results indicate that different inductors have nearly equal values of $p_{1}$, but varying values of $p_{2}$. 
To determine an appropriate value for the tolerance of series resistance, we theoretically calculated the voltage spectra by 
introducing $\pm 5\%$ variations to the parameter $p_{2}$ under input voltages of $f_0^{\text{A}} = 1~\text{V}$ and 
$f_0^{\text{A}} = 90~\text{V}$, respectively. Note that, in the theoretical calculations, the average series resistance of 
the inductors for this circuit sample is not set to $600~\mathrm{m}\Omega$ to align with the measurement results.
As shown in Fig. \ref{fig_interface_disorder}, the variation in series resistance 
only weakly perturbs the resonant peaks. Considering that additional variations may arise during the sample 
fabrication process, we have designated the tolerance for the series resistance of the inductors used in the 
experiments as $\pm 2\%$.

\begin{figure*}[tbp]
\includegraphics[width=12.9cm]{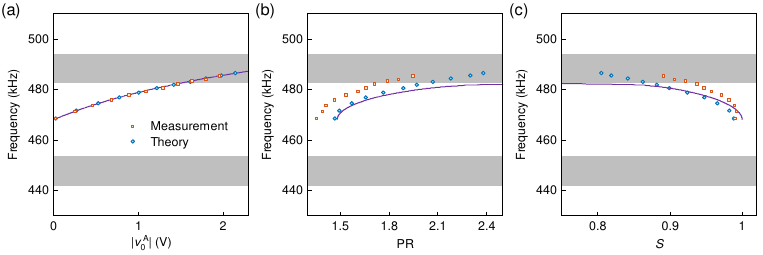}
\caption{Interface voltages $\vert v_{0}^{\text{A}} \vert$, participation ratios (PRs), and 
sublattice polarizations $S$ of the nonlinear topological interface states.
The orange squares and blue circles correspond to the experimental and theoretical 
results, respectively. The purple curves denote the result
calculated from the GP equation without the driven-dissipative terms,
and the gray areas correspond to the linear bulk bands.  }
\label{fig_interface_PR_S}
\end{figure*}

From the voltage spectra at different driving voltages, shown in Fig. \ref{fig_interface_spectra}, we extract the 
resonant frequencies and measure the voltages at all nodes of the nonlinear circuit lattice. When the
voltage spectra exhibit the bistable response, the resonant frequency is
taken from the spectrum where the driving frequency is swept from low to
high. To quantitatively study the properties of the nonlinear topological interface states,
Fig. \ref{fig_interface_PR_S} shows the interface voltages, participation ratios (PRs), and 
sublattice polarizations of the states at the different resonant frequencies.
The orange squares and blue circles correspond to the experimental and theoretical 
results, respectively. The purple curves denote the result
calculated from the GP equation without the driven-dissipative terms.
And the gray areas correspond to the linear bulk bands. 
From Fig. \ref{fig_interface_PR_S}(a), for a small interface 
voltage $\vert v_{0}^{\text{A}} \vert$, 
i.e. in the linear limit, the frequency of the topological interface state
reside in the middle of the SSH bandgap. With the increasing of $%
\vert v_{0}^{\text{A}} \vert$, the nonlinearity in the SSH circuit lattice is
enhanced, and the frequency of the topological interface state exhibits the blue
shift. Under large nonlinearities, the experimental frequencies
deviate from the theoretical data because of the increased series
resistance of the inductors and the decreased driving voltage provided by the
voltage source. To measure the localization of
the nonlinear topological interface states, we introduce participation ratio (PR) which is defined as
\begin{equation}
\text{PR} = \frac{\left[ \sum \limits_{m} \left( \vert V_{m}^{\text{A}} \vert ^{2} + \vert
V_{m}^{\text{B}} \vert ^{2} \right) \right] ^{2}} 
{\sum \limits_{m} \left( \vert V_{m}^{\text{A}} \vert
^{4} + \vert V_{m}^{\text{B}} \vert ^{4} \right) }.
\end{equation}
A high PR indicates that the wave function is spread over a large number of lattice sites, 
suggesting delocalization. Conversely, a low PR indicates that the wave function is confined to a small number 
of sites, indicating localization. For instance, when the state is uniformly distributed in an SSH lattice with $N$ unit 
cells, we have $\mathrm{PR} = 2N$. In contrast, $\mathrm{PR} = 1$ occurs when the state is strongly localized at a single site.
From Fig. \ref{fig_interface_PR_S}(b), under 
the larger input
voltages, the localization of the nonlinear topological interface state decreases.
Note that the discrepancy between the results is also
induced by the circuit dissipation. Since the experimental circuit
dissipation is usually larger than the theoretical value, the nonlinear topological interface
states observed in our experiment show smaller PRs, i.e. stronger localization. 
However, the experimental and theoretical
results show the same evolutionary trend, validating the fact that the nonlinearity
weakens the localization of the nonlinear topological interface states.
The topological interface states are chiral which are described by the 
sublattice pseudospin $S$. $S = 1$ implies the perfect sublattice localization on the
site $\text{A}$ and $S = -1$ implies the perfect sublattice localization on the site $\text{B}$. 
Figure \ref{fig_interface_PR_S}(c) shows the dependence between the frequency 
and sublattice pseudospin of the nonlinear topological interface states. From the figure, 
in the linear limit, we have $S \approx 1$ for both the experimental and theoretical
results. These results agree with the prediction from the GP equation without the
driven-dissipative terms. For the larger driving voltages, i.e., for the larger frequencies, 
the nonlinear topological interface states exhibit a decreased sublattice 
pseudospin. Again, the experimental result
deviates from the theoretical data because of the increased circuit dissipation
and decreased driving voltage. The small discrepancy near the 
linear limit is due to the inaccuracy of the measurement data recorded by
the oscilloscope. 

\begin{figure}[tbp]
\includegraphics[width=8.7cm]{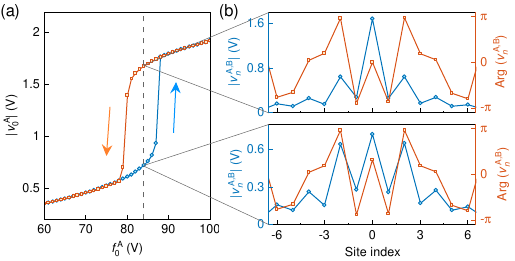}
\caption{Experimental bistable response of the nonlinear topological interface states.
(a) The bistable response between the voltage at the interface node and the external
driving voltage, where the driving frequency is $484~\text{kHz}$. 
The blue and orange curves correspond to the experimental
voltage sweeps along the two opposite directions (denoted by the blue and orange
arrows). (b) The voltage distributions at the two bistable states 
with the driving voltage indicated by the dashed line in (a). }
\label{fig_interface_bistable}
\end{figure}

To reveal the feature of the bistable response of the nonlinear topological interface
states, we experimentally fix the driving frequency to $484~\text{kHz}$, and 
measure the voltage at the interface node $v_{0}^{\text{A}}$ when the driving voltage
$f_{0}^{\text{A}}$ is swept from small to large and from large to small, respectively.
Figure \ref{fig_interface_bistable}(a) shows the bistable response between the voltage at the interface 
node and the external driving voltage. The blue and orange curves correspond 
to the experimental voltage sweep along the two opposite directions (denoted by the 
blue and orange arrows). From the the figure, 
the response curves of the interface voltages for the driving voltage sweeps along
the two opposite directions are not coincide with each other. 
The hysteresis loop shows that there is an interval for the driving voltage 
where one driving voltage corresponds to two interface
voltages. In Fig. \ref{fig_interface_bistable}(b), we show the voltage distributions at the two 
states of the bistable response where the driving voltage is $f_{0}^{\text{A}} = 84~\text{V}$ 
(indicated by the dashed line in Fig. \ref{fig_interface_bistable}(a)). Although the two voltage distributions
correspond to the same driving voltage, the state at the upper branch shows the larger
sublattice pseudospin $S$ and stronger localization (i.e., smaller PR), compared to the
state at the lower branch. 

\section{Theoretical results of the topological gap solitons\label{app_G}}

In this section, we give more theoretical results of the topological gap solitons. For 
completeness, we also introduce the topologically trivial gap solitons. Since both the 
topologically nontrivial and trivial gap solitons reside in the bulk of the lattice, to distinguish 
with the nonlinear edge states, we simply call them bulk solitons for clarity.
This section is organized as follows. In the first subsection, we introduce the the bulk solitons
in a single dimer with the onsite nonlinearity. Then the existence of the bulk solitons 
in an SSH lattice is studied in the second subsection. In the third subsection, we show
the results for the stability analysis of the bulk solitons. Finally, we discuss the physical 
interpretation of the topological gap solitons in the last subsection.

\subsection{Solitons in a single dimer}

\begin{figure}[tbp]
\includegraphics[width=8.5cm]{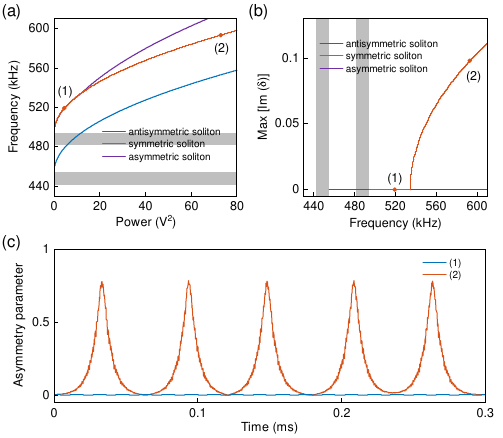}
\caption{Solitons in a single dimer with the nonsaturable nonlinearity.
(a) The symmetric, antisymmetric, and asymmetric solitons in a single dimer
with the nonsaturable nonlinearity.
(b) The maximum growth rates of the perturbed solutions of the solitons. 
The gray areas in (a)-(b) denote the linear bulk bands.
(c) The asymmetry parameters of the stable and unstable symmetric
solitons along the temporal evolutions. Noises with $\pm 1\%$ random perturbations 
are added to the initial input. The frequencies of the stable and unstable
solitons are marked in (a)-(b).}
\label{fig_bulk_AC_nonsaturable}
\end{figure}

In the AC limit, the SSH lattice reduces to a single dimer. Here we only study the dimer 
with a strong bond because the bulk solitons that we study in this paper are the 
continuations of the solitons in a circuit dimer with the coupling capacitor $C_{2} = 
560~\text{pF}$.
In the case with $C_{1} = 0$, the GP equation for the nonlinear modes reduces to
\begin{eqnarray}
E_{0}{{\psi }_{0}^{\text{A}}+g\left( {{\psi }_{0}^{\text{A}}}\right) {\psi }_{0}^{\text{A}}}+J_{2}{%
\psi }_{0}^{\text{B}}&=&\bar{\omega}{\psi _{0}^{\text{A}}},  \\
E_{0}{{\psi }_{0}^{\text{B}}+g\left( {{\psi }_{0}^{\text{B}}}\right) {\psi }_{0}^{\text{B}}}+J_{2}{{%
\psi }_{0}^{\text{A}}}&=&\bar{\omega}{\psi _{0}^{\text{B}}} .
\end{eqnarray}%
Depending on the parameters, the equations may have the three types of solutions: symmetric,
antisymmetric and asymmetric solutions. When the solution is symmetric, i.e., ${{%
\psi }_{1}^{\text{A}}={\psi }_{0}^{\text{B}}}$, the equations further reduce to%
\begin{equation}
E_{0}{+g\left( {{\psi }_{0}^{\text{A}}}\right) }+J_{2}{=\bar{\omega},}
\end{equation}%
which implies that the symmetric soliton reside in the upper semi-infinite bandgap. 
When the solution is antisymmetric, i.e., ${{\psi }_{1}^{\text{A}}=-{\psi }_{1}^{\text{B}}}$,
the equations reduce to%
\begin{equation}
E_{0}{+g\left( {{\psi }_{0}^{\text{A}}}\right) }-J_{2}{=\bar{\omega},}
\end{equation}%
which implies that the antisymmetric soliton reside in the middle SSH gap. Specifically,
asymmetric solitons may appear due to the linear instability induced by the spontaneous 
symmetry breaking of the symmetric solitons. To reveal all these features, we use
the nonsaturable nonlinearity instead and the results for the solitons in a single 
dimer with the nonsaturable nonlinearity is shown in Fig. \ref{fig_bulk_AC_nonsaturable}.

Figure \ref{fig_bulk_AC_nonsaturable}(a) shows the bifurcations of the symmetric,
antisymmetric, and asymmetric solitons. In the linear limit, only the symmetric and
antisymmetric solitons exist. However, the asymmetric solitons appear by bifurcating
from the symmetric solitons. Such bifurcation is induced by the linear instability of
the symmetric solitons. From the maximum growth rates shown in 
Fig. \ref{fig_bulk_AC_nonsaturable}(b), the antisymmetric and asymmetric solitons 
are always linearly stable. The symmetric solitons are linearly stable near the 
linear limit and become linearly unstable for large nonlinearities. The transition
frequency that separates the stable and unstable regions corresponds to the
bifurcation frequency of the asymmetric solitons in Fig. \ref{fig_bulk_AC_nonsaturable}(a).
To prove the stability or instability of the symmetric solitons, we select the two 
solitons marked in Figs. \ref{fig_bulk_AC_nonsaturable}(a)-(b) and carry out the
temporal evolutions with $\pm 1$\% noises added to the inputs. To quantitatively
measure the stability/instability, we introduce the asymmetry parameter defined
as 
\begin{equation}
\Theta =\left\vert \frac{\left\vert \psi _{1}\right\vert -\left\vert \psi
_{2}\right\vert }{\left\vert \psi _{1}\right\vert +\left\vert \psi
_{2}\right\vert }\right\vert .
\end{equation}
For the stable solitons, $\Theta$ should equal to zero approximately along the temporal
evolution. For the unstable solitons, the symmetry between $\psi_{1}$ and $\psi_{2}$
is broken, leading to the nonzero asymmetry parameters. 
Figure \ref{fig_bulk_AC_nonsaturable}(c)
shows the asymmetry parameters of the stable and unstable symmetric solitons.
Although noises are added to the input, the stable soliton always has $\Theta \approx 0$
(blue curve). In contrast, the unstable soliton shows the oscillating values of $\Theta$,
implying that the voltage oscillates between the two circuit sites (orange curve).

Considering the actual circuit nonlinearity (saturable nonlinearity), the asymmetric
solitons do not exist and we only find the symmetric and antisymmetric solitons.
The symmetric solitions in a single dimer correspond to the topologically trivial
bulk solitons in an SSH lattice, and the antisymmetric solitons in a dimer
correspond to the topologically nontrivial bulk solitons, i.e., the topological gap solitons. 

\subsection{Existence of the bulk solitons}

We again consider an SSH lattice with the intracell hopping $J_{2}$ and
intercell hopping $J_{1}$. In the study of the nonlinear topological edge states,
the SSH lattice has an open boundary at the left edge. Here we are interested in 
the bulk solitons that reside in the bulk of the lattice. The chain also has $120$ unit cells,
i.e., $N=120$. 

\begin{figure*}[tbp]
\includegraphics[width=10.3cm]{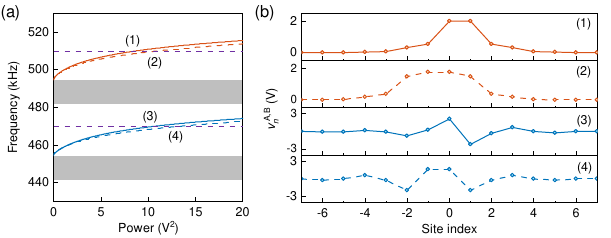}
\caption{Bulk solitons and nonlinear bulk states.
(a) Frequencies of the bulk solitons (solid curves) and nonlinear bulk states (dashed curves).
(b) Voltage distributions of the nonlinear states labeled in (a).}
\label{fig_bulk_mode}
\end{figure*}

Based on the symmetric and antisymmetric solitons in a single dimer, we get
the bulk solitons in an SSH lattice using the AC approach. To distinguish the bulk
solitons to the nonlinear bulk states, we also calculate the solutions for the nonlinear 
bulk states based on the GP equation. We focus on the two linear bulk states that
reside at the upper edges of the bulk bands. For a lattice with $N$ unit cells, the linear 
bulk states that reside at the upper edges of the bulk bands corresponds to the $N$th 
and $2N$th states. We take these linear bulk states as the initial guess solutions of the 
GP equation and obtain the nonlinear bulk states when considering the circuit
nonlinearity. The results for the bulk solitons and nonlinear bulk states are shown
in Fig. \ref{fig_bulk_mode}.

Figure \ref{fig_bulk_mode}(a) shows the frequencies of the bulk solitons (solid curves)
and nonlinear bulk states (dashed curves), and Fig. \ref{fig_bulk_mode}(b) shows the 
voltage distributions of the typical states labeled in Fig. \ref{fig_bulk_mode}(a).
The equivalent circuit power $P$ is again used as a measure of the nonlinearity strength.
The bulk solitons exhibit two branches, which reside in the 
middle SSH gap and upper semi-infinite gap, respectively.
In Fig. \ref{fig_bulk_mode}(a), the orange and blue curves correspond to the 
topologically trivial and nontrivial bulk solitons, respectively. The topologically nontrivial and
trivial bulk solitons are continuations of the antisymmetric and
symmetric solitons in a single dimer, respectively. From the first and third rows in 
Fig. \ref{fig_bulk_mode}(b), the bulk solitons mainly reside in the middle unit cell 
which contains the lattice sites $0$ and $1$. 
Specifically, for the topologically nontrivial bulk soliton in the first row, when we split the lattice 
into two parts from the middle of the middle unit cell,
the corresponding two parts of the topologically nontrivial bulk soliton have very similar profiles 
to the nonlinear topological edge states. The emergence of this type of bulk solitons 
can be understood that the circuit nonlinearity induces an interface at the middle of the 
middle unit cell and the two newly formed lattices at the two sides are both topologically nontrivial
with the intracell hopping $J_{1}$ smaller than the intercell hopping $J_{2}$. For the left 
part of the topologically nontrivial bulk soliton, it is mainly confined to the sublattice site $\text{A}$ and 
exhibits a phase jump of $\pi $ among the neighboring cells. The right part of the topologically nontrivial
bulk soliton also exhibits a phase jump but it is mainly confined to the sublattice site $\text{B}$. 
Such behavior indicates the different chiralities of the two parts of the topologically nontrivial bulk solitons: 
the left and right parts have the sublattice pseudospins with the different signs. Thus, this type 
of bulk solitons are the nonlinearity induced topologically nontrivial bulk solitons. 
While for the the topologically trivial bulk soliton in the third row,
the voltages at the two sites of the middle unit cell
are in-phase, in contrast to the out-of-phase voltage distribution of the topologically nontrivial bulk soliton.
Since the topologically trivial bulk solitons are the nonlinearity induced localization modes,
there is no phase jump among the neighboring cells. Note that here only the 
symmetric topologically trivial
bulk solitons are found because of the saturable nonlinearity of the varactor diodes.
It is worthwhile to note that, although the bulk solitons
bifurcate from the edge of linear Bloch band and converge to the bulk state in the 
linear limit, the profiles of the bulk solitons are different to those of the nonlinear
bulk states. The bulk states are extended in the linear limit, but under the circuit 
nonlinearity, they become localized with the voltages distributing mainly at the
lattice sites $-1$ and $0$, as shown by the second and fourth rows of 
Fig. \ref{fig_bulk_mode}(b).

\subsection{Stability analysis of the bulk solitons}

\begin{figure*}[tbp]
\includegraphics[width=12.9cm]{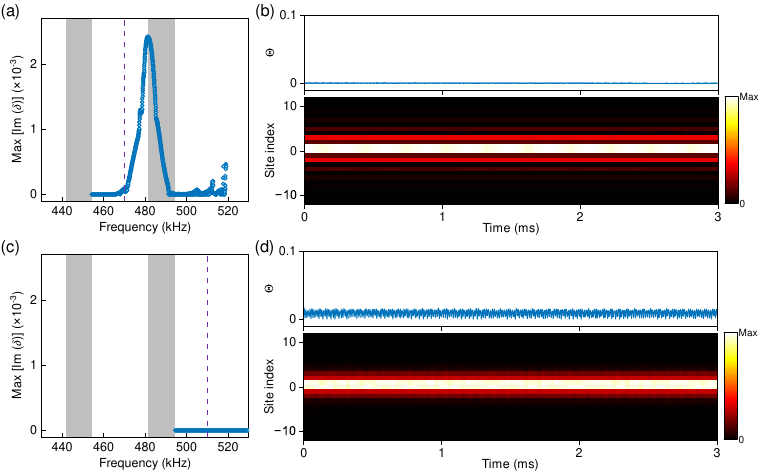}
\caption{Stability analysis of the bulk solitons.
(a) The maximum growth rates of the perturbed solutions of the
topologically nontrivial gap solitons. 
(b) The temporal evolution of the topologically nontrivial gap soliton at the frequency
indicated in (a). Noises with $\pm 2\%$ random perturbations are added to the initial input.
(c)-(d) Results for the topologically trivial gap solitons.
In (a) and (c), the gray areas denote the linear bulk bands. }
\label{fig_bulk_stability}
\end{figure*}

We then study the stability analysis of both the topologically nontrivial and trivial bulk
solitons, i.e., the gap solitons. Figure \ref{fig_bulk_stability}(a) shows the maximum 
growth rates of the topologically nontrivial gap solitons at the different frequencies. From the 
figure, the maximum growth rates are in the order of $10^{-3}$ when the soliton
frequencies are near to the lower edge of the top bulk band. This region corresponds
to the linearly unstable solitons. However, since we are interested in the localized 
topologically nontrivial
gap solitons and the maximum frequency shifts of the gap solitons are limited by 
the experimental input voltage, we carry out the temporal evolution of the topological
gap soliton at the frequency of $470~\text{kHz}$. Noises with $\pm 2\%$ random 
perturbations are added to the input amplitude. To quantitatively measure the mode 
stability/instability, we also use the asymmetry parameter $\Theta$ to characterize 
the voltage asymmetry between the sites $0$ and $1$. From the voltage distribution
in Fig. \ref{fig_bulk_stability}(b), there are no apparent variations along the temporal
evolution. The asymmetry parameters are in the order of $10^{-3}$. These results
imply that, although the unstable region exists, at least within the experimentally 
realizable parameter range, the topologically nontrivial gap solitons should be observable 
because of their weak instabilities.

The results for the stability analysis of the topologically trivial gap solitons are shown
in Figs. \ref{fig_bulk_stability}(c)-(d). Compared to the topologically nontrivial gap solitons, the
topologically trivial gap solitons have much smaller $\text{Max} \left[\text{Im}
\left( \delta \right) \right]$ (in the order of $10^{-9}$), implying that they
are much more stable. We take the topologically trivial gap soliton at $510~\text{kHz}$
as an example and study its temporal evolution, again with $\pm 2\%$ noises
added to the input. From Fig. \ref{fig_bulk_stability}(d), both the voltage distribution
and asymmetry parameters confirm that the topologically trivial gap soliton is stable.

\subsection{Physical interpretation of the topological gap solitons}

\begin{figure}[tbp]
\includegraphics[width=8.6cm]{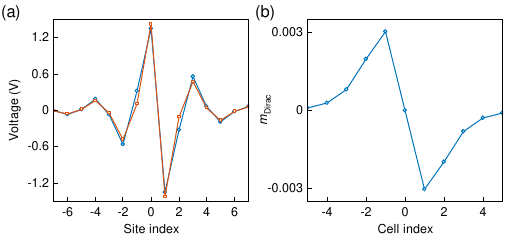}
\caption{Physical interpretation of the topologically nontrivial gap solitons.
(a) The solution of the topologically nontrivial bulk soliton (blue curve) compared with 
the state in a linear model plus an impurity potential in the bulk (orange curve).
(b) Mass term $m_{\text{Dirac}}$ of the 1D Dirac equations.}
\label{fig_bulk_interpretation}
\end{figure}

The interface, edge, and bulk solitons are related to the
solutions of the linear equations due to the self-consistent effective edges.
For the interface solitons, we consider a linear SSH model with an
additional impurity potential barrier defined as%
\begin{equation}
g_{\text{im}}=\frac{\eta C_{\text{im}}}{2\left( C_{\text{g}}+ \eta C_{\text{L}}\right) }\frac{%
\omega _{0}}{\omega _{n}}, \label{g_in}
\end{equation}%
where $C_{\text{im}}$ is the effective capacitance of the impurity
potential. The impurity potential is placed only on the interface site, i.e., the
site $\text{A}$ in the $0$th cell (hence playing the role of an effective
interface at the interface site). Considering the governing equations, the circuit
nonlinearity creates a
potential barrier at the interface, while the equations for the other sites
are the same to those in the linear limit.
For the edge solitons, we again consider a linear SSH
model with an additional impurity potential barrier defined in Eq. (\ref{g_in}).
However, for the edge solitons in a topologically nontrivial lattice,
the impurity potential is placed only on the edge site, i.e., the site 
$\text{A}$ in the $1$st cell (hence playing the role of an effective edge at the
edge site); for the edge solitons in a topologically trivial lattice, the impurity potential is placed on 
the two edge sites, i.e., the sites $\text{A}$ and $\text{B}$ in the $1$st cell. 
Then the governing equations for both the topologically nontrivial and trivial lattices can be obtained 
and solved similarly. For the bulk solitons, the impurity potential is placed on 
the two middle sites, i.e., the sites $\text{A}$ and $\text{B}$ in the $0$th cell. 
As an example, here we only compare the topologically nontrivial bulk soliton solution with 
the state in a linear model plus an impurity in the bulk, and the results are shown
in Fig. \ref{fig_bulk_interpretation}(a). From the figure, the solution of the linear model
(orange curve) is consistent to the solution of the topologically nontrivial gap soliton (blue curve),
implying that the onsite nonlinearity creates an impurity potential.

The above property can also be understood from the solution of the
GP equation. Since the bulk solitons mainly reside at the middle unit cell, we only
consider the circuit nonlinearities at the middle two sites and neglect
the nonlinearities at the other sites. With $V_{m}^{\text{A,B}} \left( T \right)
= v_{m}^{\text{A,B}} e^{-i \bar{\omega}T}$, the governing equations reduce to
\begin{eqnarray}
&\ldots& ,\\
\bar{\omega} v_{-1}^{\text{B}}&=& E_{0}v_{-1}^{\text{B}}+J_{2}{v_{-1}^{\text{A}}}+J_{1}{v_{0}^{\text{A}}},   \\
\bar{\omega} v_{0}^{\text{A}}&=& E_{0} v_{0}^{\text{A}}+g\left( v_{0}^{\text{A}} \right)v_{0}^{\text{A}}
+J_{2}{v_{0}^{\text{B}}}+J_{1}{v_{-1}^{\text{B}}},  \\
\bar{\omega} v_{0}^{\text{B}}&=& E_{0}v_{0}^{\text{B}}+g\left( v_{0}^{\text{B}} \right)v_{0}^{\text{B}}
+J_{2}{v_{0}^{\text{A}}}+J_{1}{v_{1}^{\text{A}}},  \\
\bar{\omega} v_{1}^{\text{A}}&=& E_{0} v_{1}^{\text{A}}+J_{2}{v_{1}^{\text{B}}}+J_{1}{v_{0}^{\text{B}}},  \\
&\ldots&  .
\end{eqnarray}%
Since the linear topological edge state has $\bar{\omega} = E_{0}$, the voltage 
distribution satisfies
$v_{m}^{\text{A}} = \left(- \frac{J_{1}}{J_{2}} \right)^{\vert m \vert} v_{0}^{\text{A}}$
and $v_{m}^{\text{B}}=0$ for the left part with $m \le -1$; and 
$v_{m}^{\text{B}} = \left(- \frac{J_{1}}{J_{2}} \right)^{\vert m \vert} v_{0}^{\text{B}}$
and $v_{m}^{\text{A}}=0$ for the right part with $m \le 1$. The voltages at the middle
two sites with $m=0$ satisfy 
\begin{eqnarray}
g\left( v_{0}^{\text{A}} \right)v_{0}^{\text{A}}+J_{2}{v_{0}^{\text{B}}}&=&0,  \\
g\left( v_{0}^{\text{B}} \right)v_{0}^{\text{B}}+J_{2}{v_{0}^{\text{A}}} &=&0.
\end{eqnarray}%
The topologically nontrivial gap solitons have $v_{0}^{\text{A}} = -v_{0}^{\text{B}}$ and the equations
reduce to 
\begin{equation}
g\left( v_{0}^{\text{A,B}} \right) = J_{2}. \label{bulk_reduce}
\end{equation}
Thus, if Eq. (\ref{bulk_reduce}) has solutions, the topologically nontrivial gap solitons can be explained
by the solution of the GP equation.

The topologically nontrivial bulk solitons, i.e., the topologically nontrivial gap solitons can also be interpreted 
as the Jackiw-Rebbi-type Dirac boundary mode. From Eqs. (\ref{eq-7}) and (\ref{eq-8}),
the governing equations can be rewritten as%
\begin{eqnarray}
i\frac{{dV{_{m}^{\text{A}}}}}{{dT}} &=&\left[ E_{0}+\frac{{g\left( {V_{m}^{\text{A}}}%
\right) +g\left( {V_{m}^{\text{B}}}\right) }}{2}\right] V_{m}^{\text{A}} \notag \\
&&+\frac{{g\left( {%
V_{m}^{\text{A}}}\right) -g\left( {V_{n}^{\text{B}}}\right) }}{2}{V_{m}^{\text{A}}} \notag \\
&&+J_{2}{V_{m}^{\text{B}}}%
+J_{1}{V_{n-1}^{\text{B}},} \\
i\frac{{dV{_{m}^{\text{B}}}}}{{dT}} &=&\left[ E_{0}+\frac{{g\left( {V_{m}^{\text{A}}}%
\right) +g\left( {V_{m}^{\text{B}}}\right) }}{2}\right] V_{m}^{\text{B}} \notag \\
&&-\frac{{g\left( {%
V_{m}^{\text{A}}}\right) -g\left( {V_{m}^{\text{B}}}\right) }}{2}V_{m}^{\text{B}}\notag \\
&&+J_{2}{V_{m}^{\text{A}}}+J_{1}{V_{m+1}^{\text{A}}.}
\end{eqnarray}%
These equations can be transformed to the 1D Dirac equations in the continum
limit \cite{PRL129-135501}, where the mass term is defined as%
\begin{equation}
m_{\text{Dirac}}=\frac{{g\left( {V_{m}^{\text{A}}}\right) -g\left( {V_{m}^{\text{B}}}\right) }}{2}.
\end{equation}
From Fig. \ref{fig_bulk_interpretation}(b), the mass term exhibits an inversion. Thus, the Dirac 
mass inversion leads to the formation of the Jackiw-Rebbi-type Dirac boundary modes, i.e., the
topologically nontrivial bulk solitons.

\section{Experimental measurement of the topological gap solitons\label{app_H}}

In this section, we give more experimental results of the gap solitons, including both the topologically
nontrivial and trivial ones.

\begin{figure}[tbp]
\includegraphics[width=8.5cm]{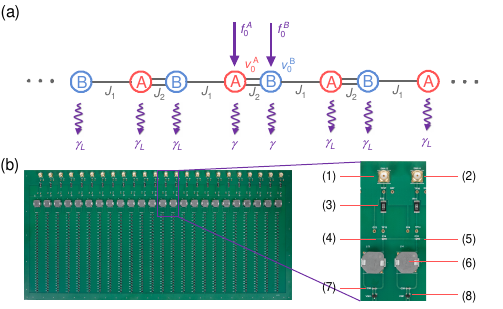}
\caption{Experimental implementation of the SSH lattice with onsite nonlinearity 
that supports the topological gap solitons.
(a) Schematic of the nonlinear SSH lattice with the two external driving 
voltages $f_{0}^{\text{A}}$ and $f_{0}^{\text{B}}$.
(b) Fabricated PCB of the driven-dissipative nonlinear circuit
lattice with the enlarged figure showing the circuit components:
(1)-(2) SMA connectors, (3) shunt resistor $R$, 
(4) coupling capacitor $C_{2}$, (5) coupling capacitor $C_{1}$, (6) grounding 
inductor $L_{\text{g}}$, (7) grounding capacitor $C_{\text{g}}$, and (8) varactor 
diode $C_{\text{v}}$. The two SMA connectors are connected to the external voltage 
sources with the voltage amplitude $f_{0}^{\text{A}}$ and $f_{0}^{\text{B}}$, respectively.}
\label{fig_bulk_PCB}
\end{figure}

Besides the nonlinear topological edge states, in this section we observe the bulk solitons
which reside in the bulk of the circuit lattice. According to the theoretical results,
both the topologically nontrivial and trivial bulk solitons are mainly localized at the
middle unit cell. For the topologically nontrivial bulk solitons, the voltages at the two sites of the
middle unit cell are out-of-phase. While for the topologically trivial bulk solitons, the
voltages at the sublattice sites $\text{A}$ and $\text{B}$ are in-phase. Due to the 
distinct properties of the topologically nontrivial and trivial bulk solitons, experimentally 
we need to excite the two types of the bulk solitons using the different driving voltages. 
As shown in Fig. \ref{fig_bulk_PCB}(a), the nonlinear SSH lattice has the intracell hopping 
$J_{2}$ and
intercell hopping $J_{1}$. The two sublattice sites of the middle unit cell are driven
by the external voltages $f_{0}^{\text{A}}$ and $f_{0}^{\text{B}}$, respectively. The 
lattice also
experiences the dissipations induced both from the series resistance of inductors and shunt
resistors. Since this lattice is a topologically trivial lattice in the linear limit, the absence of the
edge solitons avoids the coupling between the bulk and edge solitons. 
Figure \ref{fig_bulk_PCB}(b) shows the fabricated PCB of the
driven-dissipative nonlinear circuit lattice, and the inset shows the
enlarged figure with the circuit components: (1)-(2) SMA connectors, (3) shunt resistor $R$, 
(4) coupling capacitor $C_{2}$, (5) coupling capacitor $C_{1}$, (6) grounding 
inductor $L_{\text{g}}$, 
(7) grounding capacitor $C_{\text{g}}$, and (8) varactor diode $C_{\text{v}}$.
Here, the two SMA connectors are connected to the external voltage sources with the
voltage amplitude $f_{0}^{\text{A}}$ and $f_{0}^{\text{B}}$, respectively. 
The experimental circuit lattice has the parameter 
$N = 12$, which corresponds to a total of 24 lattice sites.
Note that the 
phase difference between the driving voltages can be tuned arbitrarily. Such flexibility of 
electrical circuits provides convenience to the observation of the nonlinear topological states.
In the theoretical calculations, we also set $N=120$, indicating that the lattice contains
$120$ unit cells. 

\begin{figure*}[tbp]
\includegraphics[width=17.8cm]{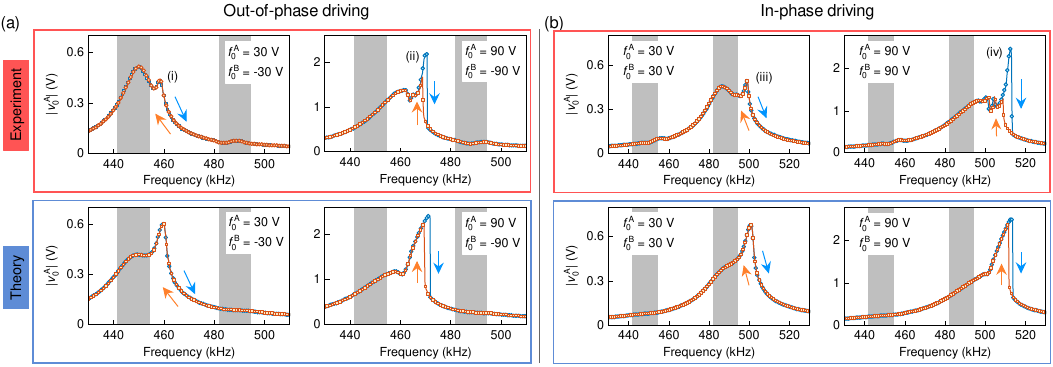}
\caption{Voltage spectra at the bulk node when the circuit is excited with the
out-of-phase or in-phase driving voltages $f_{0}^{\text{A,B}}$.
(a) Voltage spectra at the sublattice site $\text{A}$ in the middle unit cell 
when the circuit is excited with the out-of-phase driving voltages. 
(b) Voltage spectra under the in-phase driving. In (a) and (b),
the gray areas correspond to the linear bulk bands. }
\label{fig_bulk_spectra}
\end{figure*}

\begin{figure*}[tbp]
\includegraphics[width=9.4cm]{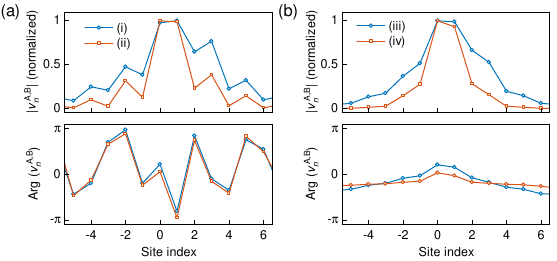}
\caption{Experimental voltage distributions at the resonant frequencies labeled 
in Fig. \ref{fig_bulk_spectra}. 
(a)-(b) correspond to the results under the out-of-phase 
driving and in-phase driving, respectively. The voltage amplitudes are normalized.}
\label{fig_bulk_voltage}
\end{figure*}

First, we impose the out-of-phase driving voltages and observe the topologically nontrivial bulk 
solitons. Figure \ref{fig_bulk_spectra}(a) shows the voltage spectra at the sublattice 
site $\text{A}$ in the middle unit cell. 
The first and second rows show the experimental and theoretical results, respectively. 
The directions of the frequency sweep are denoted by the blue and orange arrows. 
The first column in Fig. \ref{fig_bulk_spectra}(a) shows the voltage spectra 
when the driving voltages are $f_{0}^{\text{A}} = 30~\text{V}$ and $f_{0}^{\text{B}} 
= -30~\text{V}$.
We observe that one peak reside in the middle SSH gap and the voltage spectra for the
frequency sweep along the two opposite directions coincide with each other. When the
driving voltages are increased to $f_{0}^{\text{A}} = 90~\text{V}$ and $f_{0}^{\text{B}} 
= -90~\text{V}$,
the spectrum peaks become highly asymmetric and exhibit the bistable response, as shown
in the second column of Fig. \ref{fig_bulk_spectra}(a). 

In order to prove that the peaks which reside in 
the SSH gap are the signatures of the excitation of the topologically nontrivial bulk solitons,
we measure the voltage distributions in the nonlinear circuit lattice under the
different driving voltages. From Fig. \ref{fig_bulk_voltage}(a), the voltage distributions 
are mainly confined to the middle unit cell
which contains the lattice sites $0$ and $1$. When we split the lattice into the
two parts from the 
middle of the middle unit cell, i.e., the lattice site $0.5$, the left part of the voltage
distribution is mainly confined to the sublattice site $\text{A}$ and exhibits a phase jump of 
$\pi $ among the neighboring cells. The right part of the voltage distribution also exhibits a 
phase jump but it is mainly confined to the sublattice site $\text{B}$. Such phenomenon 
agrees with the behavior of the topologically nontrivial bulk solitons. 

Second, we impose the in-phase driving voltages and observe the topologically trivial bulk solitons.
In Fig. \ref{fig_bulk_spectra}(b), we observe the phenomenon similar to that shown 
in Fig. \ref{fig_bulk_spectra}(a).
When the driving voltages are $f_{0}^{\text{A}} = 30~\text{V}$ and $f_{0}^{\text{B}} 
= 30~\text{V}$, we also observe a peak from the voltage spectra, but this peak reside 
in the upper semi-infinite gap. When the driving voltages increase to $f_{0}^{\text{A}} 
= 90~\text{V}$ and $f_{0}^{\text{B}} = 90~\text{V}$,
the peak in the upper semi-infinite gap exhibits the bistable response, as shown
in the second column. From the normalized amplitudes and phases
of the voltage distributions shown in Fig. \ref{fig_bulk_voltage}(b), the topologically
trivial bulk solitons are
also mainly localized at the lattice site $0$ and $1$, but the voltages at the different sites 
have nearly equal phases, in strong contrast to the phase jump observed for the topologically nontrivial 
bulk solitons. Since the topologically trivial bulk solitons are the nonlinearity induced localized modes,
the topological trivial bulk solitons become more localized with the increasing circuit
nonlinearity.

We would like to note that, for both the topologically nontrivial and trivial bulk solitons, 
the experimental measurement results exhibit some deviations from the theoretical predictions. 
As shown in Fig. \ref{fig_bulk_spectra}, these deviations become more pronounced at higher 
input voltages, primarily due to increased series resistance of the inductors and decreased driving voltages 
provided by the voltage source under large nonlinearities. Additionally, as illustrated in Fig. \ref{fig_bulk_voltage}, 
the experimental voltage amplitudes deviate from perfectly symmetric profiles, which may arise from three 
main factors: imperfections in the circuit lattice caused by component errors that result in a lack of symmetry 
about the center of the unit cell, unequal amplitudes of the two driving voltages, and limitations in the 
temporal resolution of the oscilloscope, which may contribute to errors in the resonant frequencies we obtained.

\begin{figure*}[tbp]
\includegraphics[width=15cm]{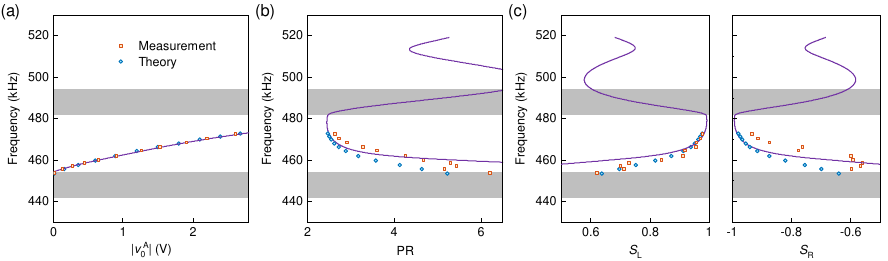}
\caption{Bulk voltages $\vert v_{0}^{\text{A}} \vert$, participation ratios (PRs), and 
local sublattice polarizations $S_{\text{L,R}}$ of the topological gap solitons.
The orange squares and blue circles correspond to the experimental and theoretical 
results, respectively. The purple curves denote the result
calculated from the GP equation without the driven-dissipative terms,
and the gray areas correspond to the linear bulk bands.  }
\label{fig_bulk_PR_S}
\end{figure*}

To quantitatively compare the experimental and theoretical results of the topologically nontrivial
gap solitons, we extract the resonant frequencies from the voltage spectra by following
the same procedure. Figure \ref{fig_bulk_PR_S}(a) shows the dependence between the 
frequency and voltage $v_{0}^{\text{A}}$. The result from the GP equation without the 
driven-dissipative terms is also plotted for comparison. The experimental
result (orange squares) agrees well with the theoretical result (blue circles) and the 
result from the GP equation without the driven-dissipative terms. This observation validates 
the fact that the topologically nontrivial gap solitons bifurcate from the edge of the
linear Bloch band and converge to the bulk state in the linear limit.
Meanwhile, since the topologically nontrivial bulk solitons are nonlinearity induced, 
with the increasing of the frequency, i.e., with the enhancement of the nonlinearity, 
the voltage distribution becomes more localized with the smaller PRs, as shown in
Fig. \ref{fig_bulk_PR_S}(b). To quantitatively measure the sublattice pseudospin of the 
topologically nontrivial bulk solitons, we split the circuit lattice into the two parts from the middle of the 
middle unit cell, and the local sublattice pseudospins are calculated for the left and right 
parts, respectively. The result is shown in Fig. \ref{fig_bulk_PR_S}(c). From the first column, 
the left parts of the topologically nontrivial bulk solitons are mainly confined to the sublattice site 
$\text{A}$. While the right parts are mainly confined to the sublattice site $\text{B}$, 
as shown in the second column. Meanwhile, with the increasing of the frequency, the 
absolute values of both $S_{\text{L}}$ and $S_{\text{R}}$ approach $1$.
This implies that, a stronger nonlinearity results in a greater voltage localization on the 
sublattice site $\text{A}$ for the left part and $\text{B}$ for the right part. Such property
validates that the topologically nontrivial bulk solitons are the nonlinearity induced topological states.

\section{Theoretical results of the self-induced topological edge states\label{app_I}} 

In this section, we give more theoretical results of the self-induced topological 
edge states. For completeness, we also introduce the topologically trivial edge
solitons. The topologically trivial chain consists of $120$ unit cells, i.e., $N = 120$. 
In the linear limit, there are no localized states present at either end of the chain.
Conventionally, the topologically trivial edge solitons are called as the surface solitons \cite{RMP83-247,PR463-1}.
Here, since both the self-induced topological edge states and topologically trivial edge 
solitons reside at the edge of the topologically trivial lattice, we simply call them edge solitons 
for clarity. This section is organized as follows. In the first subsection, 
we introduce the existence of the edge solitons. Then in the second subsection, 
we show the results for the stability analysis of the edge solitons. Finally, we discuss the 
physical interpretation of the self-induced topological edge states in the last subsection.

\subsection{Existence of the edge solitons}

We study the edge solitons that reside at the edge of a topologically trivial lattice, where 
the varactor diodes in the circuit are modeled as the nonsaturable and saturable 
nonlinearities, respectively. We solve the GP equation 
without the driven-dissipative terms using the Newton's method. 
In contrast to the nonlinear topological edge states in a topologically nontrivial lattice where the 
topological edge state in the linear limit can be taken as the initial guess solution, no edge 
states exist in the linear limit for a topologically trivial
lattice and thus soliton solutions have to be found using the AC approach. For simplicity, 
we only study the two cases with $C_{1} = 0$ and $C_{2} = 0$, respectively. 
For the case with $C_{2} = 0$, the first site is decoupled from the lattice and the remaining structure 
becomes the discrete dimers with the intracell coupling governed by $C_{1}$.
We are only interested in the solution where the first site has a nonzero value. 
Starting from the GP equation without the driven-dissipative terms, in the case with 
$C_{2} = 0$ we get the following equation
\begin{equation}
E_{0}{+g\left( {{\psi }_{1}^{\text{A}}}\right) =\bar{\omega}.}
\end{equation}%
By solving this equation, we can get the value of ${{\psi }_{1}^{\text{A}}}$. Then the 
solutions for the edge solitons can be obtained by gradually increasing $C_{2}$ to the
original value. The case with $C_{1} = 0$ is equivalent to the fully dimerized limit and 
the lattice reduces to the discrete dimers with the intracell coupling governed by 
$C_{2}$. Similarly, we are only interested in the solutions where the first dimer has 
the nonzero values. The GP equation reduces to
\begin{eqnarray}
E_{0}{{\psi }_{1}^{\text{A}}+g\left( {{\psi }_{1}^{\text{A}}}\right) {\psi }_{1}^{\text{A}}}+J_{2}{%
\psi }_{1}^{\text{B}}&=&\bar{\omega}{\psi _{1}^{\text{A}}},  \\
E_{0}{{\psi }_{1}^{\text{B}}+g\left( {{\psi }_{1}^{\text{B}}}\right) {\psi }_{1}^{\text{B}}}+J_{2}{{%
\psi }_{1}^{\text{A}}}&=&\bar{\omega}{\psi _{1}^{\text{B}}} .
\end{eqnarray}%
By solving this equation, multiple solutions may exist.

\begin{figure*}[tbp]
\includegraphics[width=10.3cm]{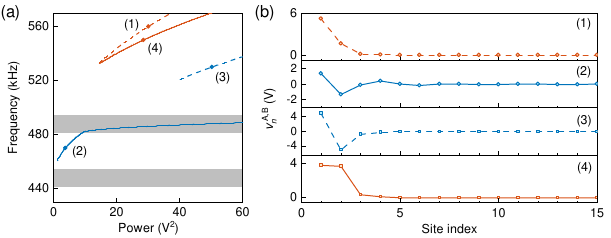}
\caption{Edge solitons in a topologically trivial lattice with the nonsaturable nonlinearity.
(a) Frequencies of the edge solitons. The solid blue curve denotes the 
self-induced topological states, and the solid orange, dashed blue, and dashed
orange curves correspond to the symmetric, antisymmetric, and asymmetric
topologically trivial edge solitons, respectively.
(b) Voltage distributions of the edge solitons labeled in (a).}
\label{fig_self_nonsaturable}
\end{figure*}

We seek for the soliton solutions in a topologically trivial SSH lattice. When the varactor 
diodes are modeled as the nonsaturable nonlinearity, we show the frequencies
of the edge solitons in Fig. \ref{fig_self_nonsaturable}(a). In the AC limit, one 
solution exists in the case with $C_{2} = 0$. This solution corresponds to the 
topologically trivial soliton which reside in the upper semi-infinite gap, as shown 
by the dashed orange curve in Fig. \ref{fig_self_nonsaturable}(a). A typical 
profile of this type of topologically trivial solitons is shown in the first row of 
Fig. \ref{fig_self_nonsaturable}(b). Its profile is asymmetric and its voltage 
mainly distributes on the first leftmost site. In the case with $C_{1} = 0$, we 
find that there are five solutions: two antisymmetric solutions, one symmetric 
solution, and two asymmetric solutions. One of the two antisymmetric solutions 
corresponds to the topologically nontrivial soliton which reside in the SSH gap, as shown by 
the sold blue curve in Fig. \ref{fig_self_nonsaturable}(a). This type of topologically nontrivial
edge solitons is the same to the topological edge solitons discussed in the main 
text and a typical profile is shown in the second row of Fig. \ref{fig_self_nonsaturable}(b).
Another antisymmetric solution corresponds to the topologically trivial edge soliton 
which reside in the upper semi-infinite gap, as shown by the dashed blue
curve in Fig. \ref{fig_self_nonsaturable}(a). It has an antisymmetric voltage distribution 
as shown in the third row of Fig. \ref{fig_self_nonsaturable}(b). The symmetric solution 
corresponds to the topologically trivial edge soliton that we discussed in the main text, 
as shown by the solid orange curve in Fig. \ref{fig_self_nonsaturable}(a). It has a 
symmetric profile as shown in the fourth row of Fig. \ref{fig_self_nonsaturable}(b).
While for the asymmetric solutions, one asymmetric solution is equivalent to
the solution in the case with $C_{2} = 0$, and another asymmetric solution also has 
an asymmetric profile but its voltage mainly distributes on the second site, i.e., 
the site $\text{B}$ of the leftmost unit cell.

\begin{figure*}[tbp]
\includegraphics[width=14.6cm]{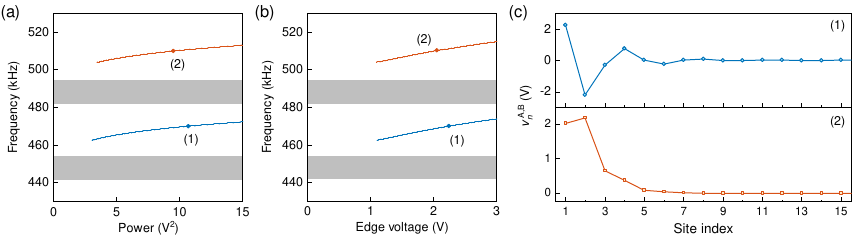}
\caption{Edge solitons in a topologically trivial lattice with saturable nonlinearity. 
(a)-(b) Frequencies of the edge solitons, with solid blue curves representing the 
self-induced topological edge states and solid orange curves corresponding 
to the topologically trivial edge solitons. 
(c) Voltage distributions of the edge solitons indicated in (a) and (b).}
\label{fig_self_mode}
\end{figure*}

When the varactor diodes are modeled as the saturable nonlinearity, there is no 
solution in the case with $C_{2} = 0$ because of the weak nonlinearity. In the case 
with $C_{1} = 0$, we find two solutions: one symmetric solution and one antisymmetric 
solution. The symmetric solution corresponds to the topologically trivial edge solitons 
and the antisymmetric solution corresponds 
to the topologically nontrivial edge solitons, i.e., the self-induced topological edge states at
the edge of a topologically trivial lattice.
The frequencies of the edge solitons in a circuit lattice with the saturable nonlinearity
are shown in Figs. \ref{fig_self_mode}(a)-(b). To compare with the theoretical
and experimental excitation spectra, we also plot the dependence between the frequency 
and edge voltage $v_{1}^{\text{A}}$. From Figs. \ref{fig_self_mode}(a)-(b),
the edge solitons do not exist in the linear limit. However,
when the power/voltage is large enough, two branches
for the edge solitons emerge where one branch reside in the SSH gap and the other one
reside in the upper semi-infinite gap.
The soliton profiles are shown in 
Fig. \ref{fig_self_mode}(c). Note that although the topologically trivial edge solitons are
the continuations of the symmetric solution in the AC limit, the voltages at the first two
sites shown in Fig. \ref{fig_self_mode}(c) is slightly asymmetric.

Based on the soliton profiles shown in Fig. \ref{fig_self_mode}(c), when we neglect the voltage 
at the first site, the topological edge soliton exhibits an antisymmetric voltage distribution across the two 
leftmost sites and resembles the linear topological edge states. Specifically, this topological edge soliton features 
a phase jump of $\pi$ between neighboring cells and is primarily confined to the sublattice site $\text{B}$, 
accompanied by a decaying tail that approaches zero.
Physically, this type of edge soliton can be understood as a nonlinearity-induced topological edge soliton in a 
topologically trivial lattice. The circuit nonlinearity introduces a defect at site $\text{A}$ of the leftmost unit cell, 
enabling the newly formed lattice to support the existence of a topological edge state that extends from 
site $\text{B}$ of the leftmost unit cell to the infinite right end.
Considering these properties, the antisymmetric solution obtained in the AC limit evolves into the topological 
edge soliton (for a more detailed discussion, see Appendix \ref{app_I}, Section 3). 
In contrast, for the topologically trivial edge soliton in a topologically trivial lattice, the voltage is predominantly 
localized on the two sites of the leftmost unit cell, without a phase jump between neighboring cells. 
These characteristics indicate that this type of edge soliton represents nonlinearity-induced localized modes, 
and thus, the symmetric solution obtained in the AC limit corresponds to the topologically trivial edge soliton.

\subsection{Stability analysis of the edge solitons}

\begin{figure*}[tbp]
\includegraphics[width=12.9cm]{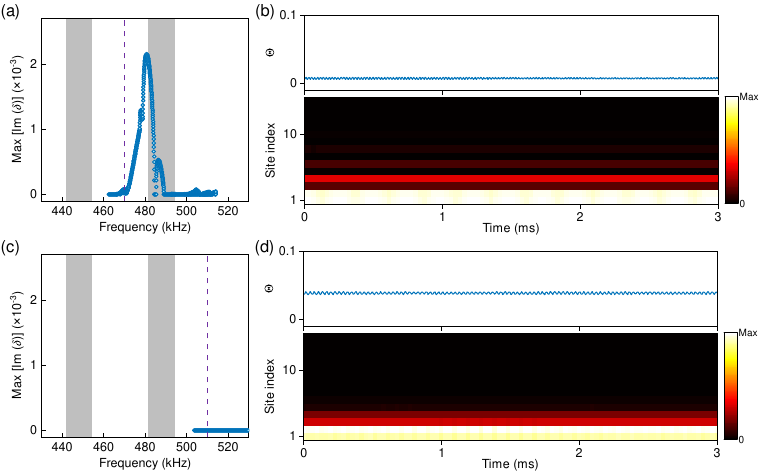}
\caption{Stability analysis of the edge solitons.
(a) The maximum growth rates of the perturbed solutions of the self-induced topological edge states. 
(b) The temporal evolution of the self-induced topological edge state at the 
frequency indicated in (a). Noises with $\pm 2\%$ random perturbations 
are added to the initial input.
(c)-(d) Results for the topologically trivial edge solitons.
In (a) and (c), the gray areas denote the linear bulk bands. }
\label{fig_self_stability}
\end{figure*}

We then study the stability analysis of both the topologically nontrivial and trivial edge
solitons. Figure \ref{fig_self_stability}(a) shows the maximum 
growth rates of the self-induced topological edge states at the different 
frequencies. From the 
figure, the maximum growth rates are in the order of $10^{-3}$ when the soliton
frequencies are near to the lower edge of the top bulk band. This region corresponds
to the linearly unstable solitons. However, since we are interested in the localized 
self-induced topological edge states
and the maximum frequency shifts of the solitons are limited by 
the experimental input voltage (see Fig. \ref{fig_self_mode}(b)), we carry out 
the temporal evolution of the self-induced topological edge states
at the frequency of $470~\text{kHz}$. Noises with $\pm 2\%$ random 
perturbations are added to the input amplitude. To quantitatively measure the mode 
stability/instability, we again use the asymmetry parameter $\Theta$ to characterize 
the voltage asymmetry between the sites $1$ and $2$, i.e., the first two sites of the
topologically trivial SSH lattice. From the voltage distribution shown
in Fig. \ref{fig_self_stability}(b), there are no apparent variations along the temporal
evolution. The asymmetry parameters are in the order of $10^{-3}$. These results
imply that, although the unstable region exists, at least within the experimentally 
realizable parameter range, the self-induced topological edge states should be 
observable because of their weak instabilities.

The results for the stability analysis of the topologically trivial edge solitons are shown
in Figs. \ref{fig_self_stability}(c)-(d). Compared to the self-induced topological edge
states, the topologically trivial edge solitons have much smaller $\text{Max} \left[\text{Im}
\left( \delta \right) \right]$ (in the order of $10^{-9}$), implying that they
are much more stable. We take the topologically trivial edge soliton at $510~\text{kHz}$
as an example and study its temporal evolution, again with $\pm 2\%$ noises
added to the input. From Fig. \ref{fig_self_stability}(d), both the voltage distribution
and asymmetry parameters confirm that the topologically trivial edge soliton is stable.

\subsection{Physical interpretation of the self-induced topological edge states}

The self-induced topological edge states can also be understood from the solution of the
GP equation. Since the edge solitons mainly reside at the two sites of the first unit cell, 
we only consider the circuit nonlinearities at the leftmost two sites and neglect
the nonlinearities at the other sites. With $V_{m}^{\text{A,B}} \left( T \right)
= v_{m}^{\text{A,B}} e^{-\mathrm{i} \bar{\omega}T}$, the governing equations reduce to
\begin{eqnarray}
\bar{\omega} v_{1}^{\text{A}}&=& E_{0} v_{1}^{\text{A}}+g\left( v_{1}^{\text{A}} \right)v_{1}^{\text{A}}
+J_{2}{v_{1}^{\text{B}}},  \\
\bar{\omega} v_{1}^{\text{B}}&=& E_{0}v_{1}^{\text{B}}+g\left( v_{1}^{\text{B}} \right)v_{1}^{\text{B}}
+J_{2}{v_{1}^{\text{A}}}+J_{1}{v_{2}^{\text{A}}},  \\
\bar{\omega} v_{2}^{\text{A}}&=& E_{0} v_{2}^{\text{A}}+J_{2}{v_{2}^{\text{B}}}+J_{1}{v_{1}^{\text{B}}},  \\
\bar{\omega} v_{2}^{\text{B}}&=& E_{0} v_{2}^{\text{B}}+J_{2}{v_{2}^{\text{A}}}+J_{1}{v_{3}^{\text{A}}},  \\
&\ldots&  .
\end{eqnarray}%
We let $\bar{\omega} = E_{0}$ which implies that the frequency of the self-induced topological
edge state equals to the frequency of the linear topological edge state in an SSH lattice, then 
the equations reduce to
\begin{eqnarray}
g\left( v_{1}^{\text{A}} \right)v_{1}^{\text{A}}+J_{2}{v_{1}^{\text{B}}} &=& 0,  \\
g\left( v_{1}^{\text{B}} \right)v_{1}^{\text{B}}+J_{2}{v_{1}^{\text{A}}}+J_{1}{v_{2}^{\text{A}}} &=& 0,  \\
J_{2}{v_{2}^{\text{B}}}+J_{1}{v_{1}^{\text{B}}} &=& 0,  \\
J_{2}{v_{2}^{\text{A}}}+J_{1}{v_{3}^{\text{A}}} &=& 0,  \\
&\ldots&  .
\end{eqnarray}%
Considering the profile of the self-induced topological edge state, we have
$v_{m}^{\text{B}} = \left(- \frac{J_{1}}{J_{2}} \right)^{\vert m \vert -1} v_{1}^{\text{B}}$
and $v_{m}^{\text{A}}=0$ for $m \geq 2$, and the equations further reduce to
\begin{eqnarray}
g\left( v_{1}^{\text{A}} \right)v_{1}^{\text{A}}+J_{2}{v_{1}^{\text{B}}} &=& 0,  \\
g\left( v_{1}^{\text{B}} \right)v_{1}^{\text{B}}+J_{2}{v_{1}^{\text{A}}} &=& 0.  
\end{eqnarray}%
The above two equations govern the voltage distributions at the leftmost two sites. Again
considering the profile of the self-induced topological edge state, we have $v_{1}^{\text{A}} = -v_{1}^{\text{B}}$
and 
\begin{equation}
g\left( v_{1}^{\text{A,B}} \right) = J_{2}. \label{self_reduce}
\end{equation}
Thus, if Eq. (\ref{self_reduce}) has solutions, the self-induced topological edge states can be 
interpreted by the solution of the GP equation. In other words, the self-induced topological 
edge states that reside at the edge of a topologically trivial lattice can 
be approximately mapped to the linear topological edge state of a semi-infinite SSH lattice.

\begin{figure*}[tbp]
\includegraphics[width=10.3cm]{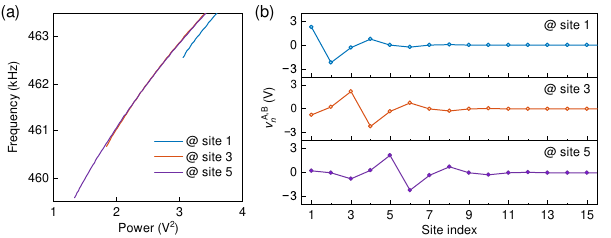}
\caption{Relation between the self-induced topological edge states and topological gap solitons.
(a) Existence curves (the dependence between the frequency and soliton power) 
of a series of solitons residing at the different positions.
(b) Voltage distributions of the different solitons at $470~\text{kHz}$.}
\label{fig_self_threshold}
\end{figure*}

To explain the relation between the self-induced topological edge states and topological gap 
solitons, we select the topological gap soliton at a given frequency and move 
it from the lattice bulk to the left edge. Due to the discrete translational symmetry, the 
soliton profile is invariant when the soliton peak is far from the edge. However, the discrete 
translational symmetry is broken at the lattice edge because the SSH lattice that we study 
has an open boundary. The soliton profile is changed due to the presence of the edge.
Figure \ref{fig_self_threshold}(a) shows the existence curves (the dependence between the
frequency and soliton power) of a series of solitons residing at the different positions, and 
Fig. \ref{fig_self_threshold}(b) correspond to the voltage distributions of the different solitons.
From Fig. \ref{fig_self_threshold}(a), a power threshold is induced when moving the topological
gap soliton towards the edge.

\section{Experimental measurement of the self-induced topological edge states\label{app_J}} 

In this section, we give more experimental results of the edge solitons
in a topologically trivial lattice, particularly the self-induced topological edge states.

Theoretically, we solve the driven-dissipative GP equation 
with the driven terms ${F_{m}^{\text{A}}}\left( t\right) =\delta _{m,1}{f_{0}^{\text{A}}}\exp \left( -%
{i\bar{\omega}T}\right) $ and $F_{m}^{\text{B}} \left( t \right) = 0$, i.e., only the edge site is excited.
Then the equations are solved by following the same procedure. Note that in order to have
a better comparison with the experimental results, in the topologically trivial lattice, the series 
resistance of inductors is set as $R_{L} = 650~\text{m}\Omega$. In the theoretical calculations, 
the lattice consists of $N = 120$ unit cells, indicating that it contains an even number of lattice sites.

\begin{figure*}[tbp]
\includegraphics[width=17cm]{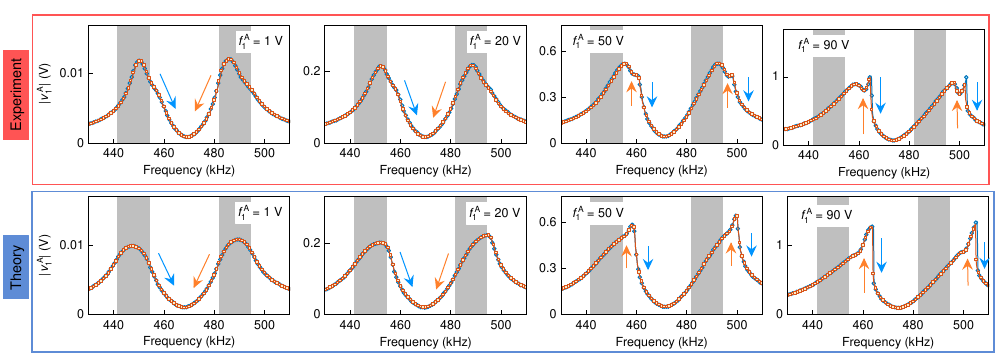}
\caption{Voltage spectra at the edge node when the topologically trivial circuit lattice is 
excited with the driving voltage $f_{1}^{\text{A}}$. 
The first and second rows show the results from the experiment and
GP equation, respectively. In all the rows, the blue and orange
curves correspond to the frequency sweep along the two opposite directions
(denoted by the blue and orange arrows).}
\label{fig_self_spectra}
\end{figure*}

We fabricate another nonlinear SSH circuit lattice with the intracell hopping $J_{2}$ and
intercell hopping $J_{1}$. The experimental circuit lattice has the parameter 
$N = 12$, which corresponds to a total of 24 lattice sites.
In the linear limit, this lattice is a topologically trivial lattice. Experimentally, 
we just change the positions of $C_{1}$ and $C_{2}$, and observe the edge solitons. We plot the 
experimental and theoretical voltage spectra at the edge node 
in Fig. \ref{fig_self_spectra}. From the fist column, when the driving voltage is small with 
$f_{1}^{\text{A}} = 1~\text{V}$, two spectrum peaks appear in the bulk bands. 
Since this lattice has no topological edge state in the linear limit, there is no peak in the SSH gap.
Meanwhile, since the nonlinearity-induced topologically trivial solitons usually exist above a power
threshold, there is also no peak in the upper semi-infinite gap.
These results imply that, in the linear limit, only the bulk states are excited
under the external driving, in contrast to the spectra for a topologically nontrivial lattice where the 
topological edge state is excited.
When the driving voltage increases to $f_{1}^{\text{A}} = 20~\text{V}$, the resonant frequencies 
of the two peaks exhibit the blue shift, as shown by the second column of 
Fig. \ref{fig_self_spectra}. 
The resonant frequencies of the two peaks reside near the band edges and there are only two 
peaks in total. 

Then we continue to enhance the circuit nonlinearity by increasing the driving voltage.
For $f_{1}^{\text{A}} = 50~\text{V}$,
the original two peaks near the band edges shrink and two new peaks appear, as shown in the
third column. The peak corresponding
to the topologically nontrivial edge soliton (i.e., the self-induced topological edge states) reside in the SSH 
gap, and the peak corresponding to the
topologically trivial edge soliton reside in the upper semi-infinite gap. The small discrepancy between
the experimental and theoretical results is due to the circuit component errors. 
When the driving voltage further increases to $f_{1}^{\text{A}} = 90~\text{V}$, from the last 
column, the peaks for the edge solitons become more
pronounced. Specifically, we observe a clear bistable response for the peak corresponding to the 
topologically trivial edge soliton. Note that, as shown in Fig. \ref{fig_self_spectra}, 
the experimental measurement results exhibit deviations from the theoretical predictions at higher 
input voltages, primarily due to increased series resistance of the inductors and reduced driving 
voltages supplied by the voltage source.

According to both the experimental and theoretical results,
the edge solitons appear when the driving voltage is approximately larger than $50~\text{V}$. 
However, the corresponding voltages at the edge sites are smaller than the theoretically predicted
values obtained from the GP equation shown in Fig. \ref{fig_self_mode}(b). Only when the driving 
voltage increases to $80~\text{V}$, the result from the driven-dissipative system agrees with
the prediction from Fig. \ref{fig_self_mode}(b). The disagreement for the
driving voltage between $50~\text{V}$ to $80~\text{V}$ may be possibly induced by the 
excitation of the higher order edge solitons, which cannot be derived simply by setting 
$C_{1} = 0$ or $C_{2} = 0$ in the AC limit. In our study, we neglect these higher 
order edge solitons. 

\begin{figure}[tbp]
\includegraphics[width=4.3cm]{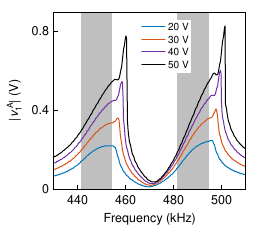}
\caption{Theoretical excitation spectra for the self-induced topological edge 
states. For simplicity, we only show the spectra when the driving
frequency is swept from low to high. The gray areas denote the linear bulk bands.}
\label{fig_self_peak}
\end{figure}

To clearly show the emergence of the peak that corresponds to the self-induced topological
edge state, Fig. \ref{fig_self_peak} shows the theoretical excitation spectra when the 
input voltage
varies from $20~\text{V}$ to $50~\text{V}$. The series resistance of inductors is now set 
as $R_{L} = 500~\text{m}\Omega$. For clarity, we only show the spectra when the driving
frequency is swept from low to high. From the figure, a new peak is gradually appeared
in the SSH gap. Such observation implies the power/voltage threshold of the existence
of the self-induced topological edge states. 

To explain the origin of the two peaks,
we also measure the voltage distributions at the corresponding resonant frequencies. 
For the peak in the SSH gap, when we neglect the voltage at the first site, the voltage 
distribution exhibits a phase jump of $\pi $ among the
neighboring cells and the voltages at the site $\text{A}$ (except the site $\text{A}$ 
in the first unit cell) are nearly zero. Thus, this peak corresponds to the self-induced
topological edge state in a topologically trivial lattice. For the peak in the upper semi-infinite
gap, the voltage is mainly localized on the two sites of the leftmost unit cell with nearly 
equal amplitudes between the two sites. From the phase distribution, the voltages at 
the different sites have nearly equal phases, in strong contrast to the phase jump observed 
for the peak in the SSH gap. Thus, the peak in the semi-infinite gap corresponds to 
the nonlinearity induced localization mode, i.e., the topologically trivial edge soliton in a 
topologically trivial lattice. Our experimental observation validates the existence of both the
topologically nontrivial and trivial edge solitons in a topologically trivial lattice,
confirming the theoretical predication.

\begin{figure*}[tbp]
\includegraphics[width=12.9cm]{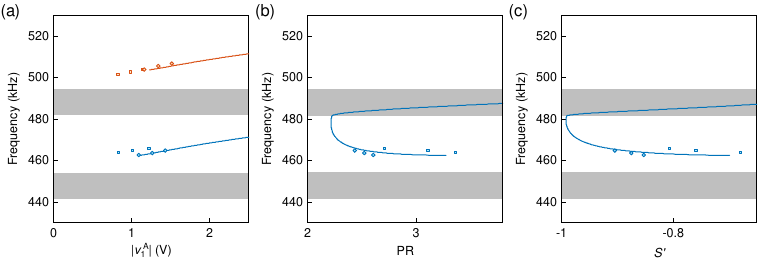}
\caption{Edge voltages $\vert v_{1}^{\text{A}} \vert$, participation ratios (PRs), and 
sublattice polarizations $S^{\prime}$ of the self-induced topological edge states.
The squares and circles correspond to the experimental and theoretical 
results, respectively. The curves denote the result
calculated from the GP equation without the driven-dissipative terms,
and the gray areas correspond to the linear bulk bands. In (a), we also plot
the results for the topologically trivial edge solitons (shown in orange).}
\label{fig_self_PR_S}
\end{figure*}

Figure \ref{fig_self_PR_S} shows the edge voltages $\vert v_{1}^{\text{A}} \vert$, participation 
ratios (PRs), and sublattice polarizations $S^{\prime}$ of the self-induced topological edge states. 
From Fig. \ref{fig_self_PR_S}(a), no edge solitons can be excited near the linear limit. Both the 
topologically nontrivial and trivial edge solitons exist when the edge voltage is above the certain threshold 
values. This property is totally different to the thresholdless excitation of the nonlinear topological
edge states. Since the increased circuit nonlinearity induces the decreased
grounding capacitance, the frequencies of the edge solitons in a topologically trivial lattice 
also exhibit the blue shift. Besides, different to the nonlinear topological edge states where the
nonlinearity weakens their localizations, the self-induced topological edge states become 
more localized with the increased nonlinearity. Since the self-induced topological edge states are 
nonlinearity induced, a stronger nonlinearity results in the
greater voltage localization, as shown in Fig. \ref{fig_self_PR_S}(b). Meanwhile,
from Fig. \ref{fig_self_PR_S}(c),
with the increasing of the frequency, the sublattice pseudospin $S^{\prime}$ approaches 
$-1$, implying a greater voltage localization on the sublattice site $\text{B}$.
Note that here $S^{\prime}$ is defined by neglecting the voltage at the first site. 
This observation further validates the origin of the self-induced topological edge states.
Considering the fact that a linear topological edge state has $S  = -1$, 
a stronger circuit nonlinearity leads to a larger defect at the site $\text{A}$ of the leftmost unit 
cell, and the self-induced topological edge state is more close to the linear topological edge state 
distributed from the site $\text{B}$ of the leftmost unit cell to the right infinite end.
For the topologically trivial edge solitons, both the experimental and theoretical results show 
that the topologically trivial edge solitons have $S  \approx 0$, regardless of the state frequencies.
This observation validates that the topologically trivial edge solitons in a topologically trivial lattice
are the nonlinearity induced localization modes and they always reside at the two sites of the first 
unit cell with the equal amplitudes.

\section{Calculation of the nonlinear Berry phase\label{app_K}}

Starting from Eqs. (\ref{eq-7}) and (\ref{eq-8}), under the periodic boundary condition, we assume that the solutions 
can still be expressed in the form of Bloch functions: ${V{_{m}^{\text{A},\text{B}}=\phi }}_{\text{A},\text{B}}
\exp \left(\mathrm{i}k m -\mathrm{i}\bar{\omega}T\right) $.
Substituting this Bloch ansatz into Eqs. (\ref{eq-7}) and (\ref{eq-8}), we obtain:
\begin{eqnarray}
\left( J_{1}+J_{2}e^{-\mathrm{i}k}\right) \phi _{\mathrm{B}}+g\left( \phi _{\mathrm{A}}\right) \phi
_{\mathrm{A}} &{=}&\left( \bar{\omega}-E_{0}\right) \phi _{\mathrm{A}},  \label{suppnew_eq1} \\
\left( J_{1}+J_{2}e^{\mathrm{i}k}\right) \phi _{\mathrm{A}}+g\left( \phi _{\mathrm{B}}\right) \phi _{\mathrm{B}}
&{=}&\left( \bar{\omega}-E_{0}\right) \phi _{\mathrm{B}}.  \label{suppnew_eq2}
\end{eqnarray}%
These equations define the corresponding Bloch Hamiltonian of a nonlinear system, which depends on the wave 
functions $\phi _{\mathrm{A,B}}$. In addition to Eqs. (\ref{suppnew_eq1})-(\ref{suppnew_eq2}), we impose the following constraint:
\begin{equation}
\left\vert \phi _{\mathrm{A}}\right\vert ^{2}+\left\vert \phi _{\mathrm{B}}\right\vert ^{2}=w,
\label{suppnew_eq3}
\end{equation}%
which is analogous to the normalization condition of eigenstates in linear systems, 
ensuring that the amplitudes of the wave functions remain constant regardless of the wavenumber $k$ \cite{arxiv}. By numerically solving 
Eqs. (\ref{suppnew_eq1})-(\ref{suppnew_eq3}), we obtain both the generally complex-valued
$\phi _{\mathrm{A,B}}$ and the real-valued normalized frequency $\bar{\omega}$.

To provide better insight, we also present some analytical results. In the linear regime with $g=0$, the eigenvalue problem has been solved, 
with Fig.~\ref{fig_band_linear}(c) illustrating the band structure and Eq.~(\ref{omega_GP}) giving the expression for $\bar{\omega}$. 
The corresponding eigenvectors are%
\begin{equation}
\left( \phi _{\mathrm{A}},\phi _{\mathrm{B}}\right) =\frac{1}{\sqrt{2}}\left( 1,\frac{\bar{%
\omega}-E_{0}}{J_{1}+J_{2}e^{-\mathrm{i}k}}\right) .
\end{equation}%
For simplicity, we define $J_{1}+J_{2}e^{\pm \mathrm{i}k} =J_{0}e^{\pm \mathrm{i}\theta }$,
where $J_{0}=\left\vert J_{1}+J_{2}e^{\pm \mathrm{i}k}\right\vert$ and $\theta =\arg \left(
J_{1}+J_{2}e^{\mathrm{i}k}\right) $. Then the eigenvalues and eigenvectors can be
written as%
\begin{eqnarray}
\bar{\omega} &=&E_{0}\pm J_{0}, \\
\left( \phi _{\mathrm{A}},\phi _{\mathrm{B}}\right) &=&\frac{1}{\sqrt{2}}\left( 1,\pm
e^{\mathrm{i}\theta }\right) .
\end{eqnarray}
In the nonlinear regime, using the same definitions of $J_{0}$ and $\theta $, Eqs. (\ref{suppnew_eq1})-(\ref{suppnew_eq3}) can be
rewritten as%
\begin{eqnarray}
\left[ \bar{\omega}-E_{0}-g\left( \phi _{\mathrm{A}}\right) \right] \phi _{\mathrm{A}}
&=&J_{0}e^{-\mathrm{i}\theta }\phi _{\mathrm{B}},  \label{suppnew_eq4} \\
\left[ \bar{\omega}-E_{0}-g\left( \phi _{\mathrm{B}}\right) \right] \phi _{\mathrm{B}}
&=&J_{0}e^{\mathrm{i}\theta }\phi _{\mathrm{A}},  \label{suppnew_eq5} \\
\left\vert \phi _{\mathrm{A}}\right\vert ^{2}+\left\vert \phi _{\mathrm{B}}\right\vert ^{2}
&=&w.  \label{suppnew_eq6}
\end{eqnarray}%
Multiplying the left-hand side of Eq.~(\ref{suppnew_eq4}) by the right-hand side of Eq.~(\ref{suppnew_eq5}), and vice versa, one obtains
\begin{equation}
\left[ \bar{\omega}-E_{0}-g\left( \phi _{\mathrm{A}}\right) \right] \phi _{\mathrm{A}}^{2}=%
\left[ \bar{\omega}-E_{0}-g\left( \phi _{\mathrm{B}}\right) \right] e^{-2\mathrm{i} \theta
}\phi _{\mathrm{B}}^{2}.  \label{suppnew_eq7}
\end{equation}%
Note that this equation holds because $J_{0}e^{\mathrm{i}\theta }$ is generally nonzero. 

To match the arguments on both sides of Eq. (\ref{suppnew_eq7}), the wave functions $\phi _{\mathrm{A,B}}$ can be expressed as
\begin{eqnarray}
\phi _{\mathrm{A}} &=&r_{\mathrm{A}}e^{\mathrm{i}\phi },  \label{suppnew_eq8} \\
\phi _{\mathrm{B}} &=&\pm r_{\mathrm{B}}e^{\mathrm{i}\left( \phi +\theta \right) },  \label{suppnew_eq9}
\end{eqnarray}%
where $r_{\mathrm{A}}$ and $r_{\mathrm{B}}$ are both real and positive. With
these definitions, Eqs. (\ref{suppnew_eq8})-(\ref{suppnew_eq9}) are further reduced to%
\begin{eqnarray}
\left[ \bar{\omega}-E_{0}-g\left( r_{\mathrm{A}}\right) \right] r_{\mathrm{A}} &=&\pm
J_{0}r_{\mathrm{B}},  \label{suppnew_eq10} \\
\left[ \bar{\omega}-E_{0}-g\left( r_{\mathrm{B}}\right) \right] r_{\mathrm{B}} &=&\pm
J_{0}r_{\mathrm{A}},  \label{suppnew_eq11} \\
r_{\mathrm{A}}^{2}+r_{\mathrm{B}}^{2} &=&w.  \label{suppnew_eq12}
\end{eqnarray}%
Meanwhile, substituting Eqs. (\ref{suppnew_eq8})-(\ref{suppnew_eq9}) into Eq. (\ref{suppnew_eq7}) yields
\begin{equation}
\left[ \bar{\omega}-E_{0}-g\left( r_{\mathrm{A}}\right) \right] r_{\mathrm{A}}^{2}=\left[ \bar{%
\omega}-E_{0}-g\left( r_{\mathrm{B}}\right) \right] r_{\mathrm{B}}^{2},  \label{suppnew_eq13}
\end{equation}%
which admits two types of solutions: $r_{\mathrm{A}}=r_{\mathrm{B}}$ and $r_{\mathrm{A}}\neq
r_{\mathrm{B}}$, respectively. For the case $r_{\mathrm{A}}=r_{\mathrm{B}}$, considering Eqs. (\ref{suppnew_eq10})-(\ref%
{suppnew_eq12}) leads to
\begin{eqnarray}
r_{\mathrm{A}} &=&r_{\mathrm{B}}=\sqrt{\frac{w}{2}}, \\
\bar{\omega} &=&E_{0}+g\left( \sqrt{\frac{w}{2}}\right) \pm J_{0}.
\end{eqnarray}
Thus, the corresponding eigenvalues and eigenvectors are%
\begin{eqnarray}
\bar{\omega} &=&E_{0}+g\left( \sqrt{\frac{w}{2}}\right) \pm J_{0},  \label{suppnew_omega_bar} \\
\left( \psi _{\mathrm{A}},\psi _{\mathrm{B}}\right)  &=&\sqrt{\frac{w}{2}}\left( 1,\pm
e^{i\theta }\right) . \label{suppnew_psiAB}
\end{eqnarray}%
For the case $r_{\mathrm{A}}\neq r_{\mathrm{B}}$, Eq. (\ref{suppnew_eq13}) yields an alternative expression for $\bar{\omega}$:
\begin{equation}
\bar{\omega}=E_{0}+\frac{g\left( r_{\mathrm{A}}\right) r_{\mathrm{B}}^{2}-g\left( r_{\mathrm{B}}\right)
r_{\mathrm{A}}^{2}}{r_{\mathrm{A}}^{2}-r_{\mathrm{B}}^{2}}+g\left( r_{\mathrm{A}}\right) +g\left( r_{\mathrm{B}}\right) .
\label{suppnew_omega_2}
\end{equation}%
However, obtaining explicit analytical forms for $r_{\mathrm{A}}$ and $r_{\mathrm{B}}$ is challenging. 
After numerically solving Eqs.~(\ref{suppnew_eq1})-(\ref{suppnew_eq3}), we find that solutions with $r_{\mathrm{A}}\neq r_{\mathrm{B}}$
are excluded due to the saturable nonlinearity inherent in our model.

\begin{figure}[tbp]
\includegraphics[width=8.6cm]{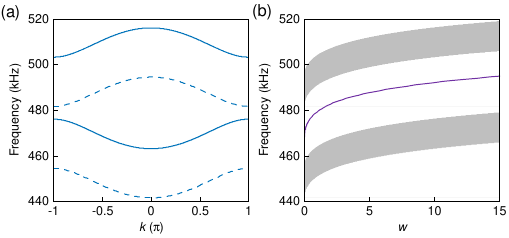}
\caption{Band structure in the nonlinear regime. 
(a) Band structure calculated at $w = 10$ (solid curves). For comparison, the band structure in the linear regime with $g = 0$ is shown as dashed curves. 
(b) Frequencies of the nonlinear topological edge states for various values of $w$. The gray shaded areas represent the corresponding nonlinear bulk bands.}
\label{fig_berry}
\end{figure}

We focus on the SSH lattice illustrated in Fig. \ref{fig_findings}(a), replacing its boundary condition with a periodic boundary. 
The intracell and intercell coupling capacitors are maintained as $C_{1} = 180~\mathrm{pF}$ and $C_{2} = 560~\mathrm{pF}$, respectively. 
Figure \ref{fig_berry}(a) shows the band structure for $w =10$. Compared to the linear band structure presented in Fig.~\ref{fig_band_linear}(c), 
the band structure in the nonlinear regime (denoted by the solid curves) displays a vertical shift, which is consistent with the prediction of Eq.~(\ref{suppnew_omega_bar}). 
For reference, the linear band structure with $g = 0$ is represented by dashed curves.
However, differing from the results reported in Ref. \cite{arxiv}, as $w$ increases, the energy (frequency) shift of our nonlinear band structure saturates, 
and no nonlinearity-induced flat bands corresponding to Eq. (\ref{suppnew_omega_2}) emerge. These discrepancies arise because the nonlinear terms in 
Eqs. (\ref{suppnew_eq1}) to (\ref{suppnew_eq2}) describe saturable nonlinearity. We have verified that if the saturable nonlinearity is replaced by a nonsaturable form, 
where the capacitance of the varactor diodes is modeled according to Eq. (\ref{C_approximate}), additional nonlinearity-induced flat bands indeed appear.

Following the definitions of the time-reversal and space-inversion operators introduced in Ref. \cite{arxiv}, we find that our nonlinear SSH model, 
as given by Eqs. (\ref{suppnew_eq1})–(\ref{suppnew_eq2}), also respects both time-reversal and space-inversion symmetries. Similarly, 
using the eigenvectors provided in Eq. (\ref{suppnew_psiAB}), the nonlinear Berry phase defined by
\begin{equation}
v(w) = -\frac{i}{w} \int_{-\pi}^{\pi} \left\langle u_{k} \left\vert \frac{\partial u_{k}}{\partial k} \right. \right\rangle \, dk \label{suppnew_berry}
\end{equation}
equals $\pi$ for any given value of $w$. Here, $\vert u_{k} \rangle = (\phi_{\mathrm{A}}, \phi_{\mathrm{B}})^{T}$ denotes the Bloch state, 
and $T$ is the transpose operator. According to the bulk-boundary correspondence formulated in Ref. \cite{arxiv}, a nonzero Berry phase corresponds to 
the existence of localized gapless modes satisfying $\vert v_{1}^{\mathrm{A}} \vert^{2} + \vert v_{1}^{\mathrm{B}} \vert^{2} = w$. As shown in Fig. \ref{fig_berry}(b), 
the nonlinear topological edge states we identified (denoted by the solid curve) consistently lie within the mid-gap described by Eq. (\ref{suppnew_psiAB}). 
The gray shaded areas represent the corresponding nonlinear bulk bands.
This implies that the nonlinear topological edge states presented in Figs. \ref{fig2}(a)–(c) can be fully characterized within the theoretical framework proposed in Ref. \cite{arxiv}.

By exchanging the values of the intracell and intercell couplings, we find that the nonlinear band structure remains identical to that shown in Fig.~\ref{fig_berry}(a). 
However, the nonlinear Berry phase defined in Eq.~(\ref{suppnew_berry}) becomes zero, indicating the absence of localized gapless modes. When two SSH lattices 
with $w = \pi$ and $w = 0$ are connected, a domain wall is formed, analogous to the linear regime. Consequently, the emergence of the nonlinear topological edge states 
observed in Figs.~\ref{fig2}(d)-(e) can also be explained within this framework.


\begin{thebibliography}{99}
\bibitem{RMP82-3045} M. Z. Hasan and C. L. Kane, Colloquium: Topological insulators, Rev. Mod. Phys. 82, 3045 (2010).

\bibitem{RMP83-1057} X.-L. Qi and S.-C. Zhang, Topological insulators and superconductors, Rev. Mod. Phys. 83, 1057 (2011).

\bibitem{QF2-22} G.-Q. Zhao, S. Li, W. B. Rui, C. M. Wang, H.-Z. Lu, and X. C. Xie, 3D quantum Hall effect in a topological 
nodal-ring semimetal, Quantum Front. 2, 22 (2023).


\bibitem{NRP4-184} O. Breunig and Y. Ando, Opportunities in topological insulator devices, Nat. Rev. Phys. 4, 184 (2022).

\bibitem{QF3-21} F. Zhan, R. Chen, Z. Ning, D.-S. Ma, Z. Wang, D.-H. Xu, and R. Wang, Perspective: Floquet engineering 
topological states from effective models towards realistic materials, Quantum Front. 3, 21(2024).


\bibitem{NRP1-281} G. Ma, M. Xiao, and C. T. Chan, Topological phases in acoustic and mechanical systems, Nat. Rev. Phys. 1, 281 (2019).

\bibitem{NRM7-974} H. Xue, Y. Yang, and B. Zhang, Topological acoustics, Nat. Rev. Mater. 7, 974 (2022).

\bibitem{RMP91-015005} N. R. Cooper, J. Dalibard, and I. B. Spielman, Topological bands for ultracold atoms, Rev. Mod. Phys. 91, 015005 (2019).

\bibitem{NRP5-483} Z.-K. Lin, Q. Wang, Y. Liu, H. Xue, B. Zhang, Y. Chong, and J.-H. Jiang, Topological phenomena at defects in acoustic, 
photonic and solid-state lattices, Nat. Rev. Phys. 5, 483 (2023).

\bibitem{NRP3-520} B. Xie, H.-X. Wang, X. Zhang, P. Zhan, J.-H. Jiang, M. Lu, and Y. Chen, Higher-order band topology, Nat. Rev. Phys. 3, 520 (2021).

\bibitem{RMP96-021002} T. Shah, C. Brendel, V. Peano, and F. Marquardt, Colloquium: Topologically protected transport in engineered 
mechanical systems, Rev. Mod. Phys. 96, 021002 (2024).

\bibitem{nphoton8-821} L. Lu, J. D. Joannopoulos, and M. Solja\v{c}i\'{c}, Topological photonics, Nat. Photon. 8, 821 (2014).

\bibitem{nphoton11-763} A. B. Khanikaev and G. Shvets, Two-dimensional topological photonics, Nat. Photon. 11, 763 (2017).

\bibitem{PQE55-52} X.-C. Sun, C. He, X.-P. Liu, M.-H. Lu, S.-N. Zhu, and Y.-F. Chen, Two-dimensional topological photonic systems, Progress in Quantum Electronics 55, 52 (2017).

\bibitem{RMP91-015006} T. Ozawa, H. M. Price, A. Amo, N. Goldman, M. Hafezi, L. Lu, M. C. Rechtsman, D. Schuster, J. Simon, O. Zilberberg, and I. Carusotto, Topological Photonics, Rev. Mod. Phys. 91, 015006 (2019).

\bibitem{LSA9-1} M. Kim, Z. Jacob, and J. Rho, Recent advances in 2D, 3D and higher-order topological photonics, Light: Science \& Applications 9, 1 (2020).

\bibitem{PR1093-1} H. Yang, L. Song, Y. Cao, and P. Yan, Circuit realization of topological physics, Phys. Rep. 1093, 1 (2024).

\bibitem{arXiv:2502.18563} H. Sahin, M. B. A. Jalil, and C. H. Lee, Topolectrical Circuits $-$ Recent Experimental Advances and Developments, 
APL Electronic Devices 1, 021503 (2025).

\bibitem{QF1-10} M. Yang, J.-S. Xu, C.-F. Li, and G.-C. Guo, Simulating topological materials with photonic synthetic dimensions in 
cavities, Quantum Front. 1, 10 (2022).

\bibitem{QF3-26} Y. Liu, K. Li, W. Liu, Z. Zhang, Y. Cheng, and X. Liu, Observation of chiral Landau levels in two-dimensional 
acoustic system, Quantum Front. 3, 26 (2024).

\bibitem{APR7-021306} D. Smirnova, D. Leykam, Y. Chong, and Y. Kivshar, Nonlinear topological photonics, Appl. Phys. Rev. 7, 021306 (2020).

\bibitem{NP20-905} A. Szameit and M. C. Rechtsman, Discrete nonlinear topological photonics, Nat. Phys. 20, 905 (2024).

\bibitem{PRA90-023813} M. J. Ablowitz, C. W. Curtis, and Y.-P. Ma, Linear and nonlinear traveling edge waves in optical 
honeycomb lattices, Phys. Rev. A 90, 023813 (2014).

\bibitem{PRA94-021801} Y. Lumer, M. C. Rechtsman, Y. Plotnik, and M. Segev, Instability of bosonic topological edge states in the presence 
of interactions, Phys. Rev. A 94, 021801 (2016).

\bibitem{optica3-1228} Y. V. Kartashov and D. V. Skryabin, Modulational instability and solitary waves in polariton topological insulators, 
Optica 3, 1228 (2016).

\bibitem{PRL119-253904} Y. V. Kartashov and D. V. Skryabin, Bistable Topological Insulator with Exciton-Polaritons, 
Phys. Rev. Lett. 119, 253904 (2017).

\bibitem{PRL121-163901} D. A. Dobrykh, A. V. Yulin, A. P. Slobozhanyuk, A. N. Poddubny, and Yu. S. Kivshar, Nonlinear Control of 
Electromagnetic Topological Edge States, Phys. Rev. Lett. 121, 163901 (2018).

\bibitem{PRB102-115411} T. Tuloup, R. W. Bomantara, C. H. Lee, and J. Gong, Nonlinearity induced topological physics in 
momentum space and real space, Phys. Rev. B 102, 115411 (2020).

\bibitem{OL45-6466} M. Guo, S. Xia, N. Wang, D. Song, Z. Chen, and J. Yang, Weakly nonlinear topological gap solitons in 
Su-Schrieffer-Heeger photonic lattices, Opt. Lett. 45, 6466 (2020).

\bibitem{nphys18-678} N. Pernet, P. St-Jean, D. D. Solnyshkov, G. Malpuech, N. C. Zambon, Q. Fontaine, B. Real, O. Jamadi, A. Lemaître, 
M. Morassi, L. L. Gratiet, T. Baptiste, A. Harouri, I. Sagnes, A. Amo, S. Ravets, and J, Bloch, Gap solitons in a one-dimensional
driven-dissipative topological lattice, Nat. Phys. 18, 678 (2022).

\bibitem{PRL128-093901} Y. V. Kartashov et al., Observation of Edge Solitons in Topological Trimer Arrays, Phys. Rev. Lett. 128, 093901 (2022).

\bibitem{PRE104-054206} Y.-P. Ma and H. Susanto, Topological edge solitons and their stability in a nonlinear 
Su-Schrieffer-Heeger model, Phys. Rev. E 104, 054206 (2021).

\bibitem{PRB104-235420} M. Ezawa, Nonlinearity-induced transition in the nonlinear Su-Schrieffer-Heeger model and 
a nonlinear higher-order topological system, Phys. Rev. B 104, 235420 (2021).

\bibitem{LSA9-147} S. Xia, D. Juki\'{c}, N. Wang, D. Smirnova, L. Smirnov, L. Tang, D. Song, A. Szameit, D. Leykam, J. Xu, Z. Chen, 
and H. Buljan, Nontrivial coupling of light into a defect: the interplay of nonlinearity and topology, Light Sci. Appl. 9, 147 (2020).

\bibitem{science372-72} S. Xia, D. Kaltsas, D. Song, I. Komis, J. Xu, A. Szameit, H. Buljan, K. G. Makris, and Z. Chen, Nonlinear tuning of PT symmetry and non-Hermitian topological states, Science 372, 72 (2021).

\bibitem{CP8-342} R. Li, X. Kong, W. Wang, Y. Wang, Y. Jia, H. Tao, P. Li, Y. Liu, and B. A. Malomed, Observation of edge solitons and transitions 
between them in a trimer circuit lattice, Commun. Phys. 8, 342 (2025).


\bibitem{PRL117-143901} D. Leykam and Y. D. Chong, Edge Solitons in Nonlinear-Photonic Topological Insulators, Phys. Rev. Lett. 117, 143901 (2016).

\bibitem{ACSPhoton7-735} S. K. Ivanov, Y. V. Kartashov, A. Szameit, L. Torner, and V. V. Konotop, Vector Topological Edge Solitons in Floquet Insulators, ACS Photonics 7, 735 (2020).

\bibitem{ncommun11-1902} Z. Zhang, R. Wang, Y. Zhang, Y. V. Kartashov, F. Li, H. Zhong, H. Guan, K. Gao, F. Li, 
Y. Zhang, and M. Xiao, Observation of edge solitons in photonic graphene, Nat. Commun. 11, 1902 (2020).

\bibitem{PRX11-041057} S. Mukherjee and M. C. Rechtsman, Observation of Unidirectional Solitonlike Edge States in Nonlinear Floquet Topological Insulators, Phys. Rev. X 11, 041057 (2021).

\bibitem{PRA103-053507} S. K. Ivanov, Y. V. Kartashov, M. Heinrich, A. Szameit, L. Torner, and V. V. Konotop, Topological dipole Floquet solitons, Phys. Rev. A 103, 053507 (2021).

\bibitem{ACSPhoton8-1077} Z. Shi, M. Zuo, H. Li, D. Preece, Y. Zhang, and Z. Chen, Topological Edge States and Solitons on a Dynamically Tunable Domain Wall of Two Opposing Helical Waveguide Arrays, ACS Photonics 8, 1077 (2021).

\bibitem{PRB106-195423} M. Ezawa, Nonlinearity-induced chiral solitonlike edge states in Chern systems, Phys. Rev. B 106, 195423 (2022).

\bibitem{nphys17-995} M. S. Kirsch, Y. Zhang, M. Kremer, L. J. Maczewsky, S. K. Ivanov, Y. V. Kartashov, L. Torner, D. Bauer, A. Szameit, and M.
Heinrich, Nonlinear second-order photonic topological insulators, Nat. Phys. 17, 995 (2021).

\bibitem{LSA10-164} Z. Hu, D. Bongiovanni, D. Juki\'{c}, E. Jajti\'{c}, S. Xia, D. Song, J. Xu, R. Morandotti, H. Buljan, and Z. Chen, Nonlinear
control of photonic higher-order topological bound states in the continuum, Light Sci. Appl. 10, 164 (2021).

\bibitem{CSF207-118044} R. Li, W. Wang, Y. Jia, Y. Liu, P. Li, and B. A. Malomed, Nonlinear quadrupole topological insulators, 
Chaos, Solitons \& Fractals 207, 118044 (2026).


\bibitem{PRL111-243905} Y. Lumer, Y. Plotnik, M. C. Rechtsman, and M. Segev, Self-Localized States in Photonic Topological Insulators, 
Phys. Rev. Lett. 111, 243905 (2013).

\bibitem{PRL118-023901} D. D. Solnyshkov, O. Bleu, B. Teklu, and G. Malpuech, Chirality of Topological Gap Solitons in Bosonic 
Dimer Chains, Phys. Rev. Lett. 118, 023901 (2017).

\bibitem{PRA98-013827} A. N. Poddubny and D. A. Smirnova, Ring Dirac solitons in nonlinear topological systems, Phys. Rev. A 98, 013827 (2018).

\bibitem{LPR13-1900223} D. A. Smirnova, L. A. Smirnov, D. Leykam, and Y. S. Kivshar, Topological Edge States and Gap Solitons 
in the Nonlinear Dirac Model, Laser \& Photonics Reviews 13, 1900223 (2019).

\bibitem{arxiv1904-10312} J. L. Marzuola, M. Rechtsman, B. Osting, and M. Bandres, Bulk soliton dynamics in bosonic topological insulators, 
arXiv:1904.10312  (2019).

\bibitem{science368-856} S. Mukherjee and M. C. Rechtsman, Observation of Floquet solitons in a topological bandgap, Science 368, 856 (2020).

\bibitem{CP5-275} R. Li, X. Kong, D. Hang, G. Li, H. Hu, H. Zhou, Y. Jia, P. Li, and Y. Liu, Topological bulk solitons in a nonlinear photonic Chern insulator, Commun. Phys. 5, 275 (2022).

\bibitem{nanophotonics14-769} C. J\"{o}rg, M. J\"{u}rgensen, S. Mukherjee, and M. C. Rechtsman, Optical control of topological end states via soliton formation in a 1D lattice, Nanophotonics 14, 769 (2025).

\bibitem{RMP83-247} Y. V. Kartashov, B. A. Malomed, and L. Torner, Solitons in nonlinear lattices, Rev. Mod. Phys. 83, 247 (2011).

\bibitem{PR463-1} F. Lederer, G. I. Stegeman, D. N. Christodoulides, G. Assanto, M. Segev, and Y. Silberberg, Discrete solitons in 
optics, Phys. Rep. 463, 1 (2008).

\bibitem{QF4-9} L. Wang, Z. Yan, Y. Zhu, and J. Zeng, Gap solitons and vortices in two-dimensional spin-orbit-coupled Bose-Einstein 
condensates loaded onto moiré optical lattices, Quantum. Front. 4, 9 (2025).

\bibitem{PRL42-1698} W. P. Su, J. R. Schrieffer, and A. J. Heeger, Solitons in Polyacetylene, Phys. Rev. Lett. 42, 1698 (1979).

\bibitem{nelectron1-178} Y. Hadad, J. C. Soric, A. B. Khanikaev, and A. Al\`{u}, Self-induced topological protection in nonlinear circuit arrays, 
Nat. Electron. 1, 178 (2018).

\bibitem{ncommun10-1102} Y. Wang, L.-J. Lang, C. H. Lee, B. Zhang, and Y. D. Chong, Topologically enhanced harmonic generation in a nonlinear transmission line metamaterial, Nat. Commun. 10, 1102 (2019).

\bibitem{PRL123- 053902} F. Zangeneh-Nejad and R. Fleury, Nonlinear Second-Order Topological Insulators, Phys. Rev. Lett. 123, 053902 (2019).

\bibitem{PNAS118-e2106411118} T. Kotwal, F. Moseley, A. Stegmaier, S. Imhof, H. Brand, T. Kie\ss ling, R. Thomale, H. Ronellenfitsch, and J. Dunkel, Active topolectrical circuits, Proc. Natl. Acad. Sci. U.S.A. 118, e2106411118 (2021).

\bibitem{PRResearch5-L012041} H. Hohmann, T. Hofmann, T. Helbig, S. Imhof, H. Brand,
L. K. Upreti, A. Stegmaier, A. Fritzsche, T. M\"{u}ller, U. Schwingenschl\"{o}gl, C. H. Lee,
M. Greiter, L. W. Molenkamp, T. Kie\ss ling, and R.Thomale, Observation of Cnoidal Wave
Localization in Nonlinear Topolectric Circuits, Phys. Rev. Research 5, L012041 (2023).

\bibitem{arXiv:2411.07522} H. Sahin, H. Akg\"{u}n, Z. B. Siu, S. M. Rafi-Ul-Islam, J. F. Kong, M. B. A. Jalil, and C. H. Lee, 
Protected Chaos in a Topological Lattice, Advanced Science 12, e03216 (2025).

\bibitem{PRL127-184101} D. Bongiovanni, D. Juki\'{c}, Z. Hu, F. Luni\'{c}, Y. Hu, D. Song, R. Morandotti, Z. Chen, and H. Buljan, Dynamically Emerging Topological Phase Transitions in Nonlinear Interacting Soliton Lattices, Phys. Rev. Lett. 127, 184101 (2021).

\bibitem{PRB84-195452} P. Delplace, D. Ullmo, and G. Montambaux, Zak phase and the existence of edge states in graphene, Phys. Rev. B 84, 195452 (2011).

\bibitem{ncommun6-6710} C. Poli, M. Bellec, U. Kuhl, F. Mortessagne, and H. Schomerus, Selective enhancement of topologically induced interface states in a dielectric resonator chain, Nat. Commun. 6, 6710 (2015).

\bibitem{ncommun3-882} T. Kitagawa, M. A. Broome, A. Fedrizzi, M. S. Rudner, E. Berg, I. Kassal, A. Aspuru-Guzik, E. Demler, and A. G. White, Observation of topologically protected bound states in photonic quantum walks, Nat. Commun. 3, 882 (2012).

\bibitem{RMP78-179} O. Morsch and M. Oberthaler, Dynamics of Bose-Einstein condensates in optical lattices, Rev. Mod. Phys. 78, 179 (2006).

\bibitem{NRP1-185} Y. V. Kartashov, G. E. Astrakharchik, B. A. Malomed, and L. Torner, Frontiers in multidimensional self-trapping of nonlinear fields 
and matter, Nat. Rev. Phys. 1, 185 (2019).

\bibitem{book1} J. K. Asb\'{o}th, L. Oroszl\'{a}ny, and A. P\'{a}lyi, A Short Course on Topological Insulators, 
Vol. 919 (Springer International Publishing, Cham, 2016).

\bibitem{PRB93-15512} Y. Hadad, A. B. Khanikaev, and A. Al\`{u}, Self-induced topological transitions and edge states supported by 
nonlinear staggered potentials, Phys. Rev. B 93, 155112 (2016).

\bibitem{ncommun13-3379} D. Zhou, D. Z. Rocklin, M. Leamy, and Y. Yao, Topological invariant and anomalous edge modes of strongly 
nonlinear systems, Nat. Commun. 13, 3379 (2022).

\bibitem{FP18-33311} J. Tang, F. Ma, F. Li, H. Guo, and D. Zhou, Strongly Nonlinear Topological Phases of Cascaded Topoelectrical Circuits, 
Front. Phys. 18, 33311 (2023).

\bibitem{ncommun16-422} K. Sone, M. Ezawa, Z. Gong, T. Sawada, N. Yoshioka, and T. Sagawa, Transition from the topological to the chaotic in the nonlinear Su-Schrieffer-Heeger model, Nat. Commun. 16, 422 (2025).

\bibitem{AP356-383} T. A. Loring, $K$-theory and pseudospectra for topological insulators, Annals of Physics 356, 383 (2015).

\bibitem{nano11-4765} A. Cerjan and T. A. Loring, An operator-based approach to topological photonics, Nanophotonics 11, 4765 (2022).

\bibitem{ncommun14-3071} W. Cheng, A. Cerjan, S.-Y. Chen, E. Prodan, T. A. Loring, and C. Prodan, Revealing topology in metals using experimental protocols inspired by $K$-theory, Nat. Commun. 14, 3071 (2023).

\bibitem{PRL133-116602} K. Bai, J.-Z. Li, T.-R. Liu, L. Fang, D. Wan, and M. Xiao, Arbitrarily Configurable Nonlinear Topological Modes, 
Phys. Rev. Lett. 133, 116602 (2024).

\bibitem{nphys20-1164} K. Sone, M. Ezawa, Y. Ashida, N. Yoshioka, and T. Sagawa, Nonlinearity-induced topological phase transition 
characterized by the nonlinear Chern number, Nat. Phys. 20, 1164 (2024).

\bibitem{ncommun13-2392} W. Zhang, H. Yuan, H. Wang, F. Di, N. Sun, X. Zheng, H. Sun, and X. Zhang, Observation of Bloch oscillations dominated by effective anyonic particle statistics, Nat. Commun. 13, 2392 (2022).

\bibitem{arxiv} K. Sone and Y. Hatsugai, Topological-to-topological transition induced by on-site nonlinearity in a one-dimensional topological insulator, Phys. Rev. Research 8, L012045 (2026).

\bibitem{PRL122-247702} T. Hofmann, T. Helbig, C. H. Lee, M. Greiter, and R. Thomale, Chiral Voltage Propagation and Calibration in a Topolectrical Chern Circuit, Phys. Rev. Lett. 122, 247702 (2019).

\bibitem{IEEE40-1511} Wonjoo Suh, Zheng Wang, and Shanhui Fan, Temporal coupled-mode theory and the presence of non-orthogonal modes in lossless multimode cavities, IEEE J. Quantum Electron. 40, 1511 (2004).

\bibitem{ncommun11-1436} N. A. Olekhno, E. I. Kretov, A. A. Stepanenko, P. A. Ivanova, V. V. Yaroshenko, E. M. Puhtina, D. S. Filonov, B. Cappello, L. Matekovits, and M. A. Gorlach, Topological edge states of interacting photon pairs emulated in a topolectrical circuit, Nat. Commun. 11, 1436 (2020).

\bibitem{PR129-959} H. A. Gersch and G. C. Knollman, Quantum Cell Model for Bosons, Phys. Rev. 129, 959 (1963).

\bibitem{PRL80-2189} L. Amico and V. Penna, Dynamical Mean Field Theory of the Bose-Hubbard Model, Phys. Rev. Lett. 80, 2189 (1998).

\bibitem{nphys8-267} I. Bloch, J. Dalibard, and S. Nascimbène, Quantum simulations with ultracold quantum gases, Nat. Phys. 8, 267 (2012).

\bibitem{RMP82-1225} C. Chin, R. Grimm, P. Julienne, and E. Tiesinga, Feshbach resonances in ultracold gases, Rev. Mod. Phys. 82, 1225 (2010).

\bibitem{NSR8-nwaa192} R. Li, B. Lv, H. Tao, J. Shi, Y. Chong, B. Zhang, and H. Chen, Ideal type-II Weyl points in topological circuits, National Science Review 8, nwaa192 (2021).

\bibitem{PRL129-135501} G. Liu, J. Noh, J. Zhao, and G. Bahl, Self-Induced Dirac Boundary State and Digitization in a Nonlinear Resonator Chain, Phys. Rev. Lett. 129, 135501 (2022).

\end{thebibliography}
\end{document}